%% file: ms.tex
\documentclass[modern]{aastex631}
\usepackage{showyourwork}

\include{style}

\def\timeInferY{under 1 second\xspace}
\def\timeInferYB{about 1 second\xspace}
\def\timeInferYBS{about 30 seconds\xspace}
\def\timeInferYBSLOSNR{about 30 seconds\xspace}
\def\timeInferTwoSpec{about 20 minutes\xspace}

\begin{document}

\title{%
    Mapping stellar surfaces\\
    III: An Efficient, Scalable, and Open-Source Doppler Imaging Model
}

\author[0000-0002-0296-3826]{Rodrigo Luger}
\email{rluger@flatironinstitute.org}
\affil{Center~for~Computational~Astrophysics, Flatiron~Institute, New~York, NY}
\author[0000-0001-9907-7742]{Megan Bedell}
\affil{Center~for~Computational~Astrophysics, Flatiron~Institute, New~York, NY}
\author[0000-0002-9328-5652]{Daniel Foreman-Mackey}
\affil{Center~for~Computational~Astrophysics, Flatiron~Institute, New~York, NY}
\author[0000-0002-1835-1891]{Ian J. M. Crossfield}
\affil{Department~of~Physics~and~Astronomy, Kansas University, Lawrence, KS}
\author[0000-0002-3852-3590]{Lily L. Zhao}
\affil{Center~for~Computational~Astrophysics, Flatiron~Institute, New~York, NY}
\author[0000-0003-2866-9403]{David W. Hogg}
\affil{Center~for~Computational~Astrophysics, Flatiron~Institute, New~York, NY}

\begin{abstract}
    The study of stellar surfaces can reveal information about the chemical composition, interior structure, and magnetic properties of stars.
    It is also critical to the detection and characterization of extrasolar planets, in particular those targeted in extreme precision radial velocity (EPRV) searches, which must contend with stellar variability that is often orders of magnitude stronger than the planetary signal.
    One of the most successful methods to map the surfaces of stars is Doppler imaging, in which the presence of inhomogeneities is inferred from subtle line shape changes in high resolution stellar spectra.
    In this paper, we present a novel, efficient, and closed-form solution to the problem of Doppler imaging of stellar surfaces.
    Our model explicitly allows for incomplete knowledge of the local (rest frame) stellar spectrum, allowing one to learn differences from spectral templates while simultaneously mapping the stellar surface. 
    It therefore works on blended lines, regions of the spectrum where line formation mechanisms are not well understood, or stars whose spots have intrinsically different spectra from the rest of the photosphere.
    We implement the model within the open source \starry stellar modeling framework, making it fast, differentiable, and easy to use in both optimization and posterior inference settings.
    As a proof-of-concept, we use our model to infer the surface map of the brown dwarf WISE 1049-5319B, finding close agreement with the solution of \citet{Crossfield2014}.
    We also discuss Doppler imaging in the context of EPRV studies and describe an interpretable spectral-temporal Gaussian process for stellar spectral variability that we expect will be important for EPRV exoplanet searches.
    This paper was prepared using the \showyourwork scientific article workflow, which exposes the data, source code, and package dependencies for all of the paper's figures, allowing readers to reproduce the results in this paper exactly and with minimal effort.
\end{abstract}

\section{Introduction}

High resolution spectra encode a vast amount of information about the structure, dynamics, and chemical composition of stars.
Spectra of fast-rotating stars, in particular, also encode the two-dimensional brightness map of the photosphere via subtle line shape changes as inhomogeneities rotate from the blueshifted side to the redshifted side of the star.
The process of inferring the map of a stellar ``surface'' given high resolution spectra taken at different rotational phases is known as \emph{Doppler imaging}, a technique that dates back to at least \citet{Deutsch1958}.
Most modern Doppler imaging studies, however, are based on the methodology developed in \citet{Khokhlova1976}, \citet{Goncharskii1977}, and \citet{Goncharskii1982}, and updated in \citet{Vogt1983} and \citet{Vogt1987}. These studies formalised the technique of forward-modeling the line shape changes due to arbitrary brightness variations rotating in and out of view. The latter study, in particular, presented a now common technique for solving the inverse problem---inferring the map of the star given a spectroscopic dataset---under certain regularizing assumptions.

Since those pioneering studies several decades ago, Doppler maps have now been obtained for a variety of targets, ranging from giant and subgiant stars \citep{Donati1999,Strassmeier1999,Roettenbacher2017} to massive main sequence and pre-main sequence stars \citep{Collier1994,Hatzes1995} to cool dwarfs \citep{Rosen2015,Zaire2021}. A Doppler map has even been obtained for a brown dwarf in the solar neighborhood \citep{Crossfield2014}. Future observations with extremely large telescopes may also allow one to obtain Doppler maps of directly imaged extrasolar planets \citep{Snellen2014,Crossfield2014b}.

As more and more objects are imaged in this way, it becomes tempting to interpret Doppler images in the broader context of stellar physics and evolution.
For instance, one may be interested in understanding trends in the population-level magnetic properties of giant stars, or in how those properties change as a function of spectral type or stellar age.
However, the vast majority of Doppler imaging studies produce \emph{point estimates}: that is, they usually report the maximum likelihood solution for the stellar surface map, with no formal assessment of its uncertainty.
In all but the simplest cases, maximum likelihood estimators are biased \citep[e.g.,][]{Lehman1998}, meaning any attempt to understand these estimates in bulk is prone to significant bias \citep[see, e.g.,][]{Hogg2010}.
This problem is compounded by four factors.

\textbf{First}, Doppler imaging is usually an ill-posed problem, meaning there is no unique solution for the surface map given a spectroscopic dataset.
This issue is typically well understood \citep[e.g.,][]{Piskunov1990,Rice2002}, and most studies circumvent this issue by employing restrictive priors to ensure the uniqueness of the solution. 
These include maximum entropy regularization \citep{Vogt1987} and Tikhonov regularization \citep[e.g.,][]{Piskunov1990}, which typically enforce sparsity and smoothness in the solution, respectively.
However, while useful for obtaining a maximum likelihood solution, both choices can lead to significant bias in the inferred map, as they can both suppress features and introduce spurious artifacts.
In fact, a recent comparison of Doppler imaging, photometric light curve inversion, and interferometry revealed qualitatively diffferent maps of the the same star \citep{Roettenbacher2017}, due to the various degeneracies afflicting each method.
\textbf{Second}, the Doppler imaging technique is generally more sensitive to longitudinal than latitudinal variations, often leading to spurious, latititudinally elongated features in the maximum likelihood solution \citep[e.g.,][]{Unruh1995,Crossfield2014b}.
\textbf{Third}, the Doppler imaging technique requires the availability of accurate spectral templates for the local rest frame spectrum of the star.
Several studies have shown that errors in line templates or the presence of unresolved line blends in the observed spectrum can introduce signficant artifacts in the map solution \citep[e.g.,][]{Rice1989,Unruh1995}.
Moreover, changes in the local spectrum over the surface of the star (such as different line depths or the presence of different absorption lines within spots) can similarly bias the inferred map if these are not properly modeled.
While some past studies have solved for variations in the local line profiles \citep{Khokhlova1976,Goncharskii1977,Goncharskii1982}, these usually make the assumption that the intensity variation is small across the surface of the star; that is, they do not simultaneously solve for the Doppler map of the surface brightness variations.
\textbf{Fourth}, and finally, Doppler imaging results reported in the literature use a variety of custom data analysis codes, most of which are not open source 
\citep[][to name a few]{Collier1997,Hackman2001,Carroll2012,Rosen2015,AsensioRamos2021}; 
many studies also utilize data that is often not included in the resulting publication or made easily available online. 
This makes it difficult to reproduce results and compare Doppler images of different stars on the same footing. 
As argued above, detailed choices in a Doppler imaging analysis will have a significant effect on the reported point-estimate result, so this inability to compare uniformly analyzed observations is a significant hindrance to interpretation.

In this paper, the third in a series dedicated to mapping stellar surfaces \citep{Luger2021a,Luger2021b}, we develop a novel algorithm for Doppler imaging that addresses each of these issues.
The various pitfalls of Doppler imaging outlined above necessitate an algorithm that is efficient enough to be used in a Bayesian posterior inference context, where one can formally address the degeneracies at play and quantify their effect on the uncertainties of the inferred stellar maps.
They also necessitate an algorithm that is flexible, capable of learning deviations from the template or capturing spatial variations in the rest-frame spectrum.
And most of all, they necessitate a user-friendly, well-documented, and open source implementation, which is currently lacking the literature.
Our algorithm addresses these various issues by building on our previous work \citep{Luger2019}, where we derived analytic solutions to the light curve mapping problem by expanding the stellar surface map in the spherical harmonic basis.
While spherical harmonics are routinely used in Doppler imaging problems \citep{Deutsch1958,Deutsch1970,Falk1974,Donati2006,Kochukhov2014}, we present a new closed-form solution to the forward model for Doppler imaging when working in this basis, making our model both precise and computationally efficient.
We also expose a fundamental linearity of the model in terms of both the surface map and the rest frame spectrum, allowing us to learn both using a fast bilinear solver.
Finally, we implement our algorithm within the open source \starry code package with accompanying documentation, examples, and tutorials.\footnote{\url{https://github.com/rodluger/starry}}
All of the figures in this paper are also open source; clickable \GitHubIconBlue icons next to each figure link to the exact version of the \texttt{Python} script used to generate them.
To ensure the full reproducibility of this paper, we prepared it using the \showyourwork workflow\footnote{\url{https://github.com/rodluger/showyourwork}}, which exposes the datasets, source code, and all package dependencies needed to exactly reproduce the results.

The paper is organized as follows: in \S\ref{sec:the_problem}~and~\S\ref{sec:the_solution} we introduce the math behind the Doppler imaging problem and derive a closed-form expression for the model. 
In \S\ref{sec:linear},~\S\ref{sec:inverse},~and~\S\ref{sec:bellswhistles} we demonstrate how to re-express the model as a linear operation on the input spectrum and stellar map and derive closed-form expressions for the two conditioned on the data and priors.
In \S\ref{sec:spotstar} we apply our techniques to a mock problem, and in \S\ref{sec:luhman16b} we apply it to a real spectral dataset of the brown dwarf WISE 1049-5319B.
We further discuss our results in \S\ref{sec:discussion}, and in \S\ref{sec:gp} we focus on an immediate extension of our work: the development of an interpretable spectral-temporal Gaussian process for stellar spectral variability.
Finally, we discuss the implementation of our Doppler imaging algorithm within the \starry framework in \S\ref{sec:starry}, and present a summary of the paper in \S\ref{sec:conclusions}. 

\section{The problem}
\label{sec:the_problem}
Let $I(\lnlam, \x, t)$ be the specific intensity observed at log wavelength $\lnlam \equiv \ln\frac{\lambda}{\lambda_\mathrm{r}}$ and at sky-projected Cartesian position $\x = (x, y, z)$ on the surface of the star at time $t$, where $\lambda_\mathrm{r}$ is an arbitrary reference wavelength, such as the central wavelength in the observed spectrum.
We may express this intensity as
\begin{align}
    \label{eq:the_problem:Ixi}
    I(\lnlam, \x, t) & =
    I\Big(\lnlam_0 + \D(\x), \x, t\Big)
\end{align}
where $\lnlam_0$ is the log wavelength in the rest frame and $\D$ is the Doppler shift in log wavelength space:
\begin{align}
    \label{eq:the_problem:D}
    \D(\x)
     & =
    \frac{1}{2}\ln\left(
    \frac{1 + \beta(\x)}{1 - \beta(\x)}
    \right)
\end{align}
where $\beta = v(\x) / c$ is the ratio of the radial velocity at a point on the surface of the star to the speed of light.
In keeping with the literature, we take positive values of $v$ (and $\D$) to mean redshifts.

A common approach to computing the Doppler-shifted spectrum is to evaluate the spectrum at the rest frame wavelength $\lnlam_0$ and interpolate back to the grid in $\lnlam$. 
This is practical when computing the spectrum at a single \emph{point} on the surface, but not ideal when one is interested in the \emph{integral} over the visible surface of the star $\Surf$, which is typically all we can observe:
\begin{align}
    \label{eq:the_problem:F}
    F(\lnlam, t)
     & \equiv
    \iint\limits_{\Surf(\x)}
    I(\lnlam, \x, t)
    \mathrm{d}{\Surf(\x)}
    \quad,
\end{align}
The difficulty in solving Equation~(\ref{eq:the_problem:F}) stems from the fact that $I(\lnlam, \x, t)$ is difficult to write down in closed form, given the nonlinearity of the Doppler shift.
The standard approach to solving this integral is therefore to discretize the surface of the star with a fine grid, evaluate the Doppler-shifted spectrum in each cell, and sum over the spatial axes to approximate the integral. 
Depending on the resolution of the grid, this may be either numerically inaccurate or computationally inefficient.

\section{The Solution}
\label{sec:the_solution}

\subsection{Doppler Deconvolution}

The observed spectrum is an intricate function of spatial, spectral, temporal, and velocity terms. 
The goal in this section is to deconvolve each of these terms to make solving the integral in Equation~(\ref{eq:the_problem:F}) easier.
The first thing we will do is to express Equation~(\ref{eq:the_problem:Ixi}) as a convolution:
\begin{align}
    \label{eq:deconv:convolution}
    I(\lnlam, \x, t) & =
    I(\lnlam_0, \x, t)
    *
    \delta\big(\lnlam_0 + \D(\x)\big)
\end{align}
where $\delta$ is the Dirac delta function and $*$ denotes the linear convolution operator, defined for two arbitrary functions $g$ and $h$ as the integral
\begin{align}
    \label{eq:deconv:convolution_def}
    (g * h)(t) \equiv \int_{-\infty}^\infty g(\tau) h(t - \tau) d\tau
\end{align}
for some independent coordinate $t$ and dummy parameter $\tau$.
The convolution of $I(\lnlam_0)$ with a delta function has the effect of shifting the spectrum by an amount $\D$, returning a function of the shifted (observed) log wavelength, $\lnlam = \lnlam_0 + \D$.

Next, we expand the spatial dependence of the specific intensity at the rest frame wavelength in terms of spherical harmonics on the unit disk:
\begin{align}
    \label{eq:deconv:Ixi0}
    I(\lnlam_0, \x, t)
     & =
    \sum_{l=0}^{l_\mathrm{max}}\sum_{m=-l}^{l} a_{lm}(\lnlam_0, t) Y_{lm}(\x)
    \quad ,
\end{align}
where $Y_{lm}(\x)$ is a spherical harmonic on the projected disk and $a_{lm}(\lnlam_0, t)$ is the corresponding spherical harmonic coefficient at log wavelength in the rest frame $\lnlam_0$ and time $t$. 
For convenience, we may write this equation in vector form:
\begin{align}
    \label{eq:deconv:Ivec}
    I(\lnlam_0, \x, t) & =
    \ylmbasis(\x) \,
    \almt(\lnlam_0, t)
    \quad ,
\end{align}
where
\begin{align}
    \label{eq:deconv:almt}
    \almt(\lnlam_0, t) \equiv
    \Big(
    a_{0,0}(\lnlam_0, t) \quad 
    a_{1,-1}(\lnlam_0, t) \quad 
    a_{1,0}(\lnlam_0, t) \quad 
    a_{1,1}(\lnlam_0, t) \quad
    ...
    \Big)^\top
\end{align}
is the vector of $N = (l_\mathrm{max} + 1)^2$ spherical harmonic coefficients and
\begin{align}
    \label{eq:deconv:ylmbasis}
    \ylmbasis(\x) \equiv
    \Big(
    Y_{0,0}(\x) \quad 
    Y_{1,-1}(\x) \quad 
    Y_{1,0}(\x) \quad 
    Y_{1,1}(\x) \quad
    ...
    \Big)
\end{align}
is the corresponding vector of $N$ spherical harmonics. 
We may further decompose our expression by linearizing the time dependence of the spherical harmonic coefficients:
\begin{align}
    \label{eq:deconv:R}
    \almt(\lnlam_0, t) = \R(t) \, \alm(\lnlam_0)
    \quad ,
\end{align}
where $\alm(\lnlam_0) = \bvec{a}(\lnlam_0, t=t_0)$ for some reference time $t_0$.
For rigid body rotation, the result is exact and $\R(t)$ is the $(N \times N)$ Wigner rotation matrix for real spherical harmonics
\citep[e.g.][]{AlvarezCollado1989}, which is implicitly a function of the inclination, obliquity, and rotation period of the star.

The equation for the specific intensity in the rest frame now reads
\begin{align}
    \label{eq:deconv:Ivecfull}
    I(\lnlam_0, \x, t) & =
    \ylmbasis(\x)
    \,
    \R(t)
    \,
    \alm(\lnlam_0)
    \quad ,
\end{align}
where it is clear that we have fully separated the spatial, temporal, and spectral terms. 
Inserting this into Equation~(\ref{eq:deconv:convolution}) and integrating over the visible portion of the star, we arrive at the expression for the observed spectrum:
\begin{align}
    \label{eq:deconv:F2d}
    F(\lnlam, t) & =
    \iint\limits_{\Surf(\x)}
    \ylmbasis(\x)
    \,
    \R(t)
    \,
    \almt(\lnlam_0)
    * \delta\big(\lnlam_0 + \D(\x)\big)
    \mathrm{d}\Surf(\x)
    \nonumber                \\[0.5em]
                         & =
    \iint\limits_{\Surf(\x)}
    \ylmbasis(\x)
    \delta\big(\lnlam_0 + \D(\x)\big)
    \mathrm{d}\Surf(\x)
    \,
    \R(t)
    \,
    *
    \,
    \alm(\lnlam_0)
    \quad ,
\end{align}
where we made use of the commutativity of the convolution operator and the fact that the integral is taken only over the spatial dimensions.
Note, importantly, that the convolution operation above is implicitly a vector operation---i.e., the convolution is taken for each spherical harmonic term individually and then summed over all terms.

Finally, we can simplify Equation~(\ref{eq:deconv:F2d}) by noting that the delta function in the integrand allows us to reduce the double integral to a line integral:
\begin{align}
    \label{eq:deconv:kT}
    \iint\limits_{\Surf(\x)}
    \ylmbasis(\x)
    \delta\big(\lnlam_0 + \D(\x)\big)
    \mathrm{d}\Surf(\x)
    \, \,
     & =
    \int\limits_{\Curve(\lnlam, \x)}
    \hspace*{-0.6em}\ylmbasis(\x)
    \mathrm{d}\Curve(\lnlam, \x)
    \nonumber                     \\[0.5em]
     & \equiv \kT(\lnlam)
    \quad.
\end{align}
where the path $\Curve(\lnlam, \x)$ corresponds to the set of all points on the visible disk where $\lnlam_0 + \D(\x) = 0$.
We thus have
\begin{align}
    \label{eq:deconv:F}
    F(\lnlam, t)
     & =
    \kT(\lnlam) \, \R(t)
    *
    \alm(\lnlam_0)
    \quad.
\end{align}
Equation~(\ref{eq:deconv:F}) is the deconvolution of the observed spectrum into velocity terms, temporal terms, and spectral/spatial terms, respectively. 
In the next section, we will discuss how to efficiently solve the line integral in Equation~(\ref{eq:deconv:kT}).

\subsection{Computing the kernel $\kT$}
\label{sec:kT}
In the case that the star's rotation is rigid (i.e., the effect of differential rotation on the spectrum is small) and other effects such as convective blueshift may be ignored, the radial velocity at any point on the surface is simply proportional to the distance to the axis of rotation. 
Without loss of generality, if we assume the axis of rotation lies along the $y-z$ plane, we have
\begin{align}
    \label{eq:kT:beta}
    \beta(\x) = \frac{\omega \, x}{c} \sin i
\end{align}
and
\begin{align}
    \label{eq:kT:D}
    \D(\x) & =
    \frac{1}{2}\ln\left(
    \dfrac{1 + \dfrac{\omega \, x}{c} \sin i}
    {1 - \dfrac{\omega \, x}{c} \sin i}
    \right)
    \quad ,
\end{align}
where $\omega$ is the stellar angular velocity and $i$ is the stellar inclination with respect to $\hat{y}$, where $i = 0^\circ$ corresponds to a pole-on orientation.

\begin{figure}[t!]
    \begin{centering}
        \includegraphics[width=\linewidth]{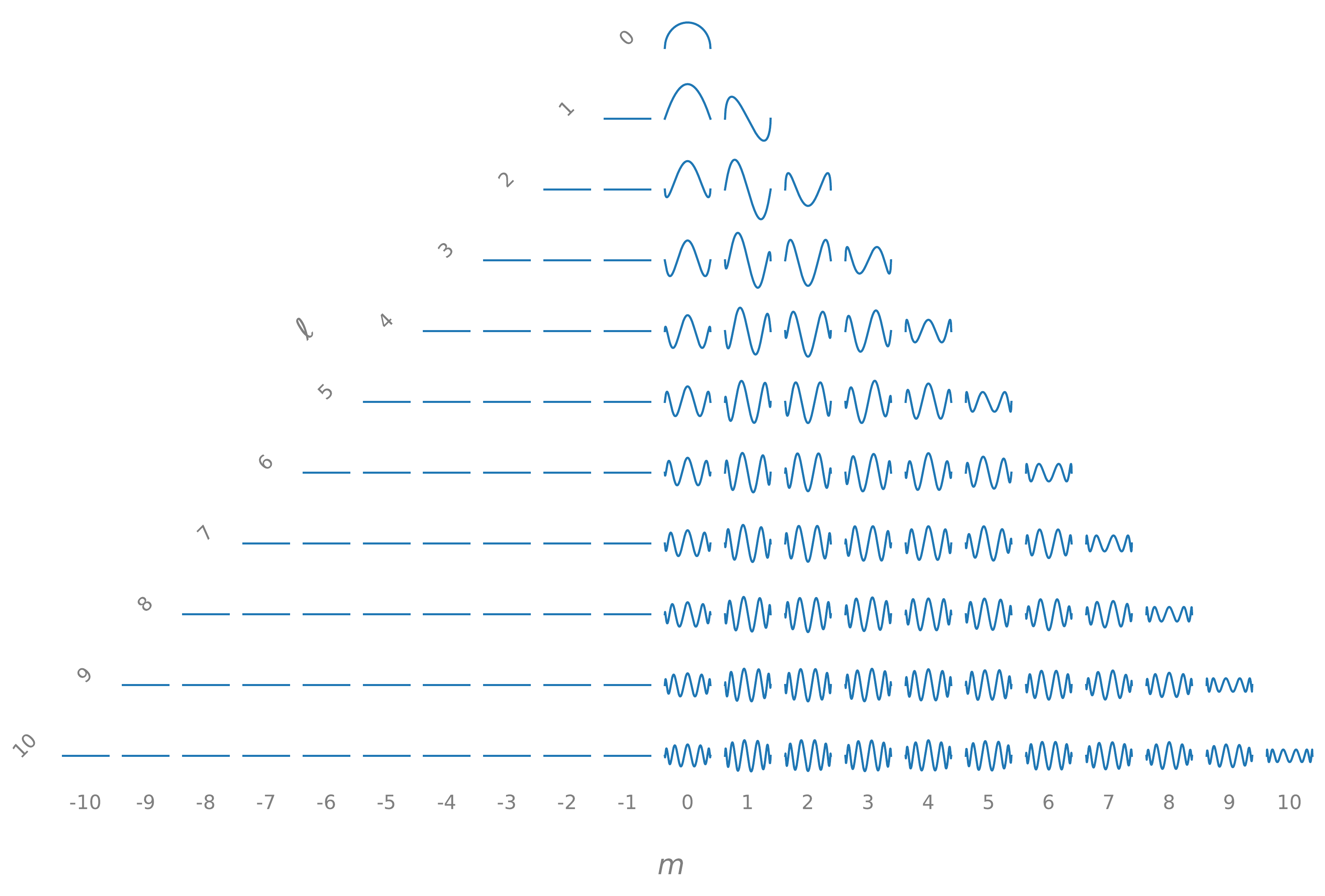}
        \caption{%
            The Doppler $\kT$ basis for a rigidly rotating star computed from Equation~(\ref{eq:kT:kT}) and Equation~(\ref{eq:kT:sTrecurrence}) up to spherical harmonic degree $l_\mathrm{max}=10$. 
            Rows correspond to the degree $l$ and columns correspond to the order $m$. 
            These functions encode the contribution of each spherical harmonic to the rotational broadening of features in the stellar spectrum. 
            Because the rotational axis is chosen to be aligned with $\hat{y}$ (along the plane of the sky), none of the $m < 0$ harmonics contribute to the observed spectrum.
        }
        \label{fig:kT}
    \end{centering}
\end{figure}

The region of integration $\Surf(\x)$ in Equation~(\ref{eq:deconv:kT}) is simply the unit disk, so $\Curve(\lnlam, \x)$ is the set of all points that satisfy both $\lnlam_0 + \D(\x) = 0$ and $x^2 + y^2 \le 1$.
Solving Equation~(\ref{eq:kT:D}) for $x$, we find that the curves $\Curve(\lnlam, \x)$ are simply vertical lines on the unit disk given by
\begin{align}
    x & =
    \Bigg(\frac{c}{\omega\sin i}\Bigg)
    \Bigg(\frac{\mathrm{e}^{-2{\lnlam_0}} - 1}
    {\mathrm{e}^{-2{\lnlam_0}} + 1}\Bigg)
    \quad .
\end{align}
The integral in Equation~(\ref{eq:deconv:kT}) is therefore just the line integral of the spherical harmonic basis over $y$:
\begin{align}
    \label{eq:kT:kT}
    \kT(\lnlam)
     & =
    \int\limits_{-\sqrt{1 - x^2}}^
    {\sqrt{1 - x^2}}
    \ylmbasis
    (x, y)
    \mathrm{d}y
    \quad .
\end{align}
In Appendix~\ref{app:kT} we derive an analytic solution to this integral and show how to recursively solve it for all terms in the spherical harmonic basis.
Figure~\ref{fig:kT} shows the $\kT$ basis computed from the formulae above and Equation~(\ref{eq:kT:kT}) up to spherical harmonic degree $l_\mathrm{max}=10$.

This completes our derivation of the analytic expression for the observed spectrum (Equation~\ref{eq:deconv:F}).

\subsection{Example}
\begin{figure}[p!]
    \begin{centering}
        \includegraphics[width=0.6\linewidth]{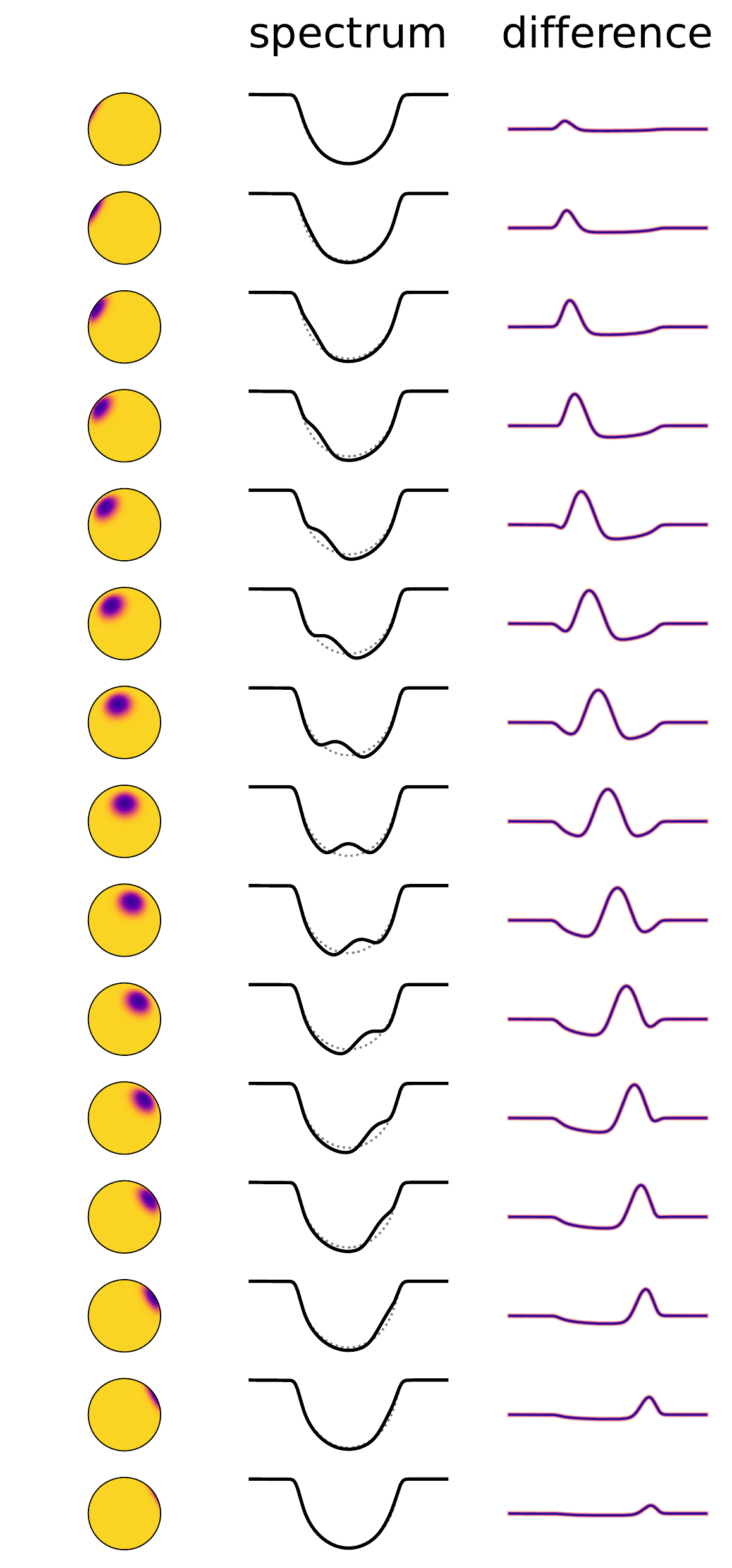}
        \caption{%
            Time evolution of a Gaussian absorption line on a rigidly rotating, spotted star computed from the analytic formulae in \S\ref{sec:kT}.
            The stellar surface is modeled as a spherical harmonic expansion up to $l_\mathrm{max}=20$, and the line shape is assumed to be the same everywhere;
            the spot simply downweights the local intensity at all wavelengths.
            As the spot rotates into view (left panel), the line shape changes slightly (center panel). 
            The difference between the line when the spot is in view (solid) and when it is on the backside of the star (dotted) are shown in the right panel, where a Gaussian-like feature can be seen tracking the spot as it rotates from the blueshifted hemisphere to the redshifted hemisphere of the star.
        }
        \label{fig:spot}
    \end{centering}
\end{figure}

Figure~\ref{fig:spot} shows an example application of the formulae derived in this section. 
We generated a synthetic stellar surface with a large, circular spot at a latitude of $30^\circ$ and computed its spherical harmonic expansion up to $l_\mathrm{max}=20$ in the form of the vector of coefficients $\bvec{y}$.
The stellar rest frame spectrum $I(\lnlam_0)$ is taken to be a single Gaussian absorption line that is spatially constant everywhere save for a multiplicative factor equal to the local spot intensity, which is described by the spherical harmonic expansion.
In the Figure, we rotated the star about an axis perpendicular to the line of sight (left panel) and computed the observed spectrum using Equation~(\ref{eq:deconv:F}) (center panel). 
The right panel shows the difference between the observed spectrum and the spectrum when the spot is not in view.
The effect of the spot on the absorption line is evident: a ``bump'' that travels from the blueshifted side of to the redshifted side of the line in sync with the change in longitude of the spot. 
This correspondence between line shape and spot location is the cornerstone of the Doppler imaging technique \citep[compare to, e.g., Figures 1 and 4 in][]{Vogt1983}.

\begin{figure}[t!]
    \begin{centering}
        \includegraphics[width=0.65\linewidth]{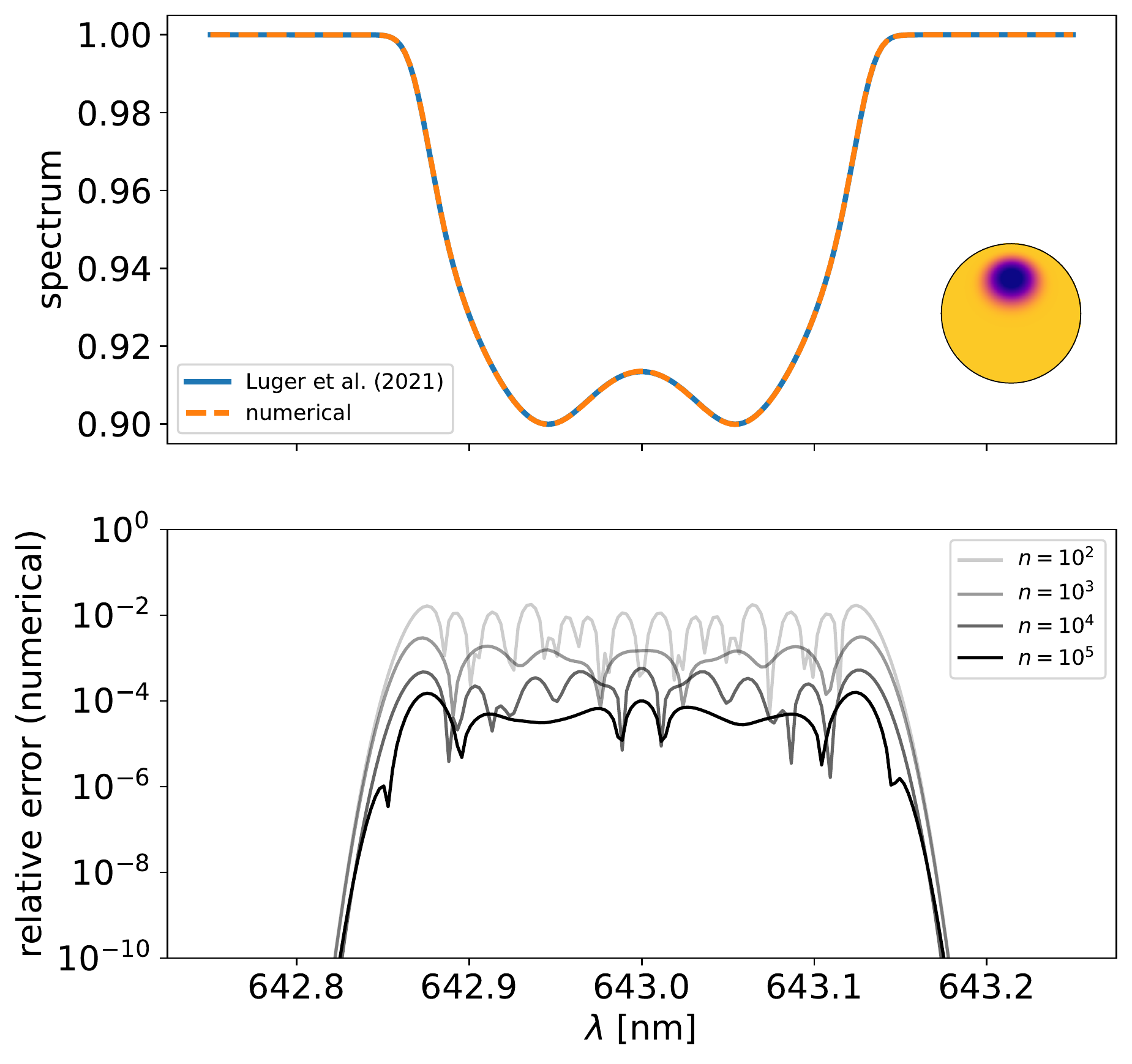}
        \caption{%
            The observed spectrum when the spot is in view computed with our algorithm (blue) and numerically (orange). 
            The problem setup is the same as that in Figure~\ref{fig:spot}. 
            The bottom panel shows the absolute value of the difference between the two spectra for different number of grid points in the numerical solution.
            As the number of grid points increases, the numerical solution approaches our solution.
            Note that here we assume the stellar surface can be expressed exactly as an $l_\mathrm{max} = 20$ spherical harmonic decomposition.
            When our algorithm is applied to real data, there will be errors due to the truncation of the expansion at finite spherical harmonic degree.
        }
        \label{fig:compare}
    \end{centering}
\end{figure}

Figure~\ref{fig:compare} shows one of the spectra in Figure~\ref{fig:spot}, this time computed using both our formulae (blue line) and the traditional technique of discretizing the stellar surface, Doppler-shifting the spectrum in each grid cell according to the local radial velocity, interpolating back to a uniform wavelength grid, and summing over the spatial dimension (dotted orange line). 
We show the residuals in the lower panel, where it is clear that as the number of grid cells increases from $10^2$ (light grey) to $10^5$ (dark grey), the numerical solution approaches the solution obtained using our approach.


\section{The forward problem}
\label{sec:linear}

\begin{figure}[ht!]
    \begin{centering}
        \includegraphics[width=0.8\linewidth]{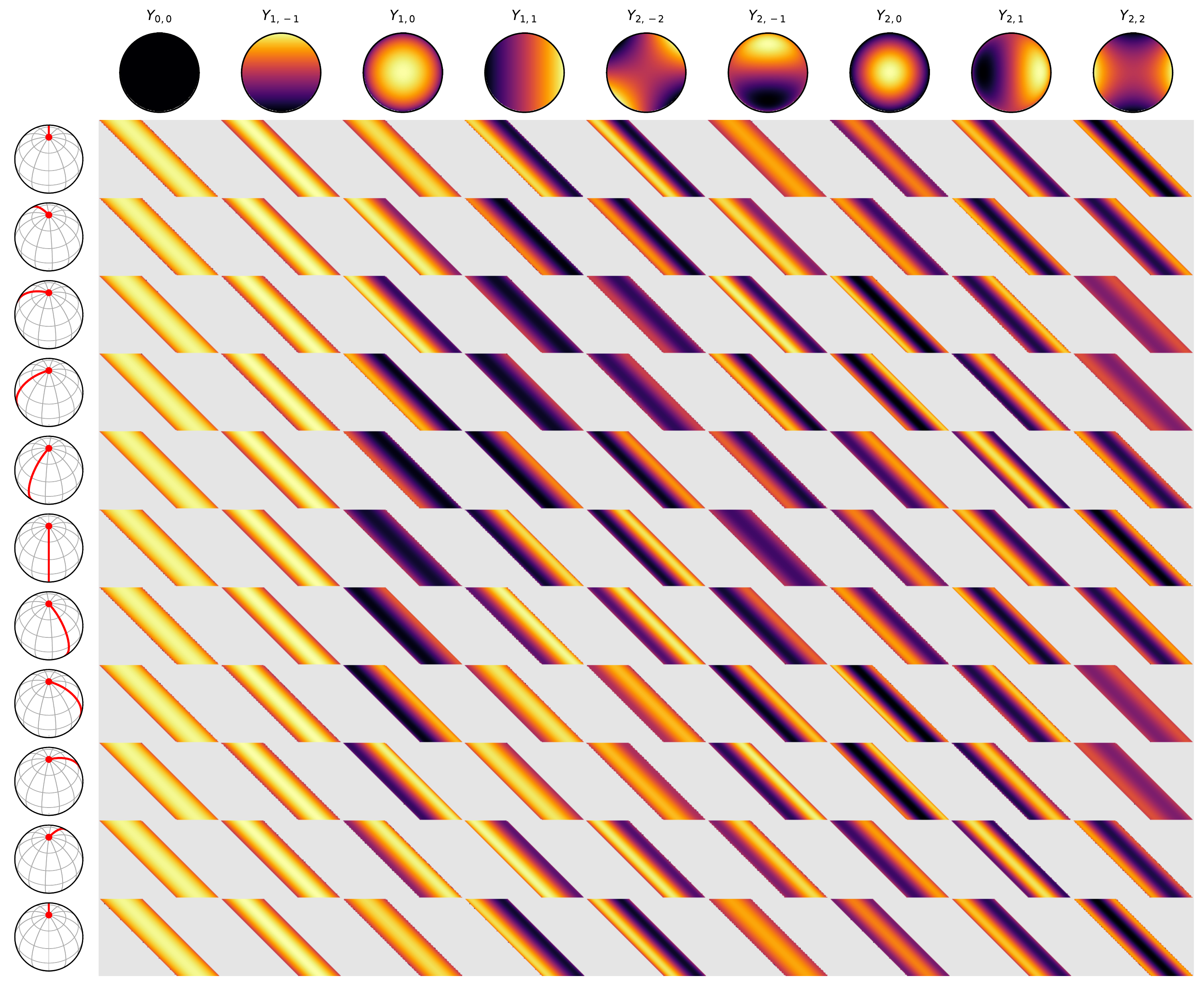}
        \caption{%
            The Doppler design matrix $\Doppler$ (Equation~\ref{eq:linear:f}), constructed from a grid of Toeplitz matrices, each of shape $K \times K'$, where $K$ is the number of wavelength bins in the broadened spectrum and $K' = K + W - 1$ is the number of bins in the latent (template) spectrum, where $W$ is the width of the convolution kernel.
            The $N$ columns of $\Doppler$ correspond to the Toeplitz matrices for each of the $N$ spherical harmonics (shown at the top);
            the $M$ rows correspond to the Toeplitz matrices rotated to each of the $M$ stellar phases observed (indicated graphically to the left of $\Doppler$).
            The color scale ranges from black (negative) to bright yellow (positive); light grey regions correspond to entries in the matrix that are identically zero.
        }
        \label{fig:linalg}
    \end{centering}
\end{figure}

Although the convolution operator (Equation~\ref{eq:deconv:convolution_def}) is defined for continuous functions, the problem of Doppler imaging deals with discrete measurements of a spectrum in bins of log wavelength. 
Provided our wavelength grid is fine enough, we may therefore approximate Equation~(\ref{eq:deconv:convolution_def}) with a discrete convolution.
For discrete arrays $\bvec{g}$ and $\bvec{h}$, we have
\begin{align}
    \label{eq:linear:convolution_def}
    \bvec{g} * \bvec{h} = \bvec{T(g)} \, \bvec{h}
\end{align}
where $\bvec{T}$ is a Toeplitz matrix, a matrix whose diagonals are constant from top left to bottom right. 
In this case, the diagonals are constructed from the values of $\bvec{g}$. 
If $\bvec{g}$ has length $L$, element $g_n$ is placed everywhere along the $k^\mathrm{th}$ diagonal of $\bvec{T}$ for $k = -L / 2 + n$; all other entries of $\bvec{T}$ are set to zero. 
Note that the matrix $\bvec{T}$ need not be square; in fact, for our purposes, it is a $(K \times K')$ matrix, where $K$ is the size of the output wavelength grid and $K' = K + W - 1$ is the size of the input wavelength grid (in the rest frame), where $W$ is the width of the convolution kernel.

Given this formulation, we may re-write Equation~(\ref{eq:deconv:F}) as a purely linear operation on $\alm(\lnlam_0)$:
\begin{align}
    \label{eq:linear:ft}
    \bvec{f}_m
     & =
    \Doppler_m
    \,
    \alm
    \quad,
\end{align}
where $\Doppler_m$ is a $(K \times N K')$ matrix constructed by horizontally concatenating the Toeplitz matrices for each of the $N$ components of the convolution kernel $\kT(\lnlam)$ transformed by the rotation matrix $\bvec{R}(t)$ evaluated at time $t = t_m$:
\begin{align}
    \label{eq:linear:Dm}
    \Doppler_m =
    \begin{pmatrix}
        \bvec{T}(\kT_0)
        \quad
         &
        \quad
        \bvec{T}(\kT_1)
        \quad
         &
        \quad
        \cdots
        \quad
         &
        \quad
        \bvec{T}(\kT_N)
        \quad
    \end{pmatrix}
    (\bvec{R}(t_m) \otimes \bvec{I}_{K'})
    \quad.
\end{align}
where $\bvec{I}_{K'}$ is the $(K' \times K')$ identity matrix and $\otimes$ denotes the Kronecker product.
The $(K \times 1)$ vector $\mathbf{f}_m$ is the model for the flux observed at each wavelength at time $t = t_m$, and the $(N K' \times 1)$ vector $\alm$ is the vector representation of the spatially-dependent spectrum of the star. 
The latter is constructed by flattening the spherical harmonic expansion of the star such that the first $K'$ terms in $\alm$ correspond to the the values of $a_{0,0}$ at each wavelength $\lnlam_0$, followed by the values of $a_{1,-1}$ at each wavelength, and so forth.

In general, our data will consist of observations made at several epochs, corresponding to different rotational phases of the star.
If we concatenate all $M$ spectra $\bvec{f}_m$ into the $(MK \times 1)$ vector $\bvec{f}$, we may write
\begin{align}
    \label{eq:linear:f}
    \bvec{f}
     & =
    \Doppler
    \,
    \alm
    \quad,
\end{align}
where $\Doppler$ is the full $(MK \times N K')$ Doppler design matrix, constructed by vertically concatenating the individual matrices $\Doppler_m$:
\begin{align}
    \label{eq:linear:D}
    \Doppler =
    \begin{pmatrix}
        \Doppler_0
        \\
        \Doppler_1
        \\
        \cdots
        \\
        \Doppler_m
    \end{pmatrix}
    \quad.
\end{align}
Equation~(\ref{eq:linear:f}) represents the full linearization of the problem, where we have expressed the quantity we can observe, $\bvec{f}$, as a linear operation on the quantity of interest, the spectral/spatial decomposition of the stellar surface, $\alm$. 
Figure~\ref{fig:linalg} shows an example of $\Doppler$.
Note, importantly, that this expression does not account for limb darkening or for the fact that spectra are often normalized to the continuum level; we account for these two effects in \S\ref{sec:norm} and \S\ref{sec:ld} below.

\section{The inverse problem}
\label{sec:inverse}
In the previous section we showed how to express the data vector $\bvec{f}$, a series of spectra obtained at different epochs, as a purely linear operation on the map vector $\alm$, the spherical harmonic decomposition of the specific intensity on the surface of the star. 
The advantage of this linearity is that, given suitable priors, it allows one to compute both the maximum likelihood solution for $\alm$ and its uncertainty \emph{analytically}. 
In particular, if one places a (multidimensional) Gaussian prior on $\alm$ with mean $\boldsymbol{\mu}_\mathbf{a}$ and covariance $\boldsymbol{\Lambda}_\mathbf{a}$, the maximum \emph{a posteriori} (MAP) solution for $\alm$ is given by the solution to the L2 regularization problem \citep[see, e.g.,][]{Luger2016}:
\begin{align}
    \label{eq:inverse:ahat}
    \bvec{\hat{a}} & =
    \boldsymbol{\Sigma}_\mathbf{\hat{a}}
    \left(
    \Doppler^\top
    {\boldsymbol{\Sigma}_\mathbf{f}}^{-1}
    \bvec{f}
    +
    {\boldsymbol{\Lambda}_{\alm}}^{-1} \boldsymbol{\mu}_\mathbf{a}
    \right)
    \quad,
\end{align}
with posterior covariance given by
\begin{align}
    \label{eq:inverse:acov}
    \boldsymbol{\Sigma}_\mathbf{\hat{a}} & =
    \left(
    \Doppler^\top
    {\boldsymbol{\Sigma}_\mathbf{f}}^{-1}
    \Doppler
    +
    {\boldsymbol{\Lambda}_{\alm}}^{-1}
    \right)^{-1}
    \quad,
\end{align}
where $\boldsymbol{\Sigma}_\mathbf{f}$ is the data covariance matrix. 
These equations require the inversion of a few fairly large matrices, but as we will see later, it allows one to obtain the map of the star ($\bvec{\hat{a}}$) and its uncertainty ($\boldsymbol{\Sigma}_{\mathbf{\hat{a}}}$) in under a second for a typical dataset.

There is, however, a catch: for most practical purposes, it is difficult to find a good Gaussian prior for $\bvec{a}$. 
Usually we will have some prior information on what the spectrum of the star is, and perhaps some prior information on the distribution of surface features such as starspots. 
The problem is that even if these priors are Gaussian, the corresponding prior on $\bvec{a}$ is \emph{not}. 
To understand this, consider the case where the stellar rest frame spectrum is spatially constant, save for an amplitude that is spatially variable. 
We can then describe the rest frame spectrum by the $(K' \times 1)$ vector $\bvec{s}$ and the amplitude by the $(N \times 1)$ vector $\bvec{y}$ representing the spherical harmonic expansion of the intensity profile of the star. 
The vector $\alm$ is then given by the (flattened) outer product of the two vectors:
\begin{align}
    \label{eq:inverse:alm}
    \alm & = \mathrm{vec}\left( \bvec{A} \right) \nonumber             \\
           & = \mathrm{vec}\left( \bvec{s} \, \bvec{y}^\top \right) \quad,
\end{align}
where $\mathrm{vec}$ denotes the vectorization operation, which in this case transforms the $(K' \times N)$ matrix $\bvec{A}$ into the $(N K' \times 1)$ column vector $\alm$ by vertically stacking all of its columns. 
In other words, the vector $\alm$ consists of the set of $K'$ intensities in each wavelength bin repeated $N$ times, each time multiplied by the $n^\mathrm{th}$ spherical harmonic coefficient in the expansion of the surface intensity. 
Returning to our point about priors, note that all of the entries of $\alm$ are the product of two independent random variables. 
Since the product of two random variables is generally non-Gaussian, even when the variables themselves are Gaussian,%
\footnote{see, e.g., \url{http://mathworld.wolfram.com/NormalProductDistribution.html}}
so too is the prior on $\alm$. 
This point makes Equation~(\ref{eq:inverse:ahat}) tricky to use in practice, since a multivariate Gaussian is not generally the most suitable prior for $\alm$.

That said, a Gaussian prior \emph{is} typically suitable for both the spectrum $\bvec{s}$ and the intensity $\bvec{y}$ (just not their product). 
Spectral templates can provide us with a physically-motivated mean for the prior on $\bvec{s}$, while the variance can be used to capture our degree of confidence in the model.
Similarly, a multivariate Gaussian prior on the spherical harmonic coefficients $\bvec{y}$ is flexible enough to capture prior information on the location and properties of starspots \citep{Luger2021b}.
Given Gaussian priors on these two quantities, it is straightforward to show that because Equation~(\ref{eq:linear:f}) is linear in $\alm$, it is also linear in both $\bvec{s}$ and $\bvec{y}$, so MAP solutions can be found analytically for $\bvec{s}$ (conditioned on $\bvec{y}$) and $\bvec{y}$ (conditioned on $\bvec{s}$).
A solution to the full problem of determining both $\bvec{s}$ and $\bvec{y}$ can then be obtained by iteratively solving these two problems.
We investigate this procedure below.

\subsection{Solution for the map}
\label{sec:solve_y}
Given a single rest-frame spectrum $\bvec{s}$, the model for the flux vector may be written as
\begin{align}
    \label{eq:inverse:fy}
    \bvec{f}
     & =
    \Doppler
    \,
    \bvec{S}
    \,
    \bvec{y}
    \quad,
\end{align}
where
\begin{align}
    \label{eq:inverse:S}
    \bvec{S} =
    \begin{pmatrix}
        \quadquad\bvec{s}\quadquad &                            &                            &                            &        \\
                                   & \quadquad\bvec{s}\quadquad &                            &                            &        \\
                                   &                            & \quadquad\bvec{s}\quadquad &                            &        \\
                                   &                            &                            & \quadquad\bvec{s}\quadquad &        \\
                                   &                            &                            &                            & \ddots
    \end{pmatrix}
\end{align}
is a $(NK' \times N)$ block matrix constructed by repeating the column vector $\bvec{s}$ $N$ times in blocks along the main diagonal.
Given prior mean $\boldsymbol{\mu}_\bvec{y}$ and covariance $\boldsymbol{\Lambda}_\bvec{y}$ on $\bvec{y}$, the MAP solution is given by
\begin{align}
    \label{eq:inverse:y1hat}
    \bvec{\hat{y}} & =
    \boldsymbol{\Sigma}_\mathbf{\hat{y}}
    \left(
    \bvec{S}^\top\Doppler^\top
    {\boldsymbol{\Sigma}_\mathbf{f}}^{-1}
    \bvec{f}
    +
    {\boldsymbol{\Lambda}_\bvec{y}}^{-1} \boldsymbol{\mu}_\bvec{y}
    \right)
    \quad,
\end{align}
with covariance
\begin{align}
    \label{eq:inverse:y1cov}
    \boldsymbol{\Sigma}_\bvec{\hat{y}} & =
    \left(
    \bvec{S}^\top\Doppler^\top
    {\boldsymbol{\Sigma}_\bvec{f}}^{-1}
    \Doppler\bvec{S}
    +
    {\boldsymbol{\Lambda}_\bvec{y}}^{-1}
    \right)^{-1}
    \quad.
\end{align}
If we choose a prior with diagonal covariance that is constant for each spherical harmonic order $m$,
\begin{align}
    \label{eq:isotropicprior}
    \boldsymbol{\Lambda}_\bvec{y} & =
    \mathrm{diag} \left(
    \sigma_0^2
    \quad\quad\quad\quad
    \sigma_1^2
    \quad\quad\quad\quad
    \sigma_1^2
    \quad\quad\quad\quad
    \sigma_1^2
    \quad\quad\quad\quad
    \sigma_2^2
    \quad\quad\quad\quad
    \sigma_2^2
    \quad\quad\quad\quad
    \sigma_2^2
    \quad\quad\quad\quad
    \sigma_2^2
    \quad\quad\quad\quad
    \sigma_2^2
    \quad\quad\quad\quad
    \cdots
    \right)
    \quad,
\end{align}
we are in fact imposing an isotropic prior with an angular power spectrum given by $C_l = \sigma_l^2$
\citep[e.g.,][]{Baldi2006}. 
This is especially useful if the typical angular scale of features on the surface of the star is known or if a Gaussian prior can be placed on it.
Alternatively, we may compute $\boldsymbol{\Lambda}_\bvec{y}$ from the Gaussian process (GP) formalism in \citet{Luger2021b} to incorporate explicit prior information on the typical latitudes, sizes, and contrasts of spots.
We discuss this point further in \S\ref{sec:discussion:priors}.

\subsection{Solution for the spectrum}
\label{sec:solve_s}
Conversely, given a spatial representation of the surface $\bvec{y}$, the model for the flux vector may be written in terms of the spectrum $\bvec{s}$ as
\begin{align}
    \label{eq:inverse:fs}
    \bvec{f}
     & =
    \Doppler
    \,
    \bvec{Y}
    \,
    \bvec{s}
    \quad,
\end{align}
where
\begin{align}
    \label{eq:inverse:Y}
    \bvec{Y} =
    \begin{pmatrix}
        y_{0,0}\, \bvec{I}_{K'} \\
        y_{1,-1}\, \bvec{I}_{K'} \\
        y_{1,0}\, \bvec{I}_{K'} \\
        y_{1,1}\, \bvec{I}_{K'} \\
        \cdots
    \end{pmatrix}
\end{align}
is a $(NK' \times K')$ matrix constructed by vertically stacking $N$ $(K' \times K')$ identity matrices, each multiplied by the corresponding coefficient of $\bvec{y}$. 
Given prior mean $\boldsymbol{\mu}_\mathbf{s}$ and covariance $\boldsymbol{\Lambda}_\mathbf{s}$ on $\mathbf{s}$, the MAP solution is given by
\begin{align}
    \label{eq:inverse:shat}
    \bvec{\hat{s}} & =
    \boldsymbol{\Sigma}_\mathbf{\hat{s}}
    \left(
    \bvec{Y}^\top\Doppler^\top
    {\boldsymbol{\Sigma}_\mathbf{f}}^{-1}
    \bvec{f}
    +
    {\boldsymbol{\Lambda}_\mathbf{s}}^{-1} \boldsymbol{\mu}_\mathbf{s}
    \right)
    \quad,
\end{align}
with covariance
\begin{align}
    \label{eq:inverse:scov}
    \boldsymbol{\Sigma}_\mathbf{\hat{s}} & =
    \left(
    \bvec{Y}^\top\Doppler^\top
    {\boldsymbol{\Sigma}_\mathbf{f}}^{-1}
    \Doppler\bvec{Y}
    +
    {\boldsymbol{\Lambda}_\mathbf{s}}^{-1}
    \right)^{-1}
    \quad.
\end{align}

\subsection{The full solution}
\label{sec:full_solve}
In the previous two sections we showed how, if the stellar rest-frame spectrum is spatially constant and known \emph{a priori} (from, say, a template spectrum), the posterior distribution for the surface map $\bvec{y}$ is \emph{analytic} (Equations~\ref{eq:inverse:y1hat} and \ref{eq:inverse:y1cov}). 
Conversely, if the surface map is known, the posterior for the spectrum is also analytic (Equations~\ref{eq:inverse:shat} and \ref{eq:inverse:scov}). 
Both sets of solutions require Gaussian priors on the quantities of interest, but these are easily interpretable. 
A Gaussian prior on the spherical harmonic coefficients of the surface map corresponds to the angular power spectrum of the surface features, whereas a Gaussian prior on the spectrum can be used to correctly propagate uncertainties in the stellar spectrum from observational uncertainties on the stellar metallicity, temperature, etc.

Importantly, the framework developed here allows one to analytically compute not only the ``optimal'' solution to the Doppler imaging problem, but also its uncertainty. 
The problem of re-constructing surface images of stars from spectra and/or light curves is fundamentally ill-posed because of intrinsic degeneracies and a large null space \citep[e.g.,][]{Cowan2017,Luger2019,Luger2021a}. 
Coupled to uncertainties in properties like the stellar inclination, rotation period, and limb darkening coefficients, as well as the finite signal-to-noise ratio of the data, this makes the process of searching for the ``optimal'' solution flawed at best, and meaningless at worst. 
Access to the full covariance matrix of the solution is therefore essential to assessing our confidence in the various features of the inferred surface maps.

However, as we mentioned above, the Doppler imaging problem can be solved analytically for either the surface map or the spectrum, but not typically both. 
If neither are known exactly \emph{a priori}, which is quite often the case, we can still find the MAP solution quickly by guessing at an initial spectrum $\bvec{s}$ and iterating between optimizing for $\bvec{y}$ and optimizing for $\bvec{s}$. 
Note, importantly, that this bilinear problem is not generally convex: the likelihood space often has many local minima, which can in principle make it difficult to find the true solution.
Nevertheless, given a good starting point for $\bvec{s}$ and sufficient regularization on $\bvec{y}$, we find that it typically converges to the correct solution; see \S\ref{sec:spotstar}.

Once the MAP solution has been found for both $\bvec{s}$ and $\bvec{y}$, the covariance of one conditioned on the other can be computed from the formulae above. 
However, this is necessarily a lower bound on the true covariance of the solution, since it neglects any covariance \emph{between} $\bvec{s}$ and $\bvec{y}$. 
To \emph{correctly} account for this, one must resort to Monte Carlo methods or other sampling techniques. 
We discuss this further in \S\ref{sec:discussion:priors}.

\section{Further considerations}
\label{sec:bellswhistles}
Before we demonstrate our formalism in practice, we consider three modifications to our model that are often required when modeling real data. 
Below we discuss what to do when the relative brightness of the star across different observations is unknown (\S\ref{sec:norm}), how to account for the effects of limb darkening (\S\ref{sec:ld}), and how to model stars whose local spectra differ within and outside of starspots (\S\ref{sec:eigen}).

\subsection{Unknown continuum normalization}
\label{sec:norm}
Consider the star depicted in Figure~\ref{fig:spot}. 
When the spot is in view, absorption lines become distorted due to the suppression of either blueshifted or redshifted light. 
However, what is not shown in the figure is the fact that overall, the star also gets \emph{dimmer}: the continuum level drops in the presence of the spot. 
Unfortunately, spectrographs are not usually designed to measure this, since the change in the star's overall brightness from one observation to the next is often dwarfed by (unknown) changes in the spectrograph stability, cloud cover, etc.
For this reason, the spectra are usually normalized to have a fixed baseline continuum, and Equation~(\ref{eq:linear:f}) is no longer a good model for the data.

One option is to collect simultaneous photometry on the target in a band similar to the wavelength range observed, and use this light curve to (un-)normalize the spectral timeseries $\bvec{f}$. 
However, when this is not an option, our model for $\bvec{f}$ (Equation~\ref{eq:linear:f}) must be amended:
\begin{align}
    \label{eq:norm:f}
    \fnorm
     & =
     \frac{
        \mathbf{f}
    } {
        \bvec{B}
        \,
        \alm
    }
    \nonumber 
    \\[1em]
    & =
    \frac{
        \Doppler
        \,
        \alm
    } {
        \bvec{B}
        \,
        \alm
    }
    \quad,
\end{align}
where $\fnorm$ is the normalized spectral timeseries, $\bvec{B}$ is the baseline design matrix, and the vector division is performed elementwise. 
If $c$ is an index in the observed spectrum corresponding to the continuum (at all epochs), the elements of $\bvec{B}$ are computed from
\begin{align}
    B_{ij} = D_{i'j}
\end{align}
where
\begin{align}
    i' = c + \bigg\lfloor\frac{i}{K}\bigg\rfloor K
    \quad.
\end{align}
In other words, the matrix $\bvec{B}$ is constructed by repeating the $c^\mathrm{th}$ row of each epoch block $\Doppler_m$ of $\Doppler$ $K$ times.
Equation~(\ref{eq:norm:f}) is therefore equivalent to computing the $M$ unnormalized spectra using Equation~(\ref{eq:linear:f}) and dividing each one by the value at index $c$.

Given this new equation, we must change how we approach the inverse problem (\S\ref{sec:inverse}) if our data vector is the normalized spectral timeseries $\fnorm$. 
When solving for the spectrum given the spherical harmonic vector (\S\ref{sec:solve_s}), we need only make the replacements 
$\bvec{f} \rightarrow \fnorm \circ (\bvec{B}\, \bvec{a})$ 
and 
$\boldsymbol{\Sigma}_\mathbf{f} \rightarrow \boldsymbol{\Sigma}_{\fnorm} \circ (\bvec{B}\, \bvec{a})^2$ 
in Equations~(\ref{eq:inverse:shat}) and (\ref{eq:inverse:scov}), where $\circ$ denotes the elementwise product.
However, when solving for the spherical harmonic vector given the spectrum (\ref{sec:solve_y}), the problem is no longer linear in $\bvec{y}$, as the vector of interest appears in both the numerator and the denominator:
\begin{align}
    \label{eq:norm:fy}
    \fnorm
     & =
    \frac{
        \bvec{D}
        \,
        \bvec{S}
        \,
        \bvec{y}
    }{
        \bvec{B}
        \,
        \bvec{S}
        \,
        \bvec{y}
    }
    \quad.
\end{align}
Unfortunately, this equation does not admit a closed-form solution for $\mathbf{y}$. 
One can, however, still solve for the maximum \emph{a posteriori} value of $\mathbf{y}$ in an efficient manner by iteratively solving for the baseline and then for the spherical harmonic coefficients.
We describe a procedure for this in \S\ref{sec:spotstar} below.

\subsection{Limb darkening}
\label{sec:ld}
\begin{figure}[t!]
    \begin{centering}
        \includegraphics[width=0.5\linewidth]{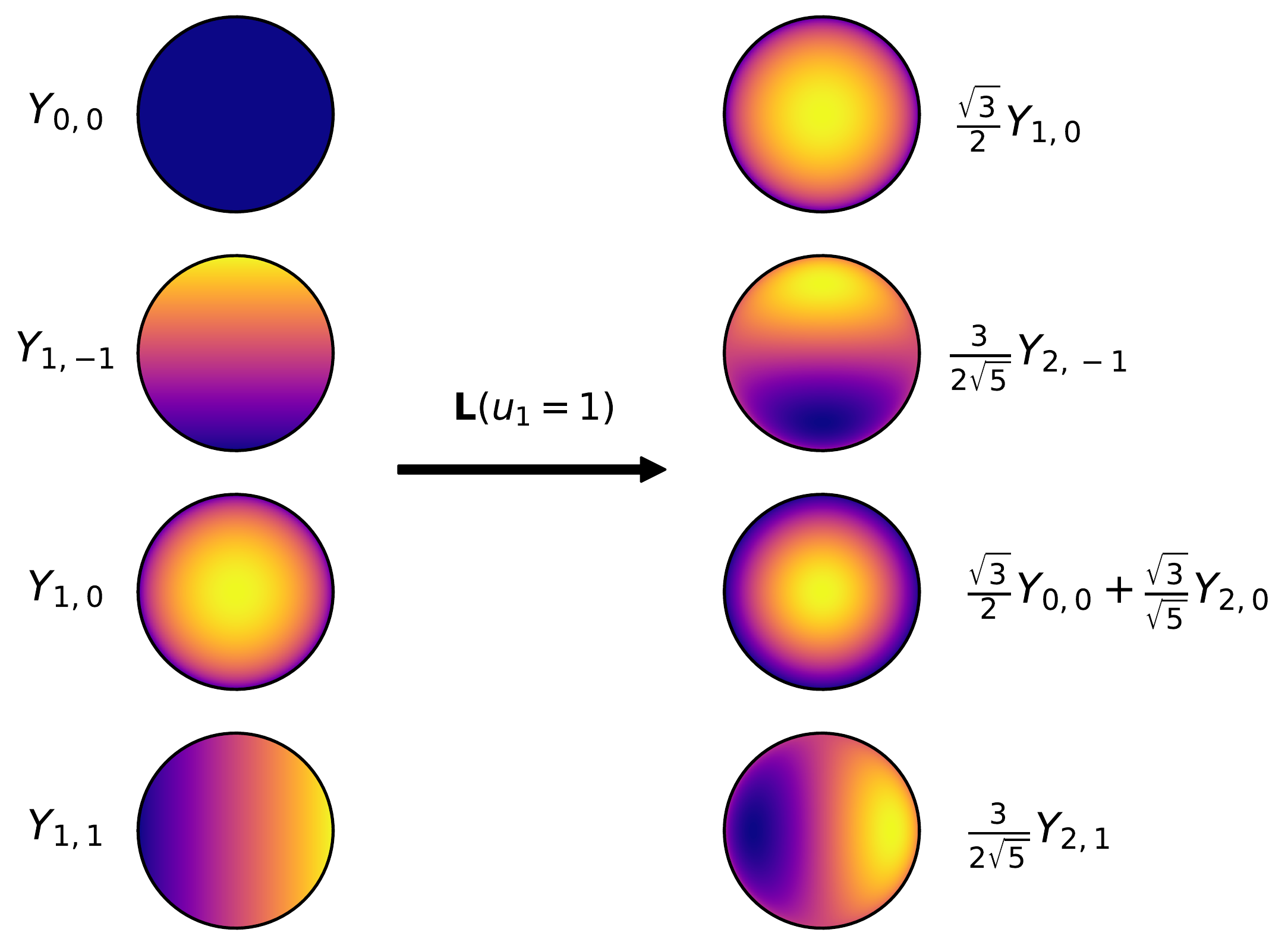}
        \caption{%
            Illustration of the limb darkening transformation given by Equation~(\ref{eq:ld:L}) for a linear limb darkening law with $u_1 = 1$. 
            Each row shows a spherical harmonic before (left) and after (right) transformation by $\bvec{L}$.
        }
        \label{fig:ld}
    \end{centering}
\end{figure}

Thus far we have avoided mention of limb darkening, whose effect on the observed spectrum can often be significant. 
A proper treatment of limb darkening requires solution of radiative transfer equations and should account for the temperature and composition gradients in the stellar atmosphere. 
However, in the spirit of data-driven inference, we can approximate the effect of limb darkening by weighting the spatial component of our map by the limb darkening profile of the star, which can either be imposed or learned from the data. 
For a polynomial limb darkening law of the form
\begin{align}
    \label{eq:ld:I}
    \frac{I(\mu)}{I(\mu = 1)} = 1 - \sum_{n=1}^{n_\mathrm{max}} u_n(1 - \mu)^n
    \quad,
\end{align}
where $\mu = z = \sqrt{1 - x^2 - y^2}$ is the radial coordinate on the projected disk and $u_n$ is a limb darkening coefficient, the effect of limb darkening on the stellar map can be expressed exactly as a linear operation on the spherical harmonic coefficient vector \citep{Luger2019,Agol2020}. 
This can be understood by noting that all terms in Equation~(\ref{eq:ld:I}) are strictly polynomials in $x$, $y$, and $z$, all of which can be expressed exactly as sums of spherical harmonics \citep{Luger2019}. 
When weighting the surface intensity by the limb darkening profile, the resulting intensity is simply a product of spherical harmonics, which is itself a sum of spherical harmonics. 
Given a limb darkening law of degree $n_\mathrm{max}$, we can always construct a matrix $\bvec{L}$ that transforms a spherical harmonic vector $\bvec{y}$ of degree $l_\mathrm{max}$ to a limb-darkened spherical harmonic vector $\bvec{y'}$ of degree $l_\mathrm{max} + n_\mathrm{max}$. 
As an example, consider a map of degree $l_\mathrm{max} = 1$ and the linear limb darkening law ($n_\mathrm{max} = 1$).
The transformation matrix from $\bvec{y}$ to $\bvec{y'}$ is
\begin{align}
    \label{eq:ld:L}
    \bvec{L} =
    \frac{1}{1 - \frac{u_1}{3}}
    \begin{pmatrix}
        \quadquad1-u_1\quadquad & 0                       & \frac{u_1}{\sqrt{3}}    & 0                       \\
        0                       & \quadquad1-u_1\quadquad & 0                       & 0                       \\
        \frac{u_1}{\sqrt{3}}    & 0                       & \quadquad1-u_1\quadquad & 0                       \\
        0                       & 0                       & 0                       & \quadquad1-u_1\quadquad \\
        0                       & 0                       & 0                       & 0                       \\
        0                       & \frac{u_1}{\sqrt{5}}    & 0                       & 0                       \\
        0                       & 0                       & \frac{2u_1}{\sqrt{15}}  & 0                       \\
        0                       & 0                       & 0                       & \frac{u_1}{\sqrt{5}}    \\
        0                       & 0                       & 0                       & 0
    \end{pmatrix}
    \quad.
\end{align}
The columns of $\bvec{L}$ are constructed from the coefficient vectors of each transformed spherical harmonic, which are in turn computed by multiplying each spherical harmonic by the spherical harmonic decomposition of the particular limb darkening law.
Figure~\ref{fig:ld} shows this transformation with $u_1 = 1$ applied to each of the first four spherical harmonics.

Now, to incorporate limb darkening into our model, we must modify how we compute the Doppler design matrix $\Doppler$ (Equation~\ref{eq:linear:D} and Figure~\ref{fig:linalg}). 
Recall that this matrix is constructed by vertically stacking the submatrices $\Doppler_m$ given by Equation~(\ref{eq:linear:Dm}), which we now compute as
\begin{align}
    \label{eq:ld:Dm}
    \Doppler_m =
    \begin{pmatrix}
        \quad
        \bvec{T}(\kT_0)
        \quad
         &
        \quad
        \bvec{T}(\kT_1)
        \quad
         &
        \quad
        \cdots
        \quad
         &
        \quad
        \bvec{T}(\kT_{N'})
        \quad
    \end{pmatrix}
    (\bvec{L} \, \bvec{R}(t_m) \otimes \bvec{I}_{K'})
    \quad,
\end{align}
where we have replaced the rotation matrix $\bvec{R}(t_m)$ with the linear transformation $\bvec{L}\,\bvec{R}(t_m)$, which first rotates the input spherical harmonic coefficient vector to the correct phase then limb darkens it. 
Note also that the terms in the convolution kernel $\kT$ are now indexed from 0 to $N' = (l_\mathrm{max} + n_\mathrm{max} + 1)^2$, where $n_\mathrm{max}$ is the degree of the limb darkening operator.

Note that it is also possible to model wavelength-dependent limb darkening in this fashion. 
The procedure is similar, except that the limb darkening coefficients $u_n$ are now functions of $\lnlam$, and each element of the matrix $\bvec{L}$ is now a vector of length $K'$. 
The submatrices $\Doppler_m$ may be written as
\begin{align}
    \label{eq:ld:Dmwav}
    \resizebox{0.91\textwidth}{!}{
        $
            \Doppler_m =
            \begin{pmatrix}
                \quad
                \bvec{T}(\kT_0)
                \quad
                 &
                \quad
                \bvec{T}(\kT_1)
                \quad
                 &
                \quad
                \cdots
                \quad
                 &
                \quad
                \bvec{T}(\kT_{N'})
                \quad
            \end{pmatrix}
            \begin{pmatrix}
                \mathrm{diag}(\bvec{L}_{0,0})
                 &
                \mathrm{diag}(\bvec{L}_{0,1})
                 &
                \cdots
                 &
                \mathrm{diag}(\bvec{L}_{0,N})
                \\
                \mathrm{diag}(\bvec{L}_{1,0})
                 &
                \mathrm{diag}(\bvec{L}_{1,1})
                 &
                \cdots
                 &
                \mathrm{diag}(\bvec{L}_{1,N})
                \\
                \cdots
                 &
                \cdots
                 &
                \ddots
                 &
                \cdots
                \\
                \mathrm{diag}(\bvec{L}_{N',0})
                 &
                \mathrm{diag}(\bvec{L}_{N',1})
                 &
                \cdots
                 &
                \mathrm{diag}(\bvec{L}_{N',N})
            \end{pmatrix}
            (\bvec{R}(t_m) \otimes \bvec{I}_{K'})
        $
    }
    \quad,
\end{align}
where $\mathrm{diag}(\bvec{v})$ denotes the $(K' \times K')$ diagonal matrix constructed by placing the vector $\bvec{v}$ along the main diagonal.

\subsection{Multiple spectral components}
\label{sec:eigen}
In our discussion so far we have focused on the particular case where the spectral/spatial decomposition of the stellar surface can be expressed as the outer product of two vectors (Equation~\ref{eq:inverse:alm}). 
As we mentioned previously, this corresponds to the case where the spectrum is constant across the surface of the star except for a spatially-variable amplitude: i.e., spots have the same spectrum as the rest of the star, but emit less flux overall. 
However, most of the formalism developed here also applies to the more general case where the spectrum at a particular point on the surface of the star is a linear combination of several different spectral components. 
In this case, the map vector $\alm$ may be written as the sum over $J$ spectra $\bvec{s}_j$, each of which is dotted into a different vector of spherical harmonic coefficients $\bvec{y}_j^\top$:
\begin{align}
    \label{eq:eigen:alm}
    \alm
     & =
    \sum_{j=0}^{J-1}\mathrm{vec}\left( \bvec{s}_j \, \bvec{y}_j^\top \right) \\
     & =
    \mathrm{vec}\left( \boldsymbol{\mathbb{S}} \, \boldsymbol{\mathbb{Y}}^\top \right) \quad,
\end{align}
where $\boldsymbol{\mathbb{S}}$ is the $(K' \times J)$ matrix whose columns are the $J$ spectral components and $\boldsymbol{\mathbb{Y}}^\top$ is the $(J \times N)$ matrix whose rows are the corresponding $J$ spherical harmonic coefficient vectors.

The model remains linear in both the spherical harmonic coefficients and the spectral components, and the procedures outlined in \S\ref{sec:inverse} may still be followed to solve the inverse problem, provided we unroll these matrices into vectors, letting $\mathbf{s} = \mathrm{vec}(\boldsymbol{\mathbb{S}})$ and $\mathbf{y} = \mathrm{vec}(\boldsymbol{\mathbb{Y}})$.
We present an example of inference on a two-component stellar spectrum in \S\ref{sec:twospec} below.

\pagebreak

\section{Application to a spotted star}
\label{sec:spotstar}

In this section we apply the formalism developed in \S\ref{sec:inverse} and \S\ref{sec:bellswhistles} to a synthetic dataset of a rapidly rotating spotted star. 
This section is motivated by the examples in \cite{Vogt1987}, who generated rotationally-broadened spectra of a rapidly-rotating star with the letters ``VOGT'' painted across the northern hemisphere. 
In that paper, the authors generated synthetic line profiles of the Fe I $\lambda 6430$ absorption line with equivalent width ${\sim}10$~km/s and from these computed rotationally broadened spectra for a star at an inclination $i=40^\circ$ rotating at projected velocity $v\sin i = 40$~km/s.
The spectra were computed at 16 equally spaced rotational phases and downsampled to 32 wavelength bins per epoch, for a total of 512 data points.

\begin{figure}[t!]
    \begin{centering}
        \includegraphics[width=0.85\linewidth]{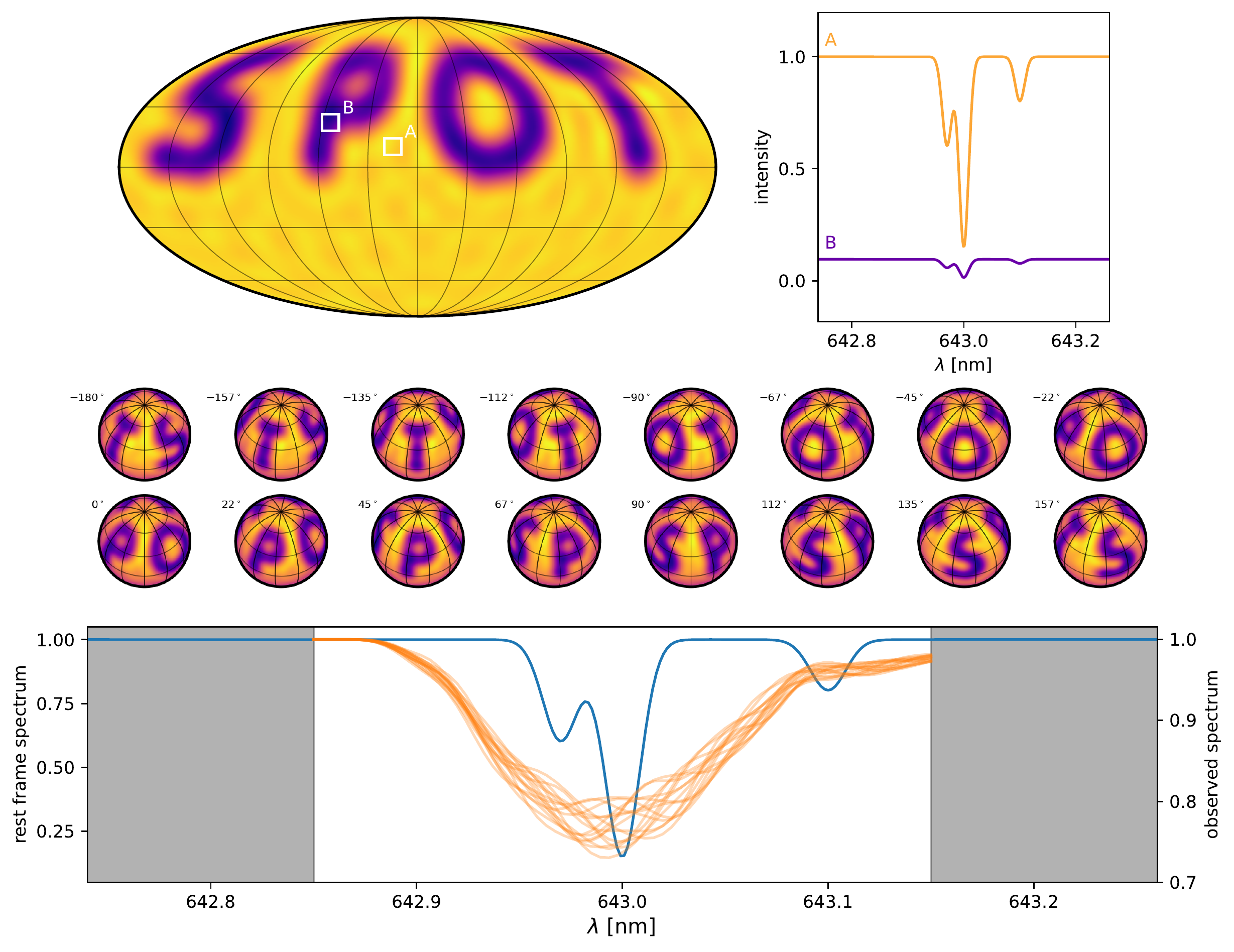}
        \caption{%
            The basic setup for the \spot problem. 
            We compute the $l_\mathrm{max} = 15$ spherical harmonic expansion of a map where the word ``SPOT'' has been placed across the northern hemisphere (top left). 
            Each point on the surface is given the same spectrum, composed of a sum of synthetic Gaussian absorption lines (blue curve in the bottom panel).
            The specific intensity at each point on the surface is equal to this spectrum weighted by the intensity of the map, which is about ten percent within the spot (point labeled \texttt{B}) and close to unity outside the spot (point labeled \texttt{A}).
            To generate the synthetic dataset (orange curves in the bottom panel), the star is inclined at $40^\circ$ with respect to the line of sight and the disk-integrated spectrum is computed at 16 equally-spaced phases between $-180^\circ$ and $180^\circ$, assuming an equatorial velocity $v = 60$ km/s (middle panel).
            The grey regions indicate wavelength regions outside the observation window, but whose features can still contribute to the observed spectrum due to rotational broadening.
        }
        \label{fig:spot_setup}
    \end{centering}
\end{figure}

We repeat the same experiment here using our new formalism, except we instead generate a stellar surface with the word ``SPOT'' written across the northern hemisphere (top left panel of Figure~\ref{fig:spot_setup}).
We do this by computing the $l_\mathrm{max} = 15$ spherical harmonic decomposition of an image of the word, generating a set of $N = (l_\mathrm{max} + 1)^2 = 256$ spherical harmonic coefficients $\bvec{y}$. 
We assume an equatorial velocity $v = 60$ km/s and an inclination $i = 40^\circ$, such that $v \sin i = 38.6$~km/s (compare to $i = 40^\circ$ and $v \sin i = 40$~km/s in \citealt{Vogt1987}).

Our input spectrum $\bvec{s}$ was generated at extremely high resolution
($R \equiv \frac{\lambda}{\Delta\lambda} = 750,000$) 
centered on the Fe I $\lambda 6430$ absorption line (blue curve in the bottom panel of Figure~\ref{fig:spot_setup}).
The spectrum consists of a (purely synthetic) Gaussian absorption line of fractional depth 50\% centered at $\lambda = 643$~nm with FWHM of $0.02$~nm, or, in velocity space, $\Delta v = \mathrm{FWHM} \times c / \lambda \approx~10$~km/s, equal to the value used in \citet{Vogt1987}. 
Unlike \citet{Vogt1987}, we add two additional small absorption lines, one of which is blended with the central absorption line.
In total, the rest frame spectrum $\bvec{s}$ consists of 
$K = 351$ 
wavelength bins in the range $642.85~\mathrm{nm} \leq \lambda \leq 643.15~\mathrm{nm}$ plus an additional
$W - 1 = 252$ 
bins of padding 
(126 on either side) 
to eliminate edge effects due to the convolution, for a total of 
$K' = 603$
wavelength bins. 
Finally, we linearly interpolate this spectrum to a wavelength grid uniform in log space to obtain the vector $\bvec{s}$ evaluated at each $\lnlam$.

Given $\bvec{y}$ and $\bvec{s}$, we use Equations~(\ref{eq:deconv:F}) and~(\ref{eq:inverse:alm}) to compute the data vector $\bvec{f}$, which we downsample to a grid of 
$K_\mathrm{obs} = 70$ 
wavelength bins in the range $642.85~\mathrm{nm} \leq \lambda \leq 643.15~\mathrm{nm}$, corresponding to a resolution of
$R = 150,000$.
We compute $\bvec{f}$ at $M = 16$ equally spaced rotational phases; these phases are shown in orthographic projection (as an observer would see the star) in the center panel of Figure~\ref{fig:spot_setup}. 
The synthetic spectra for each phase are shown as the orange curves in the bottom panel of the figure (note the different scaling, as all spectral lines become shallower due to the broadening). 
We assume a wavelength-independent, quadratic limb darkening law with $u_1 = 0.5$ and $u_2 = 0.25$. 
Finally, we divide the flux by the baseline at each epoch (Equation~\ref{eq:norm:f}) and add random Gaussian noise with covariance matrix $\boldsymbol{\Sigma}_\bvec{f} = \sigma_\bvec{f}^2 \mathbf{I}$, with $\sigma_\bvec{f} = 2\times 10^{-4}$ to simulate a very high signal-to-noise ratio (SNR) spectrum. 
The typical fractional line depth in the observed spectrum is 0.2, so our choice of $\sigma_\bvec{f}$ corresponds to SNR $\sim 10^{3}$ in each of the 70 wavelength bins of $\bvec{f}$.

In the following sections we discuss various approaches to solving the Doppler imaging problem for this dataset.
Unless otherwise noted, in all experiments we adopt a Gaussian prior for the spherical harmonic coefficients $\bvec{y}$ with mean $\boldsymbol{\mu}_\bvec{y} = \left(1 \quad\quad \bvec{0}\right)$ and covariance $\boldsymbol{\Lambda}_\bvec{y} = 10^{-4} \bvec{I}$, corresponding to a flat power spectrum prior.
When solving for the rest frame spectrum, we also place a Gaussian prior on $\bvec{s}$ with unit mean and covariance $\boldsymbol{\Lambda}_\bvec{s} = 10^{-3}\bvec{I}$.
These choices are arbitrary and primarily based on trial-and-error given knowledge of the true solutions.
We discuss strategies for choosing priors for real datasets in \S\ref{sec:discussion:priors} below.

\subsection{Known baseline and spectrum}
\label{sec:spot_y1}
In our first experiment, we attempt to infer the stellar map assuming we have perfect knowledge of the rest frame spectrum $\bvec{s}$ and the baseline, as well as the stellar inclination, rotation period, and limb darkening coefficients.
This is analogous to the case where a good physical model exists for the stellar spectrum and simultaneous photometry is available to constrain the baseline variations.
As we showed in \S\ref{sec:solve_y}, the posterior distribution for the map coefficients $\bvec{y}$ is analytic in this case, with mean given by Equation~(\ref{eq:inverse:y1hat}) and covariance given by Equation~(\ref{eq:inverse:y1cov}). 
Note that we must first scale the data vector $\fnorm$ by the baseline (which in this case is known) to obtain $\bvec{f}$.

\begin{figure}[p!]
    \begin{centering}
        \includegraphics[width=0.39\linewidth]{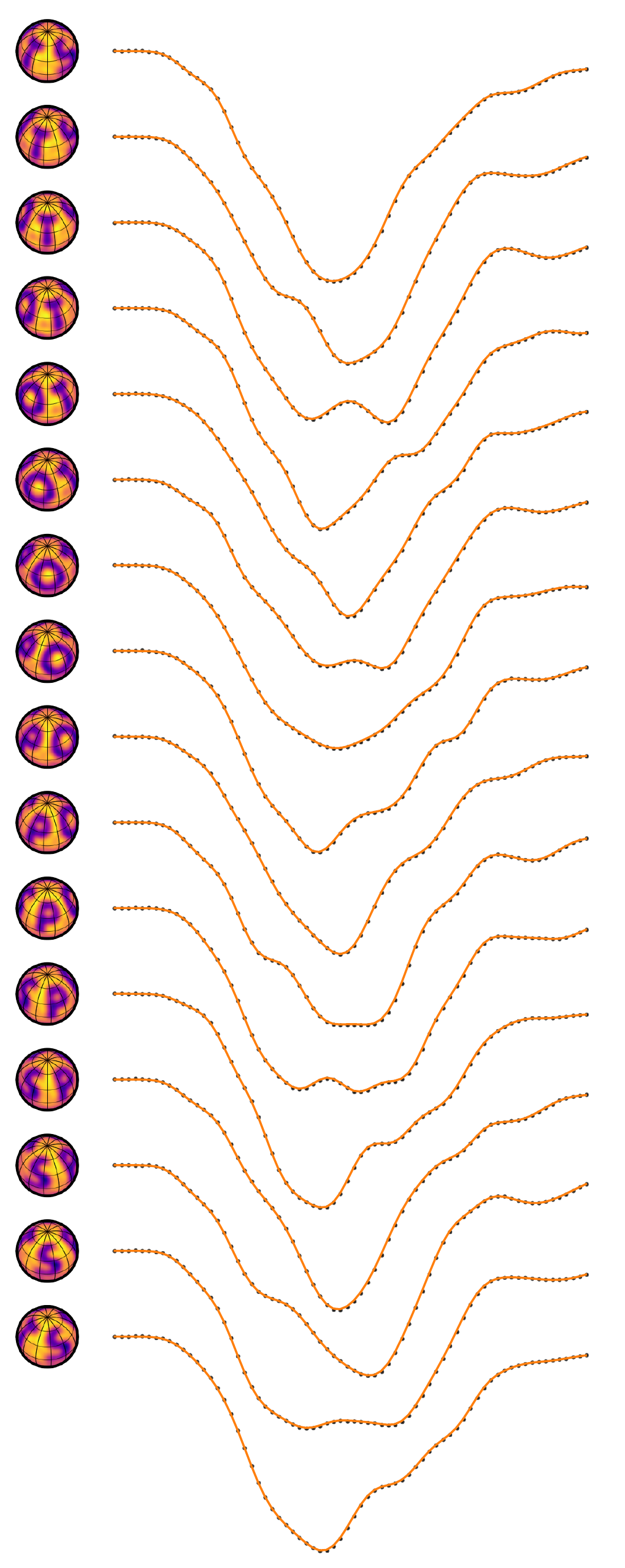}
        \includegraphics[width=0.59\linewidth]{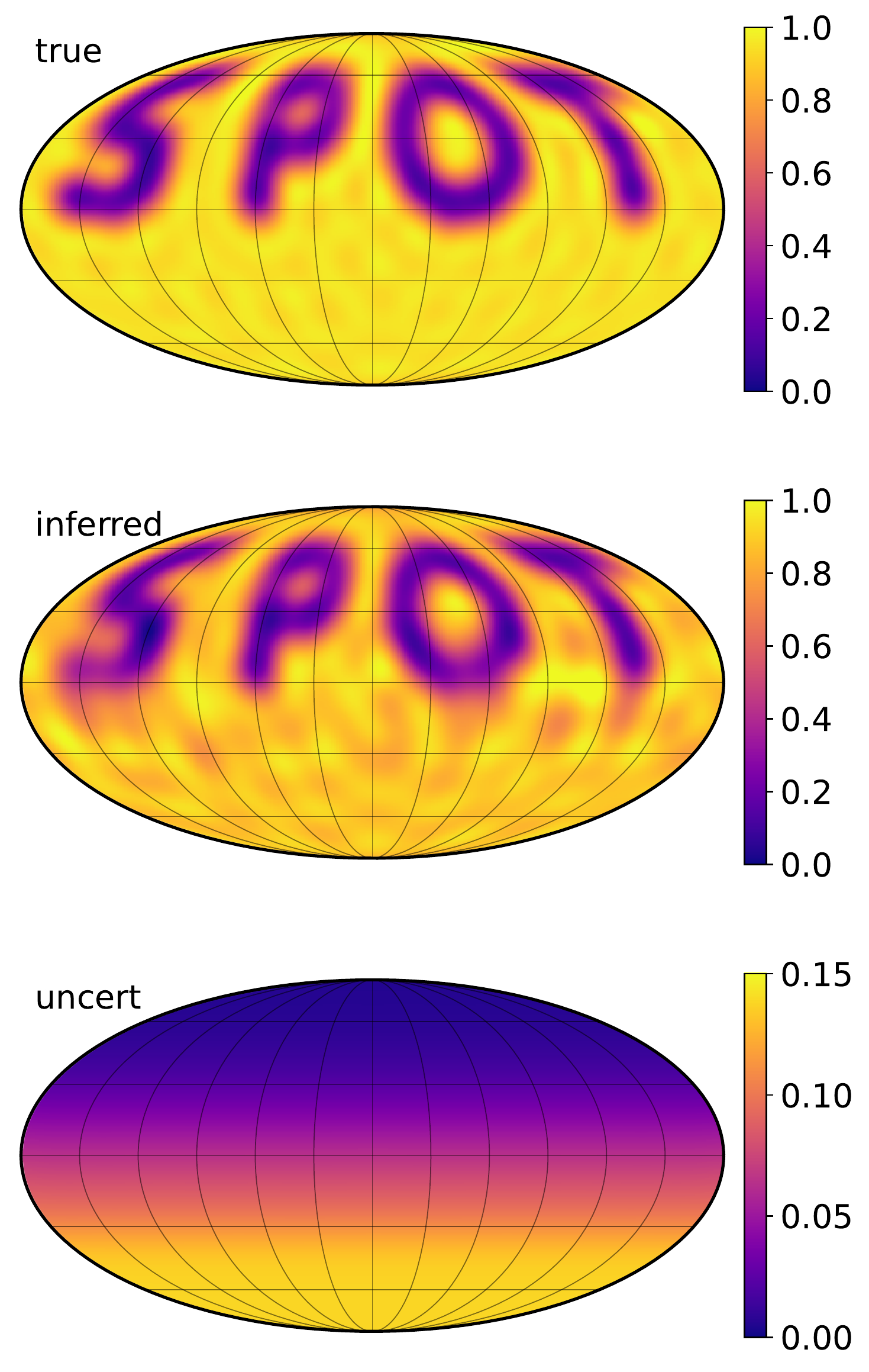}
        \caption{%
            The solution to the \spot problem at high SNR when both the rest frame spectrum and the baseline are known.
            The panel on the left shows the inferred map at each of the 16 epochs, alongside the observed spectrum (black points) and the best-fit model (orange lines). 
            The Mollweide projection of the original map is shown at the top right; the inferred map is displayed below it. 
            The bottom panel shows the posterior uncertainty on the inferred intensity, ranging from close to zero in the northern hemisphere to $\sigma_\mathrm{prior} \approx 13\%$ below a latitude of $-30^\circ$. 
            The latter is the uncertainty associated with our particular choice of prior, as the southernmost latitudes of the star are never visible to the observer.
            The optimization step in this experiment ran in \timeInferY on a two-core virtual machine.
        }
        \label{fig:spot_infer_y}
    \end{centering}
\end{figure}

Figure~\ref{fig:spot_infer_y} shows the results.
The left hand panel shows the inferred map alongside the data (black points) and best fit model (orange curves) at each of the 16 epochs.
The panels on the right show the true map (top), the inferred map (center) and the posterior pointwise uncertainty (bottom) on a Mollweide grid. 
The uncertainty on the vector of pixels $\bvec{p}$ where we evaluate the map is equal to the square root of the diagonal of the posterior covariance matrix for the pixel values, computed from
\begin{align}
    \label{eq:spotstar:uncert}
    \boldsymbol{\Sigma}_\bvec{p} = \bvec{P} \boldsymbol{\Sigma}_\bvec{\hat{y}}\bvec{P}^\top
\end{align}
where $\bvec{P}$ is the change-of-basis matrix from spherical harmonic coefficients to pixels on the grid.

The inferred map very closely matches the true map, aside from a few small artifacts in the southern hemisphere.
Note, importantly, that regions south of the equator are almost never in view to the observer (below $-40^\circ$, they are \emph{never} in view); our inferred map is therefore informed primarily by our prior in southern latitudes.
This is well captured by the uncertainty (bottom panel), which increases south of the equator, reaching a maximum at $-30^\circ$, below which the uncertainty is constant at about 13\%, a value associated with our particular choice of prior.

\subsection{Known spectrum, unknown baseline}
\label{sec:spot_y1b}
We next investigate the case where the spectrum is known exactly, but the baseline is not. 
This corresponds to the case discussed in \S\ref{sec:norm}, in which each spectrum is normalized to its continuum, and the true difference in continuum levels across spectral epochs is not precisely known.
This is likely to be the most common application of the methodology presented in this work.

Our approach is to iteratively solve Equation~(\ref{eq:norm:fy}) by re-expressing it as
\begin{align}
    \fnorm
     & =
    \frac{
        \bvec{D}
        \,
        \bvec{S}
        \,
        \bvec{y}
    }{
        \bvec{b}
    }
    \quad.
\end{align}
where the baseline $\bvec{b}$ is given by
\begin{align}
    \bvec{b}
    =
    \bvec{B}
    \,
    \bvec{S}
    \,
    \bvec{y}
    \quad.
\end{align}
We adopt a simple iterative scheme where we solve the L2 problem for $\bvec{y}$ with $\bvec{b}$ fixed, then update the value of $\bvec{b}$ and repeat.
Because of the high SNR of our data, we find that it is necessary to employ a simple tempering procedure, in which we artificially inflate the variance of the data by a factor $T$, which we slowly decrease with each step until $T = 1$. 
This has the effect of smoothing out the likelihood surface, effectively preventing the solver from getting stuck in local minima.
For this particular experiment, we set $\ln T_0 = 2$ and $\Delta \log T = -0.04$, for a total of $50$ steps of the bilinear solver.
We also add a small variance term $\sigma_b^2 = 10^{-2}$ to all entries of the data covariance $\boldsymbol{\Sigma}_{\bvec{f}}$; this has the effect of marginalizing over a small additive model term, which we find improves the convergence of the algorithm.

\begin{figure}[p!]
    \begin{centering}
        \includegraphics[width=0.39\linewidth]{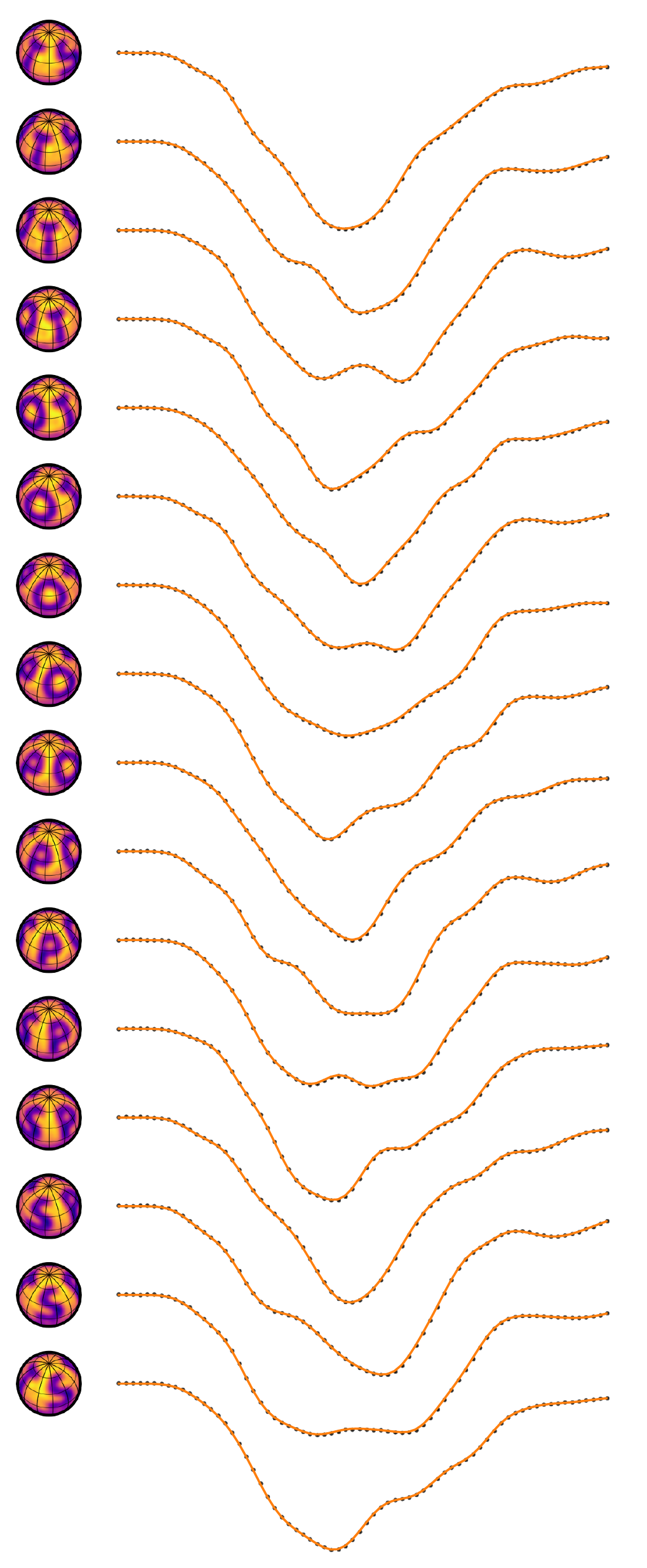}
        \includegraphics[width=0.59\linewidth]{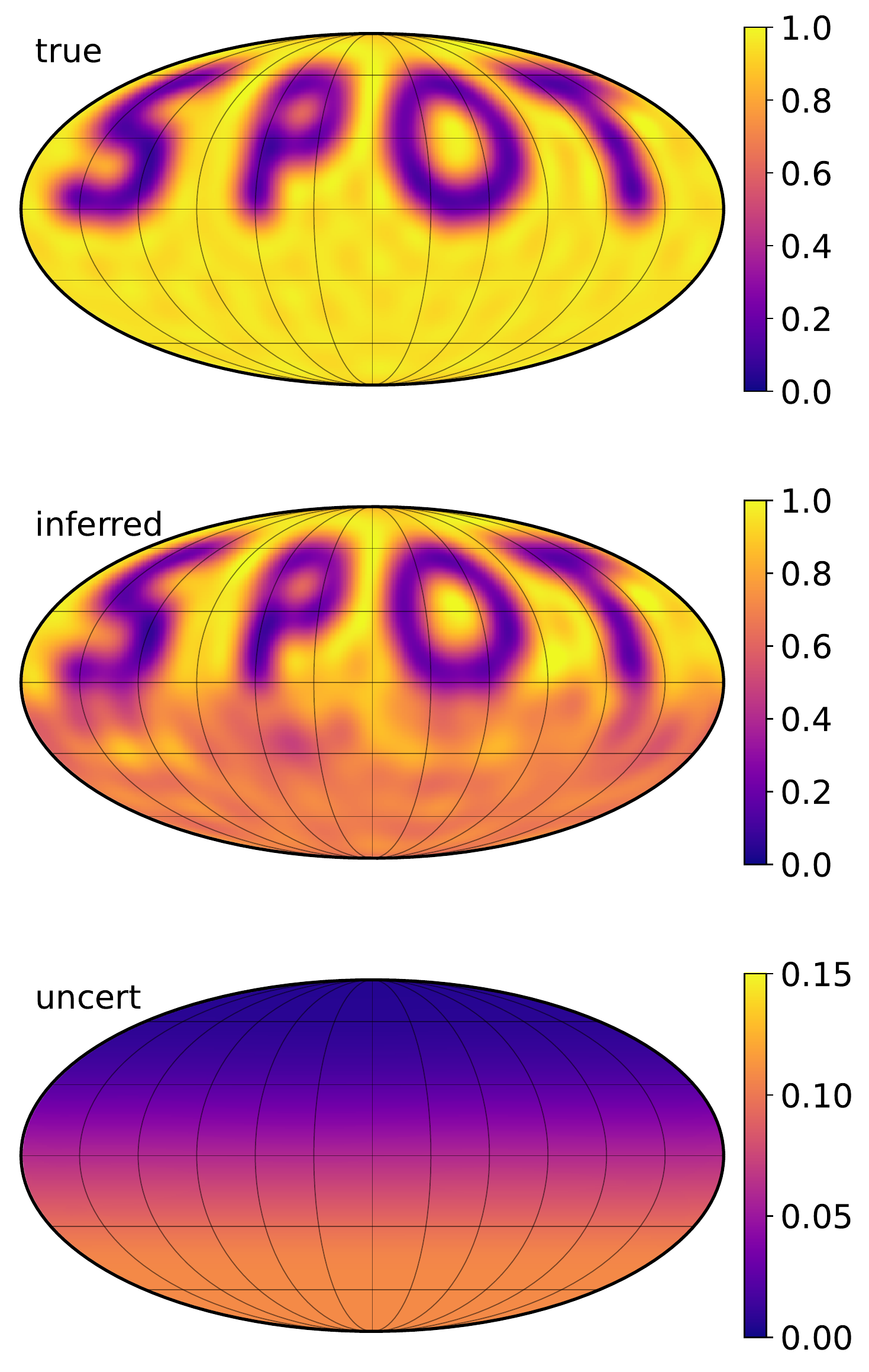}
        \caption{%
            Similar to Figure~\ref{fig:spot_infer_y}, but assuming the input spectra are continuum-normalized.
            We are able to infer the surface map just as well as in the case where the true continuum level is known (Figure~\ref{fig:spot_infer_y}). 
            As before, regions south of the equator are almost never in view, so that portion of our inferred map is dominated by the prior.
            Note that the uncertainty shown (lower panel) is a strict lower bound on the true uncertainty, as it is conditioned on our point estimate of the baseline level.
            The optimization step in this experiment ran in \timeInferYB.
        }
        \label{fig:spot_infer_yb}
    \end{centering}
\end{figure}

Our results are shown in Figure~\ref{fig:spot_infer_yb}.
As in the case where the continuum level was known exactly (Figure~\ref{fig:spot_infer_y}), we recover the stellar surface map with high fidelity in the northern hemisphere.
However, there is now a strong dipolar artifact; the background intensity level is higher in the northern hemisphere than in the southern hemisphere.
This is due to a degeneracy involving the $Y_{1,-1}$ spherical harmonic (\raisebox{-0.2em}{\includegraphics[width=1em]{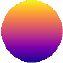}}), which, because a large portion of the southern hemisphere is never in view, cannot be constrained by the data.

Interestingly, while the map uncertainty still increases toward the south pole (bottom panel), its magnitude is smaller than that in Figure~\ref{fig:spot_infer_y}.
This is counter-intuitive, since the continuum-normalized spectrum should contain \emph{less} information than the unnormalized one.
Indeed, the uncertainties shown in the figure are conditioned on our estimate of the baseline, so they are actually a strict \emph{lower bound} on the map uncertainty.
In order to constrain the true uncertainty on the solution to this bilinear problem, we must resort to approximate sampling methods such as Markov Chain Monte Carlo (MCMC). 
We discuss this further in \S\ref{sec:discussion:priors}.

\subsection{Unknown spectrum, unknown baseline}
\label{sec:spot_y1bs}
Suppose we are handed a series of spectra in a wavelength regime where we do not have a good physical model for the line generation mechanisms, or a wavelength regime with many line blends for which we do not have reliable templates. 
Or perhaps the spectra are of a star whose rotational broadening does not dominate over other broadening mechanisms, in which case our template might not be accurate enough to allow us to tease out the small Doppler signals. 
In these cases, we may want to relax the assumption that we know the rest frame spectrum $\bvec{s}$ exactly, and instead try to \emph{learn} it from the data.

Given the same data as before, we now try to learn both the stellar map and the stellar spectrum, conditioned only on the wide Gaussian priors discussed above. 
We must follow the iterative procedure presented in
\S\ref{sec:full_solve}, adjusting for the fact that we do not know the true baseline, either (\S\ref{sec:norm}).

\begin{figure}[p!]
    \begin{centering}
        \includegraphics[width=0.39\linewidth]{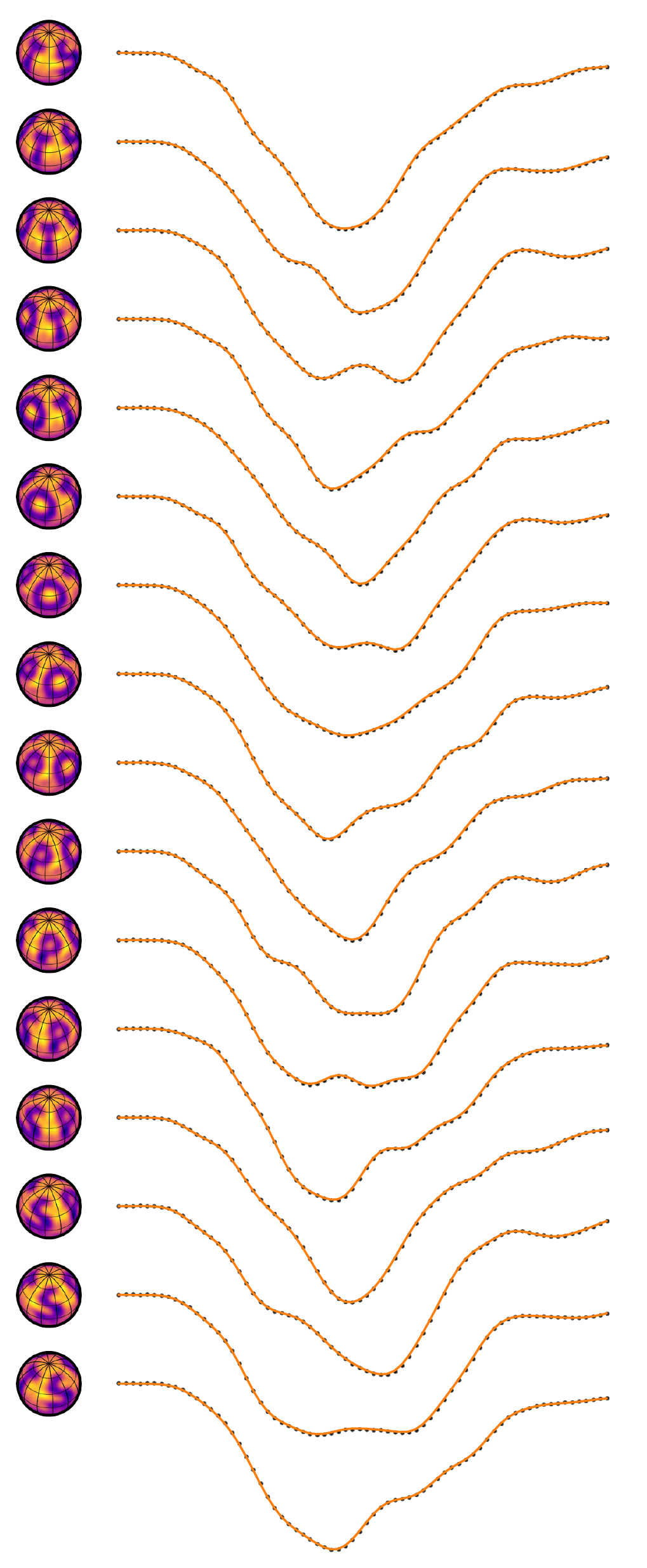}
        \includegraphics[width=0.59\linewidth]{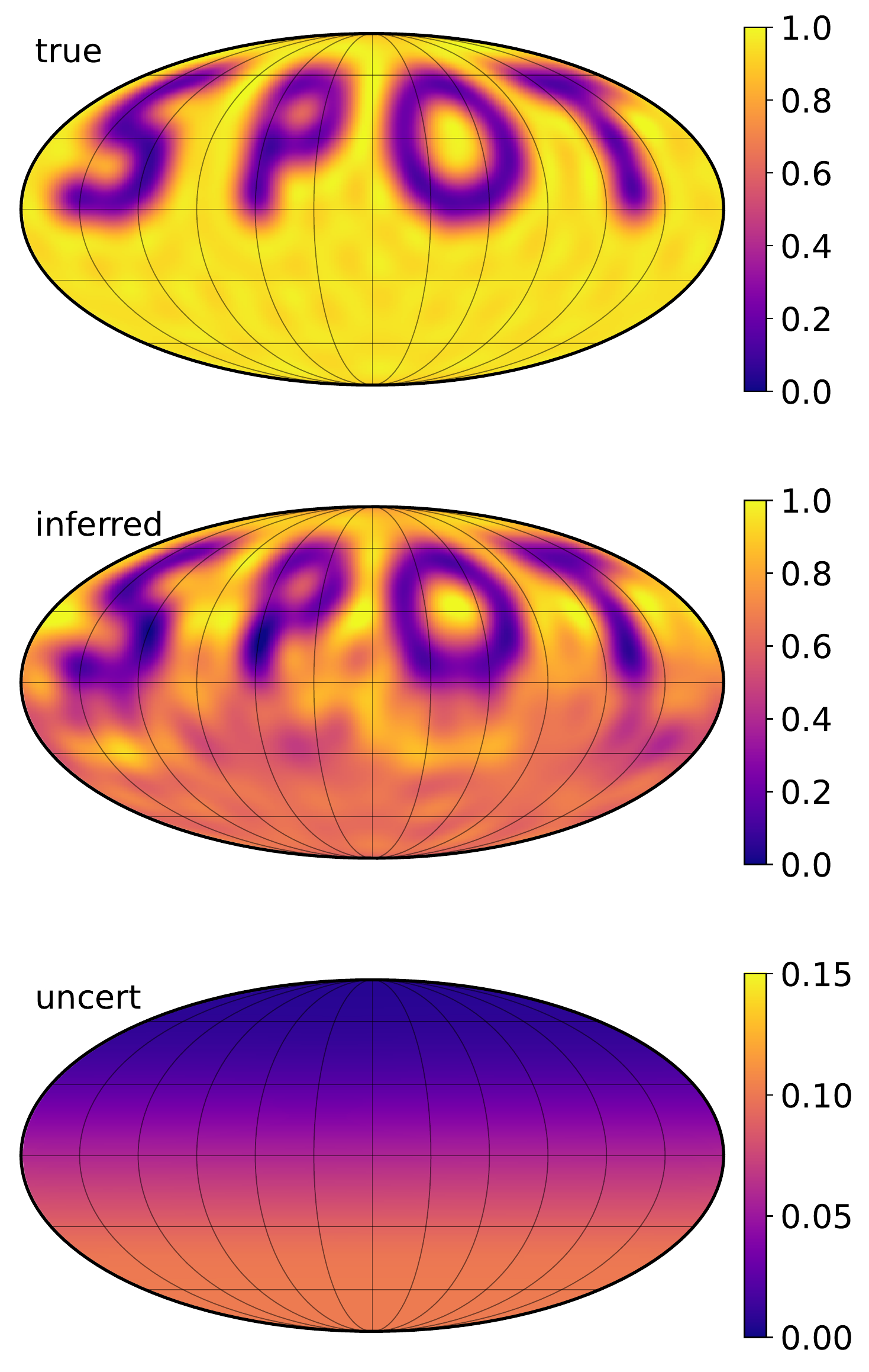}
        \vspace{1em}
        \includegraphics[width=\linewidth]{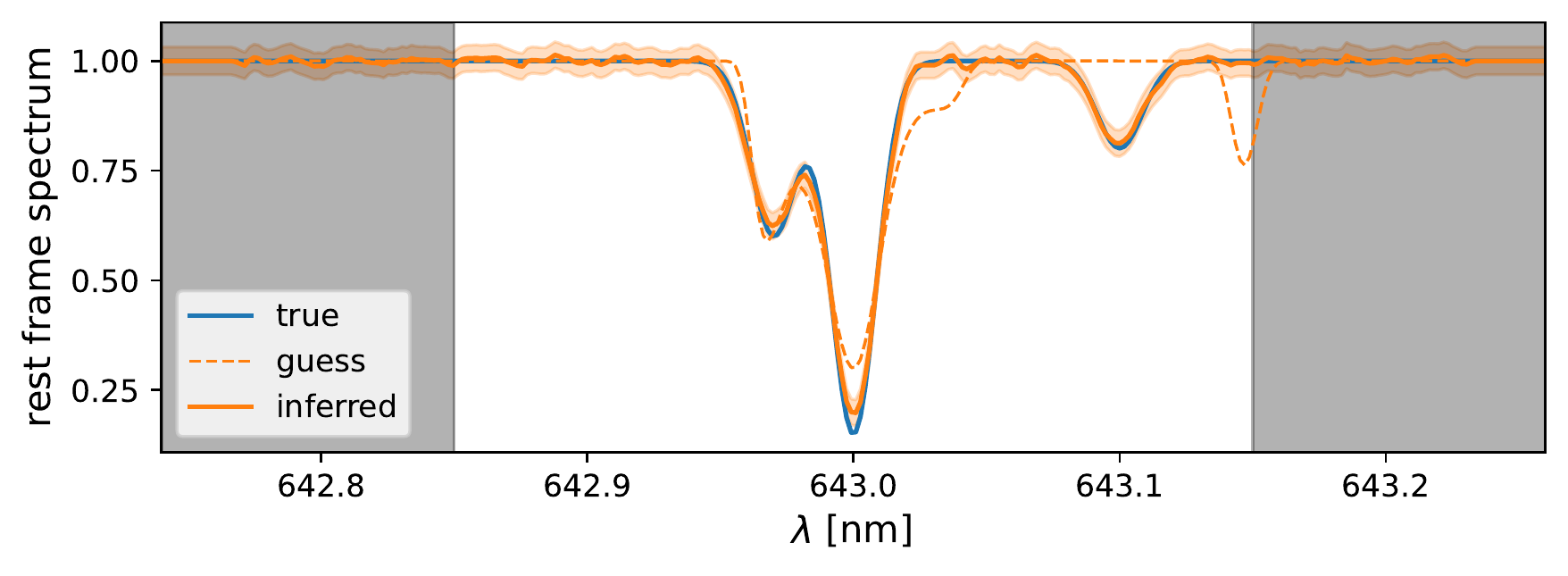}
        \caption{%
            Similar to Figure~\ref{fig:spot_infer_y}, but assuming we have no information about either the spectrum, the baseline, or the map coefficients. 
            The bottommost panel now shows the true rest frame spectrum (blue line), the initial guess based on a deconvolution (dashed orange line), and the inferred spectrum (orange line, with shaded $1\sigma$ uncertainty).
            The optimization step in this experiment ran in \timeInferYBS.
        }
        \label{fig:spot_infer_ybs}
    \end{centering}
\end{figure}

\begin{figure}[p!]
    \begin{centering}
        \includegraphics[width=0.39\linewidth]{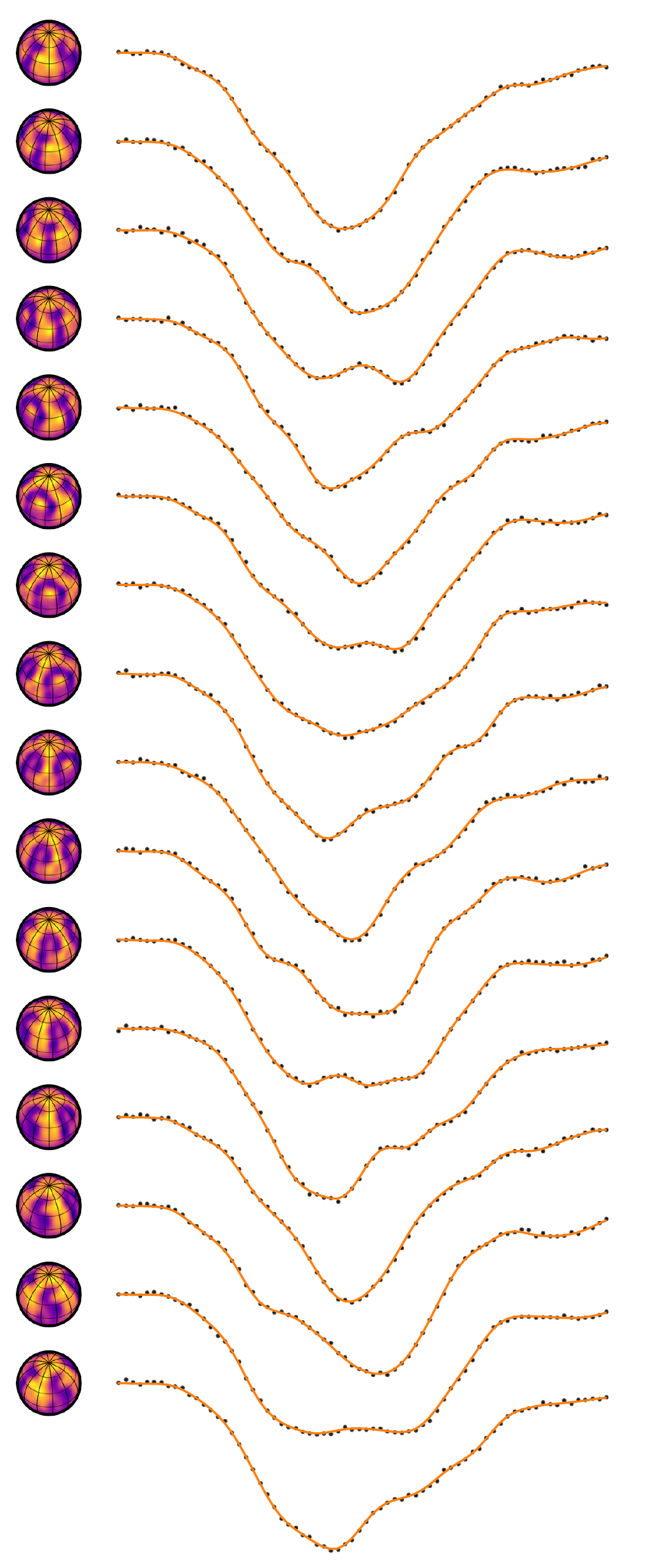}
        \includegraphics[width=0.59\linewidth]{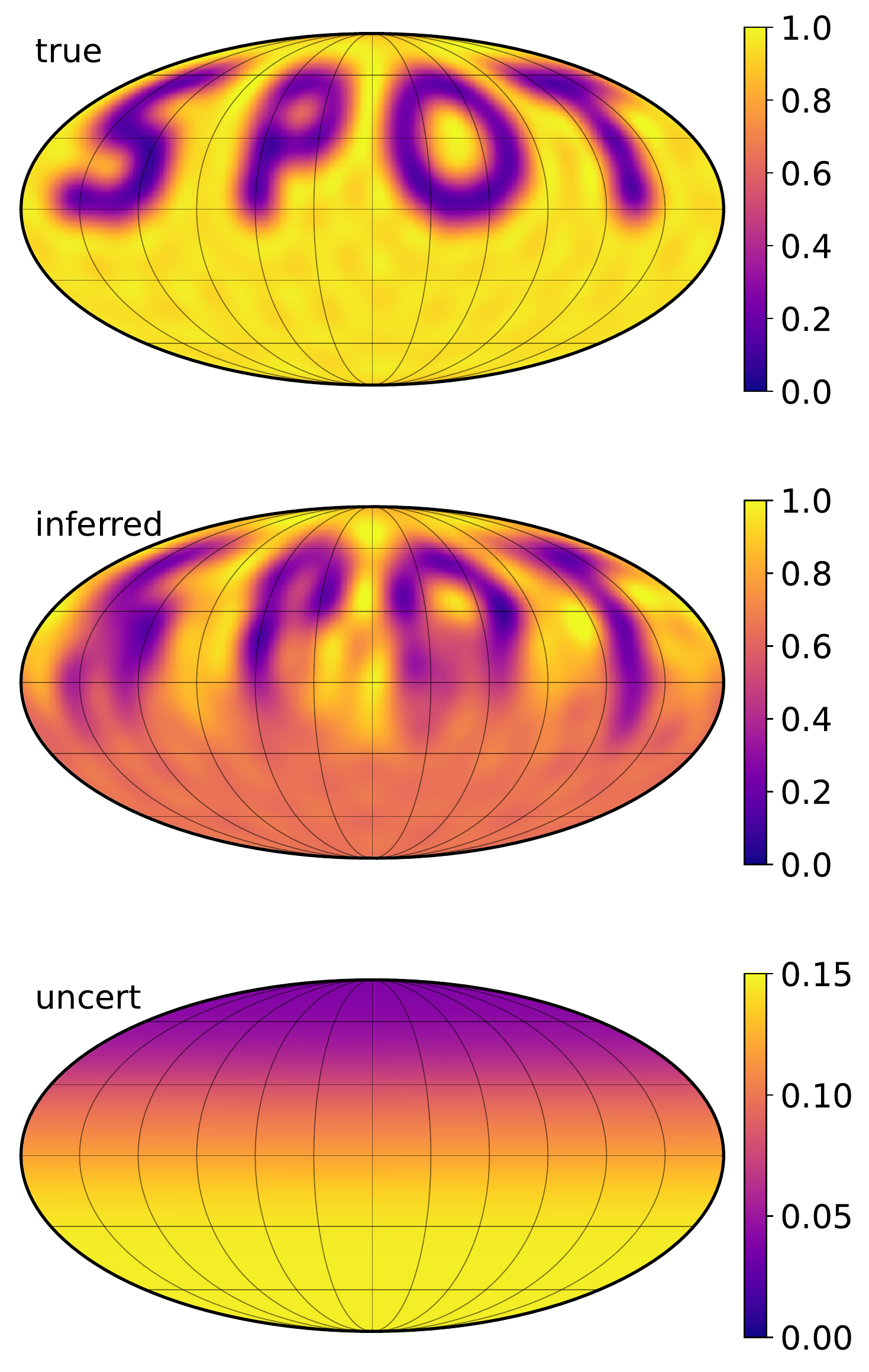}
        \includegraphics[width=\linewidth]{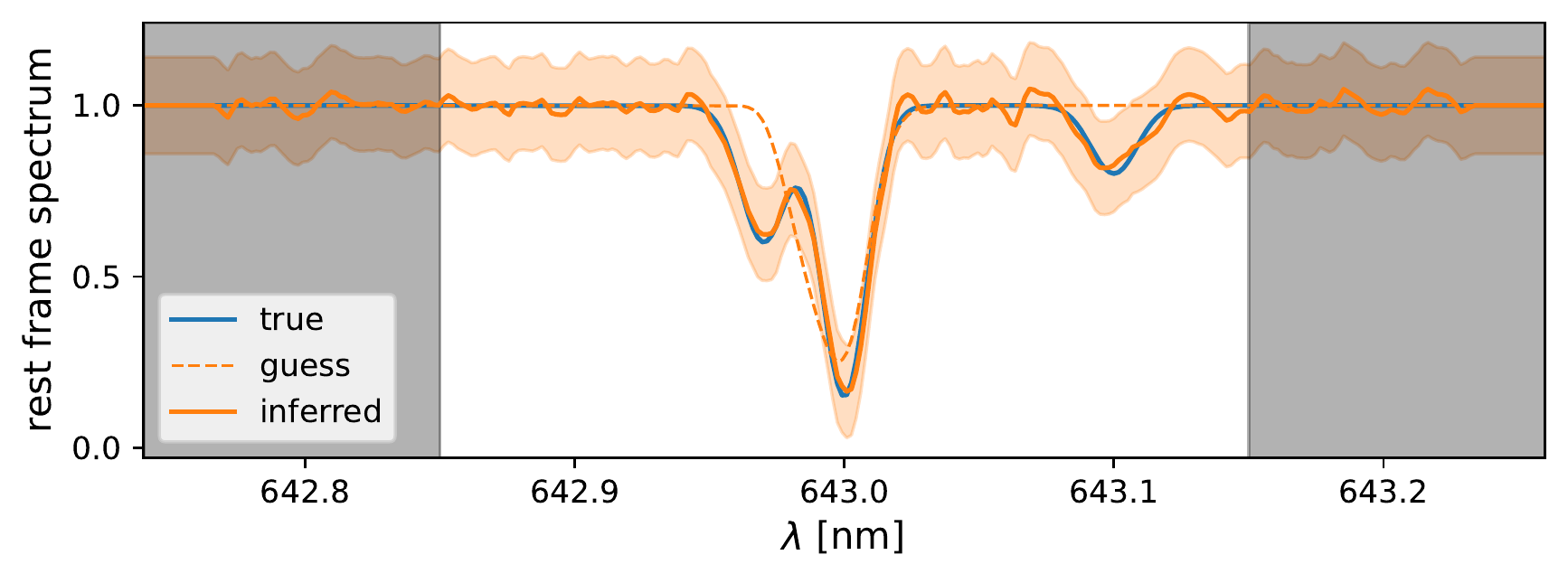}
        \caption{%
            Same as Figure~\ref{fig:spot_infer_ybs}, but for $10\times$ lower SNR.
            We still recover the \spot map and the input spectrum, albeit with higher uncertainty, as expected.
            Once again, the optimization step in this experiment ran in \timeInferYBSLOSNR.
        }
        \label{fig:spot_infer_ybs_low_snr}
    \end{centering}
\end{figure}

Because the problem is generally not convex, we need a good starting guess for either the spectrum or the map coefficients. 
We note that we can very crudely approximate the rest frame spectrum by deconvolving the average observed spectrum with the first term of the convolution kernel $\kT$ (\raisebox{-0.1em}{\includegraphics[width=2em]{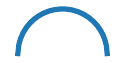}}); we do this by solving the linear problem in Equation~(\ref{eq:linear:convolution_def}) with aggressive regularization.
In practice, this is a fairly poor approximation, but we find that it suffices as an initial guess. 
We then iterate between solving for $\bvec{y}$ (\S\ref{sec:solve_y}) and solving for $\bvec{s}$ (\S\ref{sec:solve_s}), followed by a step in which we update our estimate of the baseline.
We adopt the same tempering scheme discussed above.

Figure~\ref{fig:spot_infer_ybs} shows the results; the top portion of the figure is the same as in Figure~\ref{fig:spot_infer_yb}.
A panel has been added at the bottom showing the true rest frame spectrum (blue), the initial guess based on the deconvolution (dashed orange), and the inferred spectrum with 1-$\sigma$ uncertainty shading (orange). 
We recover the \spot feature as before, albeit with more noise, particularly south of the equator. 
Once again, our map uncertainties are somewhat underestimated, as expected, since they are computed conditioned on the optimum value of the baseline and the spectrum. 
On the other hand, we recover the spectrum exceedingly well, with low uncertainty and virtually no bias anywhere.

To test our technique on more realistic data, we repeat the experiment on data with $10$ times higher uncertainty, setting $\sigma_\bvec{f} = 2\times 10^{-3}$.
After some tuning, we find it necessary to increase our prior variances slightly to get good fits to the data: we set $\boldsymbol{\Lambda}_\bvec{y} = 2\times 10^{-4} \bvec{I}$ and $\boldsymbol{\Lambda}_\bvec{s} = 2\times 10^{-2}\bvec{I}$.
Figure~\ref{fig:spot_infer_ybs_low_snr} shows the results.
We again correctly infer the \spot feature and the rest frame spectrum, although the surface map is considerably degraded, particularly at high spatial frequencies.

\subsection{Spatially-variable spectrum}
\label{sec:twospec}
In our final experiment, we relax the assumption that the stellar rest frame spectrum is spatially constant, allowing for different spectral features inside and outside of the \spot. 
This is the case for, say, chemically peculiar Ap stars, whose abundance of certain elements is significantly different within spotted regions on their surfaces. 
It is also the case in general for spots that are significantly cooler than the rest of the photosphere, where the relative abundance of different ionization states or molecular features may be different than in the hotter, unspotted regions.

The problem setup is the same as that discussed in \S\ref{sec:eigen}, where our full map vector $\bvec{a}$ is the outer product of a spectral matrix $\mathbb{S}$, whose $J$ columns are the different spectral components, and a spherical harmonic coefficient matrix $\mathbb{Y}^\top$, whose $J$ rows are the corresponding coefficient vectors. 
For simplicity, in this experiment we consider only $J = 2$ map components and further require that the surface map for the second component is the complement of the surface map for the first component.
Specifically, the first component map ($\bvec{y_0}$, the first row of $\mathbb{Y}^\top$) corresponds to the background photosphere, whose value is unity everywhere except within the \texttt{SPOT} feature, where the value is zero.
The second component map ($\bvec{y_1}$, the second row of $\mathbb{Y}^\top$) corresponds to the spot and is zero in the photosphere and unity within the spot.
The two columns of $\mathbb{S}$ are $\bvec{s_0}$ and $\bvec{s_1}$, the unknown spectral components corresponding to the photosphere and the spot, repsectively.
\begin{figure}[p!]
    \begin{centering}
        \includegraphics[width=0.39\linewidth]{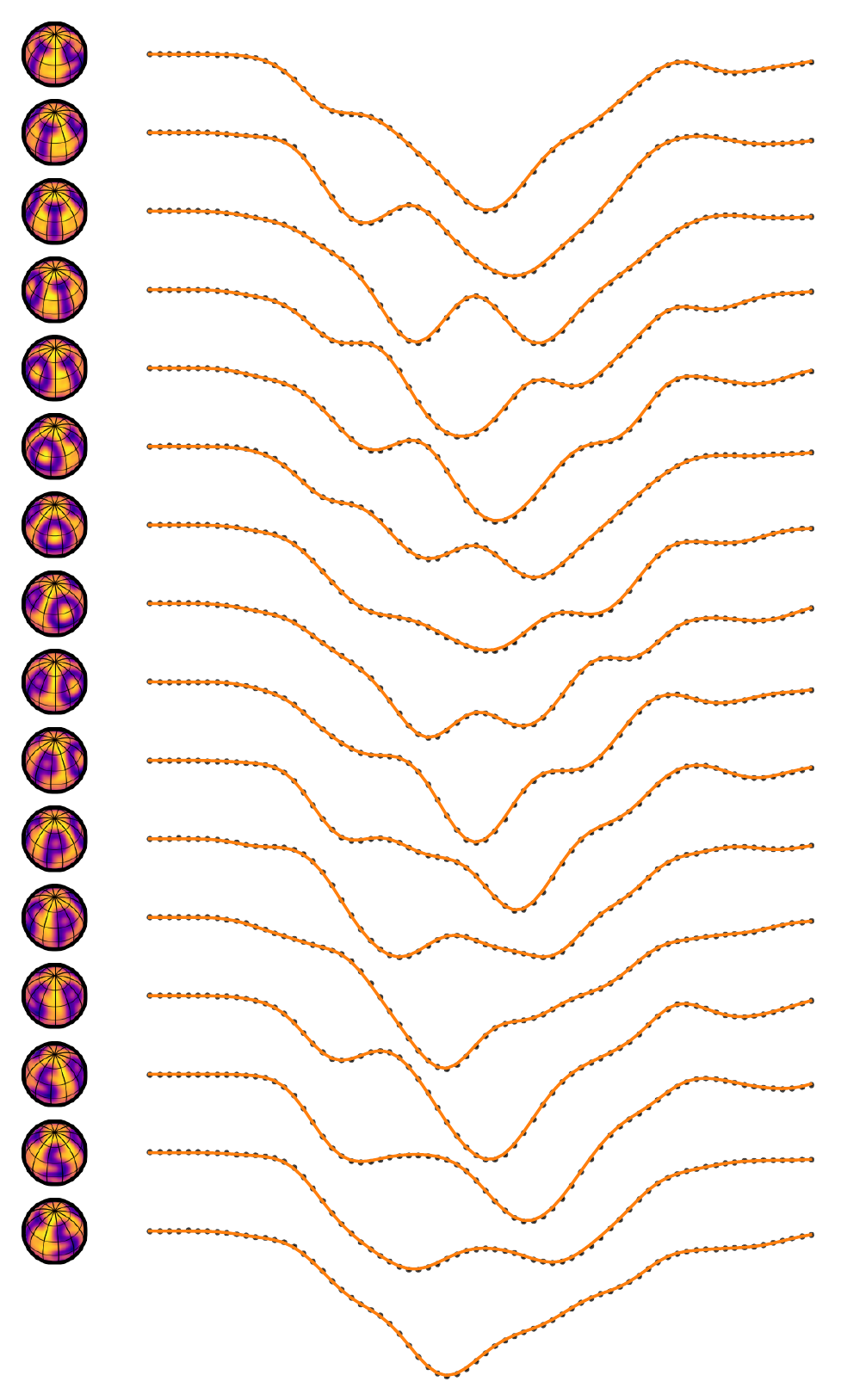}
        \includegraphics[width=0.59\linewidth]{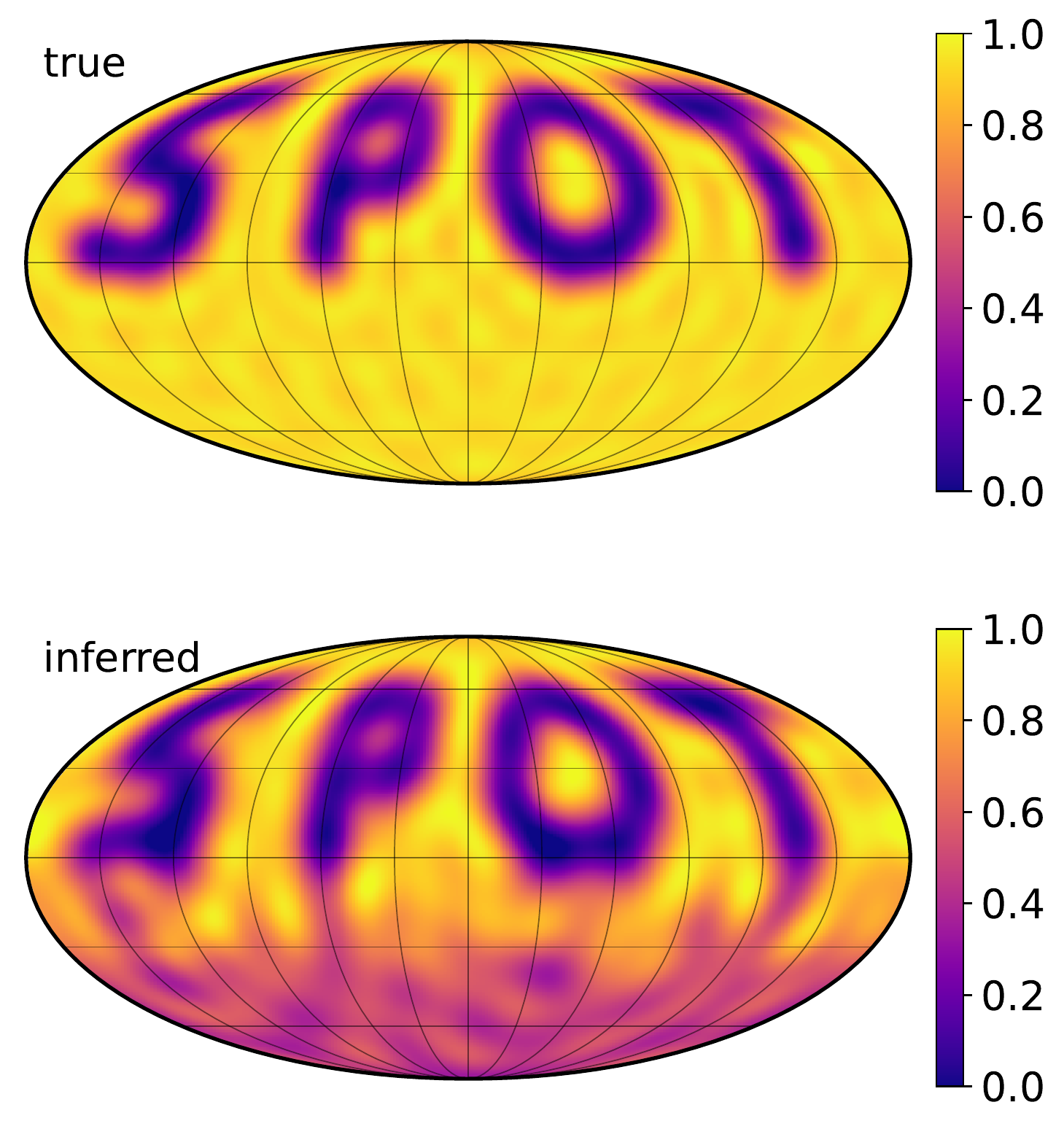}
        \includegraphics[width=\linewidth]{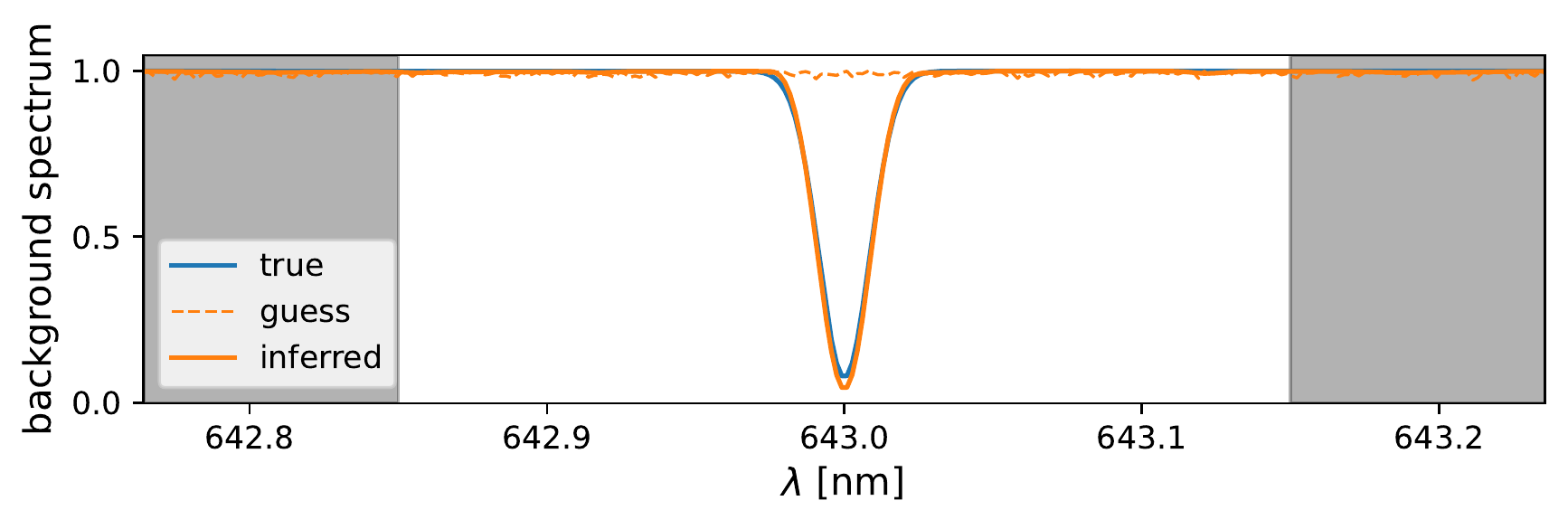}
        \includegraphics[width=\linewidth]{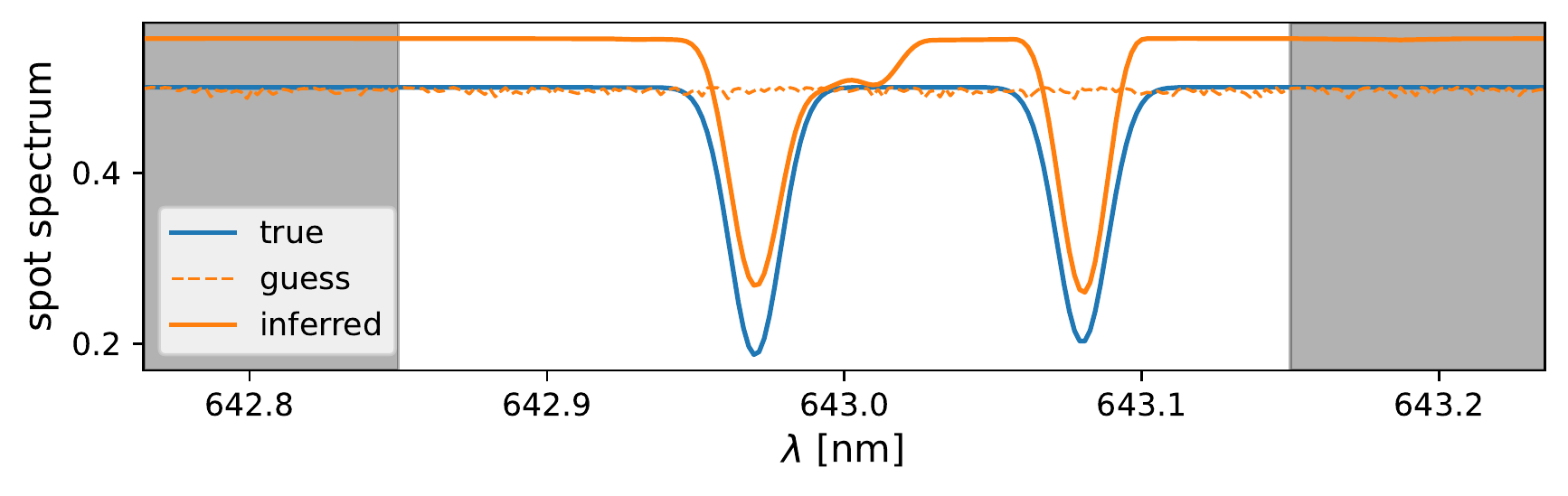}
        \caption{%
            An example of a star whose spectrum is different within the spot and outside of the spot.
            We are able to infer both the surface map of the star (upper right) and the two component spectra (lower two panels), assuming virtually no prior knowledge of either.
            Starting from a random initial guess for the maps and the spectra, the optimization step in this experiment ran in \timeInferTwoSpec.
        }
        \label{fig:twospec}
    \end{centering}
\end{figure}
We set $\bvec{s_0}$ to unity everywhere, with an absorption feature at $643$~nm similar to the one in the previous experiments.
The spot spectrum $\bvec{s_1}$ corresponds to two absorption lines on either side of this feature, and the baseline intensity is set to half that of the photospheric background.
The rest frame spectrum at any point on the surface of the star is then a linear combination of $\bvec{s_0}$ and $\bvec{s_1}$, weighted by the respective surface maps $\bvec{y_0}$ and $\bvec{y_1}$.
We generate the synthetic dataset in the same way as in \S\ref{sec:spot_y1}: 16 spectra, each with 70 wavelength bins observed with SNR $\sim 10^{3}$.

In our inference step, we again assume we have no knowledge of the map, the baseline, or either of the spectral components.
While our bilinear method (\S\ref{sec:spot_y1bs}) could in principle work here, we find that in practice it is very difficult to converge on the global optimum without either good priors or a good initial guess.
We therefore optimize our model using the nonlinear gradient-descent \texttt{Adam} optimizer \citep{Adam} within the \texttt{pymc3} probabilistic modeling framework \citep{Salvatier2016}.
Since we are no longer restricted to using Gaussian priors on the variables (a requirement of our bilinear method), we place a unit-mean Laplacian (sparsity) prior on the two spectra with scale parameter $b = 10^{-3}$ and bounded to the range $[0, 1]$.
We further place a Gaussian process (GP) prior on the spectra to enforce smoothness, adopting a squared exponential kernel with amplitude $10^{-1}$ and lengthscale $10^{-2}$~nm.
We also solve for a single scale parameter $r \sim \mathrm{Uniform}(0, 0.75)$, equal to the ratio of the continuum levels of the two spectra (whose true value is 0.5).
Finally, in order to enforce that the two component surface maps sum to unity (as described above), we optimize for the actual \emph{pixel} intensities of the photosphere map, on which we place a simple uniform prior on $[0, 1]$; the spot map is then simply the complement.
The spherical harmonic coefficients are then computed via a fast, linear spherical harmonic transform (SHT) operation at each step, as in \citet{Bartolic2021}.
In order to satisfy the Nyquist criterion, our pixel grid consists of twice as many pixels as spherical harmonic coefficients.%
\footnote{%
While it may seem strange to return to a discretized parametrization of the surface---after all the work we did to derive a closed-form solution in the spherical harmonic basis---our approach capitalizes on the strengths of both parametrizations: the ease of incorporating a nonnegativity prior in the pixel basis and the efficiency with which we can compute the model in the spherical harmonic basis.
Thanks to the linear relationship between these two bases, the transformation between them accrues virtually no overhead.}

We initialize our two maps with uniform intensities and initialize our two spectra at unity, with small random negative deviations to break the symmetry.
We then take 20,000 steps of the optimizer with a learning rate of $10^{-2}$.

The results are shown in the bottom panel of Figure~\ref{fig:twospec}.
Even though we assume no knowledge of the surface map---except its intensity is bounded in $[0, 1]$---we are able to recover the \spot feature just as well as in the previous experiments.
We also recover the two spectral components quite well.
Our solution for the photospheric component, however, is noticeably better than that for the spot component, simply because effective SNR of the former is larger due to its greater surface area and higher baseline intensity.
We also slightly overestimate the baseline level of the spot spectrum, but we correctly recover the locations, shapes, and depths of the two absorption lines.
Note, importantly, that the nonlinear optimization step does not yield uncertainties (or lower bounds, like the bilinear solver); for this, we must resort to MCMC sampling (see \S\ref{sec:discussion:priors}).

\subsection{Dependence on SNR}
\label{sec:snr}
\begin{figure}[t!]
    \begin{centering}
        \includegraphics[width=0.98\linewidth]{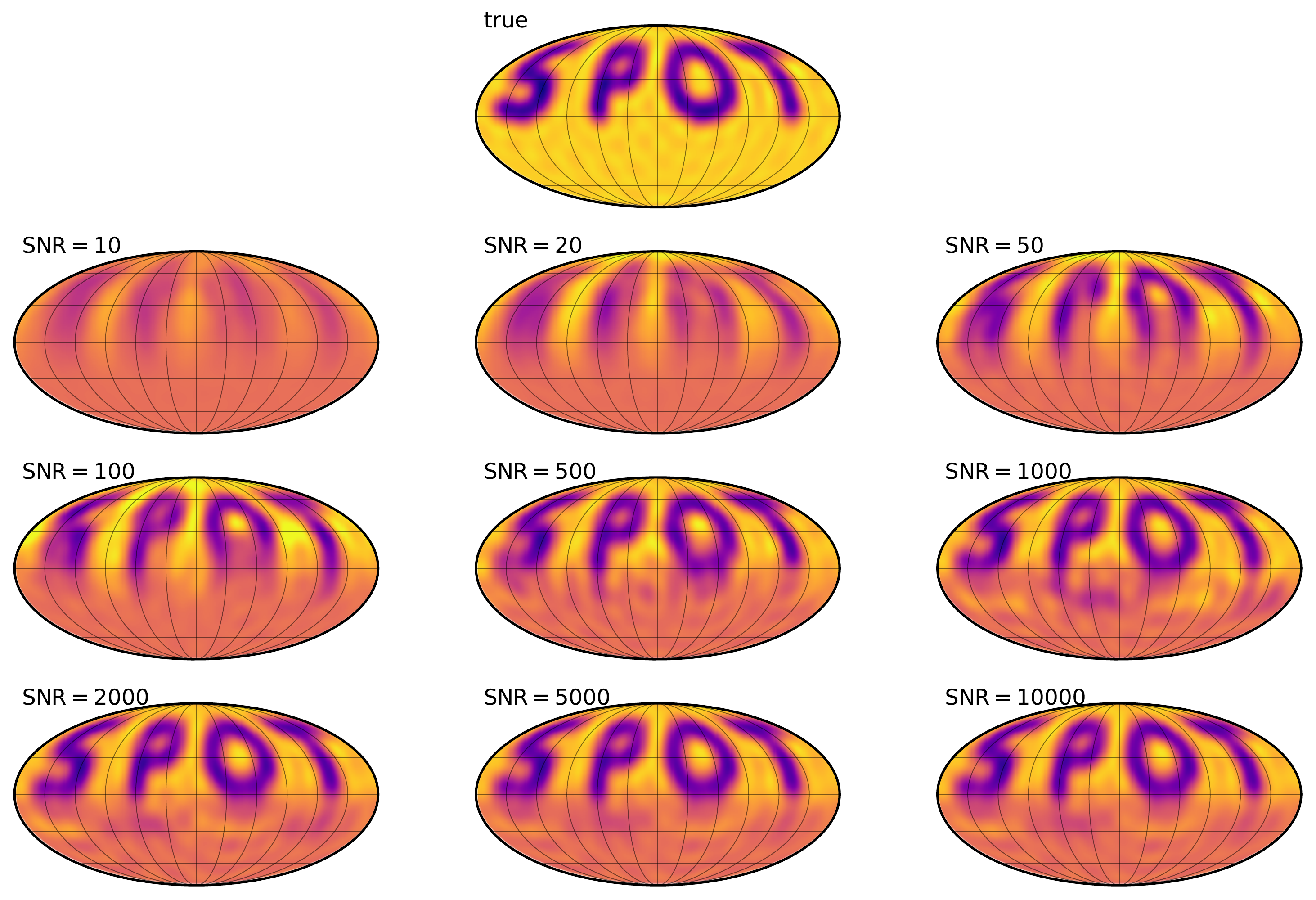}
        \caption{%
            Recovered \spot map as a function of SNR per resolution element. 
            The observed spectra are continuum-normalized and the rest frame spectrum is assumed to be known exactly.
            The SNR=1,000 panel corresponds to the solution presented in Figure~\ref{fig:spot_infer_yb}.
        }
        \label{fig:snr}
    \end{centering}
\end{figure}

Most of the examples in this paper considered very high SNR spectra.
Figure~\ref{fig:snr} shows the effect of different SNR on the quality of the inferred map. 
As before, the SNR we report is the ratio of the typical line depth in the observed spectrum to the standard deviation of the added noise term, $\sigma_\bvec{f}$.
The panels in Figure~\ref{fig:snr} show the inferred map for SNR in the range 10---10,000, where SNR $\sim 1,000$ is the fiducial case considered in the examples so far. 
The setup is the same as that in \S\ref{sec:spot_y1b}, where the spectrum is known exactly but the map coefficients and the baseline are not.

As may be expected, the quality of the map decreases with lower SNR, but a reduction of the SNR by $\sim 20\times$ still allows for reasonable recovery of the \spot feature. 
In general, as the SNR decreases further, the features begin to blur out, especially south of about $+30^\circ$ latitude, where features are more limb-darkened at this inclination and in view for a smaller fraction of the time.

As the SNR increases, however, the quality of the \spot feature does not improve appreciably, since a SNR of 500 is sufficient to recover a $l_\mathrm{max} = 15$ band-limited map. 
Interestingly, at constant spatial prior variance ($\boldsymbol{\Lambda}_\bvec{y} = 10^{-4}\bvec{I}$), the map quality actually begins to degrade with higher SNR south of the equator due to overfitting.
To circumvent this, we decrease the prior variance by a factor of 10--100 for the highest SNR cases, which penalizes excessive complexity in the solution.

\subsection{Dependence on inclination}
\label{sec:inc}
\begin{figure}[t!]
    \begin{centering}
        \includegraphics[width=0.98\linewidth]{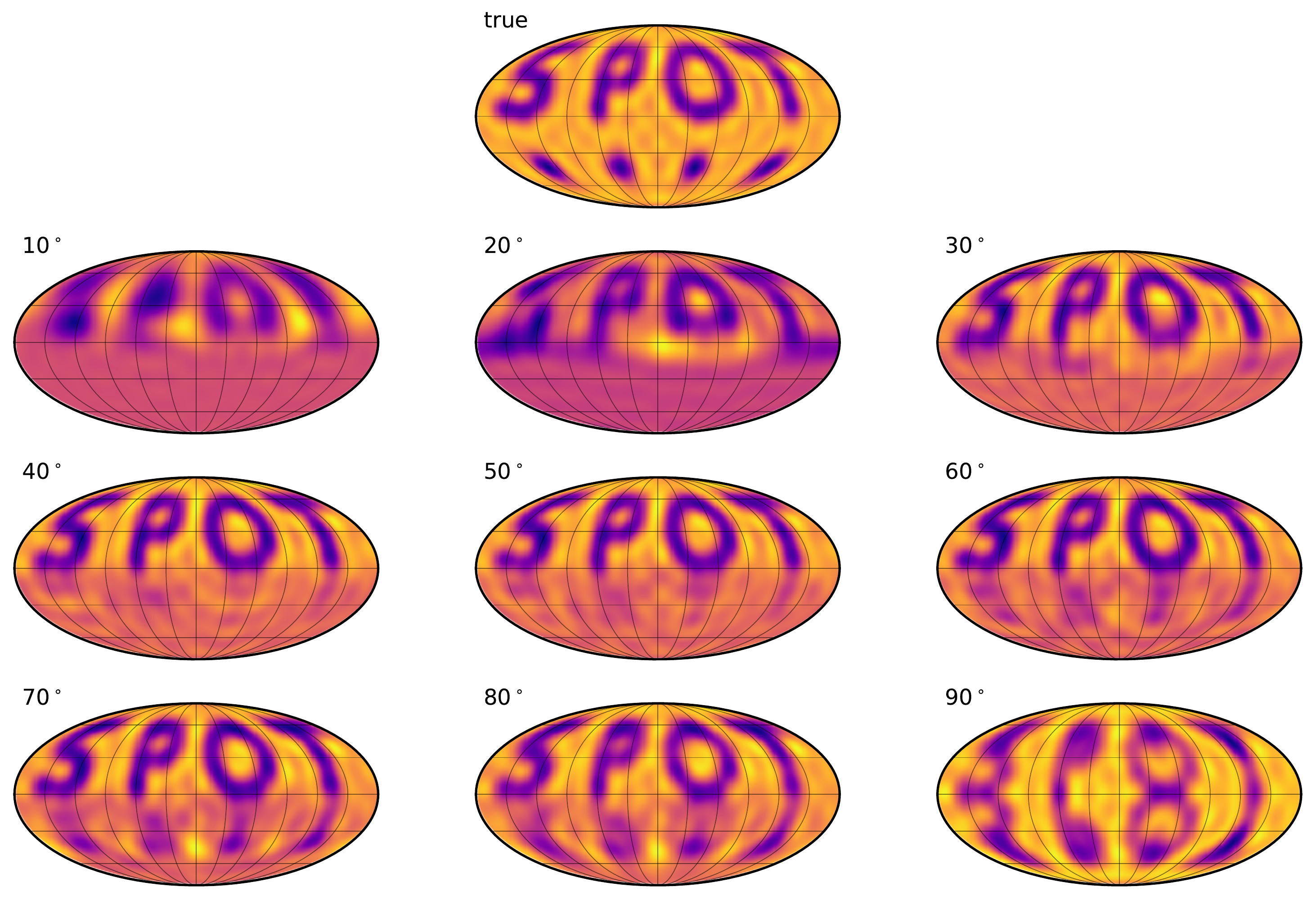} %
        \caption{%
            Recovered \texttt{SPOT} map as a function of stellar inclination. 
            The observed spectra are continuum-normalized and the rest frame spectrum is assumed to be known exactly.
            In general, it is very difficult to infer features in the hemisphere facing away from the observer (in this case, the southern hemisphere). 
            At low inclinations, those features are never in view, while at inclinations close to 90$^\circ$, it becomes difficult to differentiate between features above and below the equator.
        }
        \label{fig:inclinations}
    \end{centering}
\end{figure}
As in \citet{Vogt1987}, we also investigate the dependence of the solution on the inclination of the star.
Figure~\ref{fig:inclinations} shows the inferred map for the same problem setup as before, but this time varying the inclination between $10^\circ$ (nearly face-on) and $90^\circ$ (perfectly edge-on). 
We also add four small circular features at $-45^\circ$ latitude to test our ability to infer features in the southern hemisphere (i.e., the side of the star facing away from the observer).
We set the covariance of the spatial prior to $\boldsymbol{\Lambda}_\bvec{y} = 3\times 10^{-5}\bvec{I}$ and, as before, assume the inclination to be known exactly. 
We further keep the true equatorial velocity of the star constant at $60$~km/s across all panels, so that the figure corresponds to the same star viewed at different orientations. 
Because of this, decreasing the inclination decreases $v\sin i$, reducing the width of the convolution kernel and therefore the spatial resolution of our inferred map.
Increasing the inclination does not result in appreciable changes until an inclination of $\sim 80^\circ$, above which it becomes difficult to differentiate between features in the northern and southern hemispheres, whose projected velocities are nearly equal. 
At $i = 90^\circ$, there is a formal degeneracy between the two hemispheres, which can be seen from the mirror-image symmetry in the inferred map. 
Finally, as expected, the features in the southern hemisphere are not recovered below inclinations of about $60^\circ$, as they are never actually in view. 
At an inclination of $80^\circ$ the spots are recovered fairly well, but never with as high fidelity as features in the observer-facing hemisphere.
These results are broadly consistent with those of \citet{Vogt1987}.

\section{Application to WISE 1049-5319B}
\label{sec:luhman16b}

Having shown that our model performs well on synthetic data, we now turn to a real dataset: the VLT/CRIRES observations of WISE 1049-5319B (Luhman 16B) presented in the Doppler imaging study of \citet{Crossfield2014}.
WISE 1049-5319B belongs to a binary brown dwarf system, which, at a distance of 2 pc, consists of the two closest known brown dwarfs to the Solar system.
Despite its proximity, WISE 1049-5319B is intrinsically dim and, with a projected rotational velocity $v \sin i = 26.1$ km/s, is at the slow end of what can be imaged using the traditional Doppler imaging technique.
Therefore, unlike those of classical Doppler imaging targets, the spectral lines of WISE 1049-5319B do not individually show large scale spot modulation (Figure~\ref{fig:spot}) from one epoch to the next; nor are the spectral templates for brown dwarfs particularly well understood.
Nonetheless, we chose to model WISE 1049-5319B specifically to show the power behind our new methodology, in particular its ability to learn both the surface map and the rest frame spectrum simultaneously from the entire dataset---not just from unblended spectral lines that have reliable templates.

In the original Doppler imaging study of this target, \citet{Crossfield2014} took 56 spectra of each object (the angular resolution was high enough to obtain individual spectra of both brown dwarfs) over the four CRIRES detectors in the range $2.288-2.345 \mu$m over a time span of 5 hours, with individual exposures of 300 s.
The 5 hour baseline was enough to capture one full rotation of WISE 1049-5319B, whose period was constrained to be about 4.9 hours from its photometric variability \citep{Gillon2013}.
These spectra were then downbinned to 14 epochs to boost the SNR, corrected for tellurics and spectrograph systematics, and continuum-normalized.
Each of the 14 spectra consist of 1,024 intensity measurements in each of 4 detectors, for a total of $57,344$ data points.
For more details on the data reduction, refer to \citet{Crossfield2014}.

\begin{figure}[p!]
    \begin{centering}
        \includegraphics[width=\linewidth]{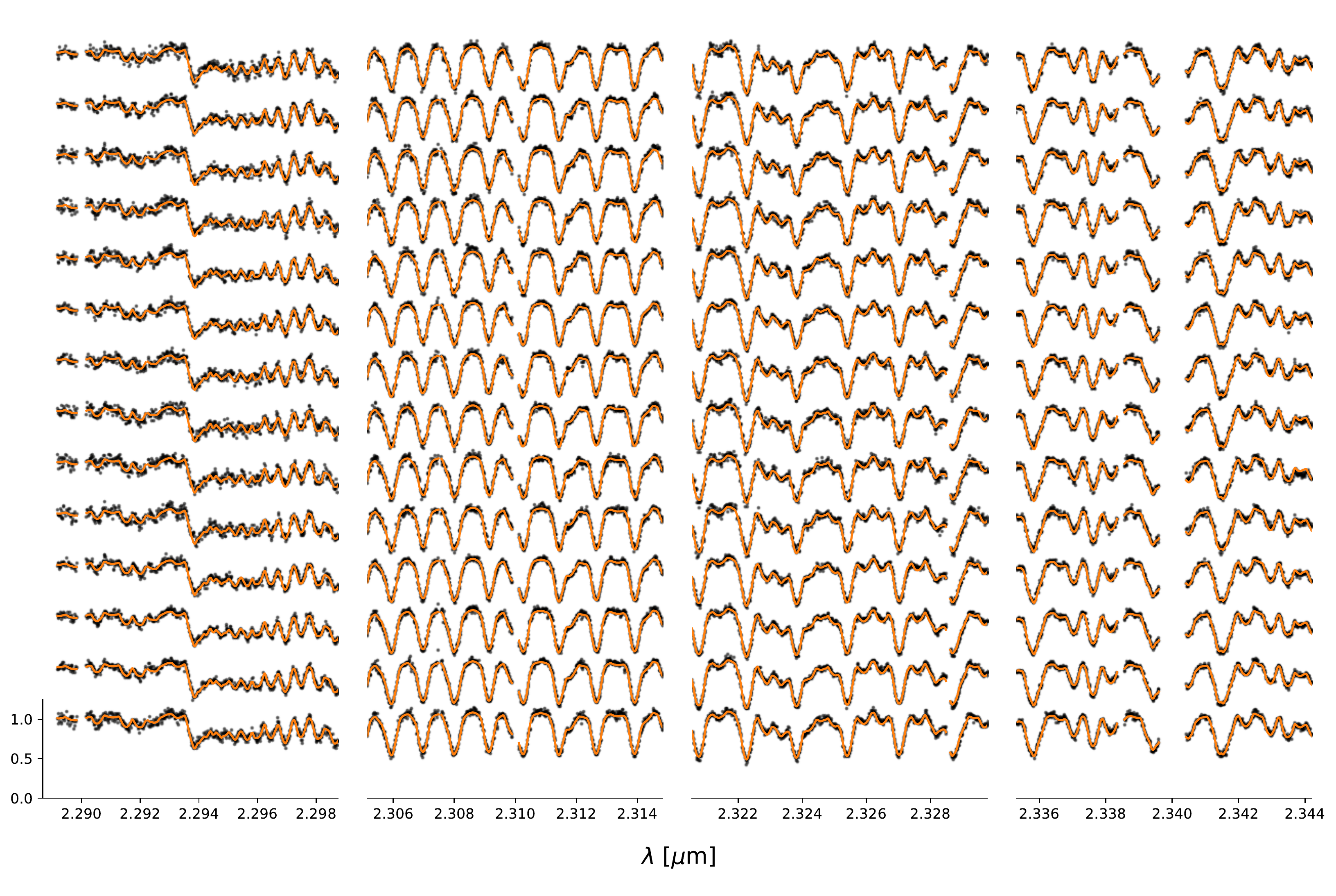} %
        \caption{%
            Calibrated and tellurics-corrected CRIRES spectra of WISE 1049-5319B from \citet{Crossfield2014} taken at 14 epochs spanning a full rotation of the brown dwarf (black points).
            Our maximum \emph{a posteriori} solution is shown as the orange curves.
        }
        \label{fig:luhman16b:data_model}
    \end{centering}
\end{figure}

The black points in Figure~\ref{fig:luhman16b:data_model} show the exact dataset used in the Doppler imaging analysis of \citet{Crossfield2014}, except for the fact that we masked additional outliers at $>4\sigma$ (identified by subtracting a version of the spectrum smoothed on a lengthscale of 20 pixels) as well as regions of the spectrum that had been visibly overcorrected for tellurics.
This figure shows the same data as the lower panel of Extended Data Figure 2 in \citet{Crossfield2014}.

\begin{figure}[p!]
    \begin{centering}
        \includegraphics[width=\linewidth]{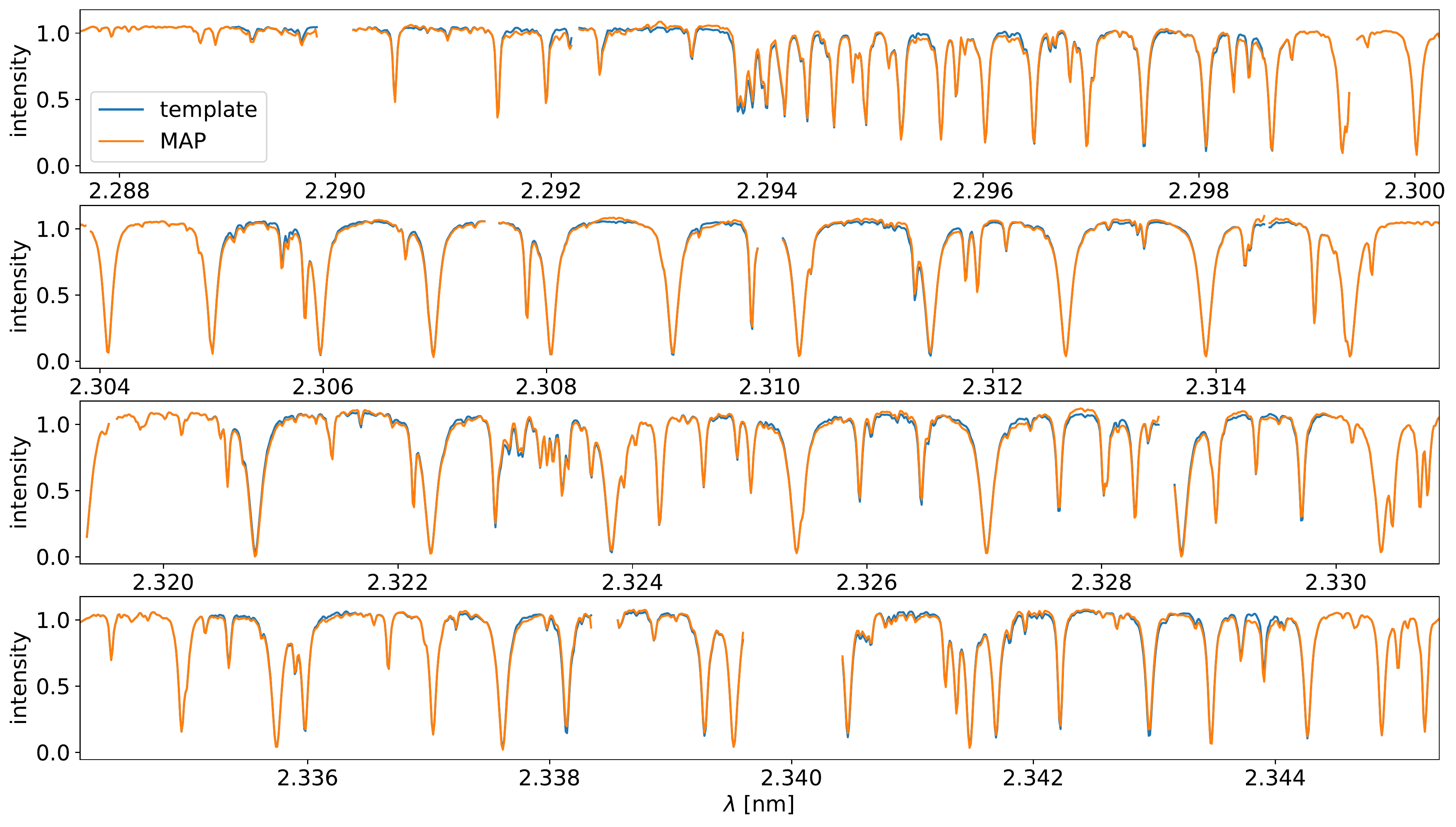} %
        \caption{%
            Unbroadened template spectra used to model the data in \citet{Crossfield2014} (blue curves). 
            These were used as a prior on the spectra inferred in this work (orange curves). 
            We infer a slightly different depth for some lines, as well as subtle changes in the continuum.
        }
        \label{fig:luhman16b:template_spectra}
    \end{centering}
\end{figure}

In order to produce Doppler images of WISE 1049-5319B, \citet{Crossfield2014} used brown dwarf template spectra from the BT-Settl library \citep{Allard2013}, finding that the spectrum with $\log_{10} g = 5.0$ and $T_\mathrm{eff} = 1450$ K provided the best match to the dataset.
This spectrum was then Doppler-shifted by the systemic radial velocity of the brown dwarf and broadened to match the spectrograph resolution.
This final template spectrum is shown as the blue curves in Figure~\ref{fig:luhman16b:template_spectra}, where we again masked regions where data was discarded.

\begin{figure}[t!]
    \begin{centering}
        \begin{minipage}[t]{0.49\linewidth}
            \centering
            \quad\quad\quad
            \textbf{Luger et al. (2021)}
        \end{minipage}
        \vspace{1em}
        \begin{minipage}[t]{0.49\linewidth}
            \centering
            \textbf{Crossfield et al. (2014)}
        \end{minipage}
        \includegraphics[width=0.495\linewidth]{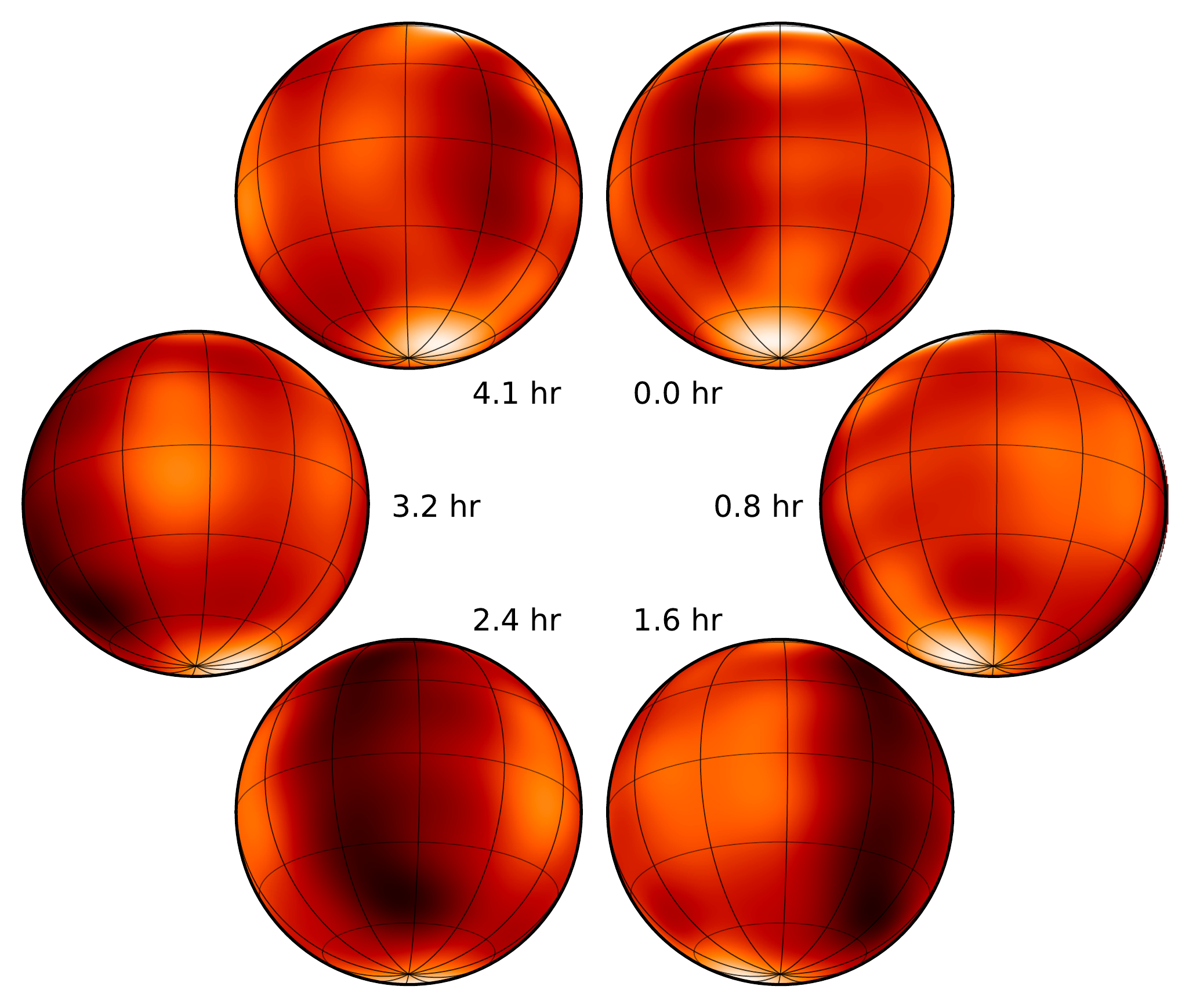} %
        \includegraphics[width=0.462\linewidth]{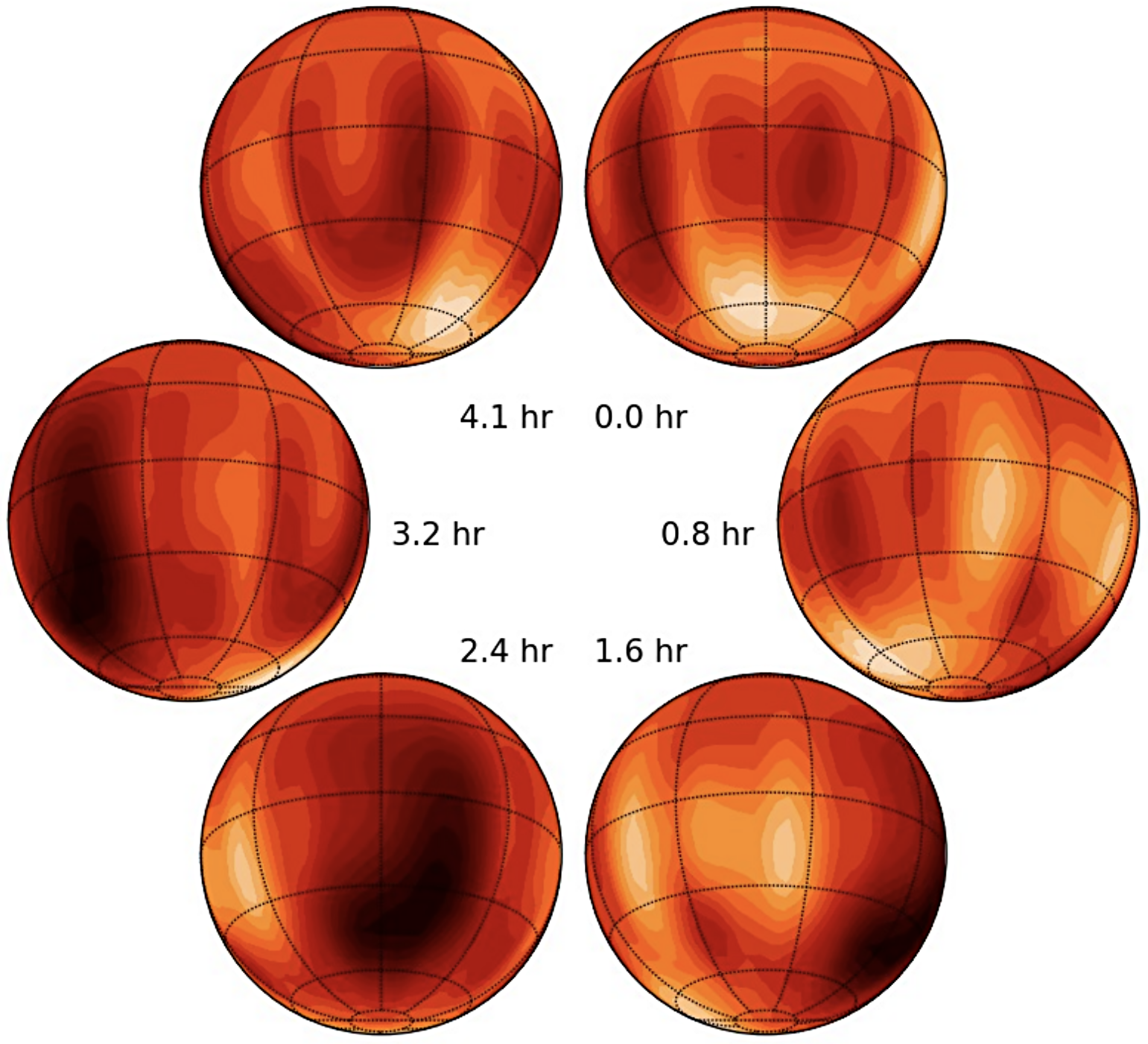} %
        \caption{%
            Maximum \emph{a posteriori} surface maps of WISE 1049-5319B inferred from the spectra in Figure~\ref{fig:luhman16b:data_model}. 
            \emph{Left:} surface map inferred using the algorithm developed in this work. 
            \emph{Right}: surface map inferred in \citet{Crossfield2014}, reproduced from Figure~2 in that paper.
            We infer the same two dominant features as in \citet{Crossfield2014b}:
            a bright circumpolar spot, seen along the sub-observer meridian at $t = 0.0$ hr,
            and a dark, latitudinally elongated band prominent at $t = 2.4$ hr.
            While the exact latitude of the bright spot and the behaviour of the dark spot equatorward of $-30^\circ$ latitude differ betwee the two solutions, both are point estimates obtained under different choices of prior and from significantly different methods.
            We take their qualitative agreement to be a successful validation of our method.
        }
        \label{fig:luhman16b:maps}
    \end{centering}
\end{figure}

\citet{Crossfield2014} then employed the common technique of Least Squares Deconvolution \citep[LSD;][]{Donati1997} to transform each of the 14 spectra in Figure~\ref{fig:luhman16b:data_model} into a single high-SNR line spanning 35 pixels, which was used as the input to their Doppler imaging code, based on that of \citet{Vogt1987}.
They discretized the stellar surface using 200 cells and solved for the optimal surface map with a maximum entropy prior as in \citet{Vogt1987}, assuming a fiducial inclination $i=70^\circ$ roughly based on the photometric rotation period and the inferred value of $v\sin i$; given the low SNR of the dataset, they were unable to place meaningful constraints on $i$.
We show their resulting surface brightness map in the right panel of Figure~\ref{fig:luhman16b:maps}, seen at six different phases spanning a full rotation.


In our analysis, we use the same data (Figure~\ref{fig:luhman16b:data_model}) and the same spectral template (Figure~\ref{fig:luhman16b:template_spectra}), but instead following the methodology derived in this work.
Our problem is similar to that in \S\ref{sec:spot_y1bs}, in which we treat both the spectrum and the surface map as unknowns given continuum-normalized spectra as input.
However, this time we use the template spectrum as the prior mean on the true rest frame spectrum $\bvec{s}$, setting the prior standard deviation to one percent (i.e., a covariance $\boldsymbol{\Lambda}_\bvec{s} = 10^{-4}\bvec{I}$) to allow the model to learn any small deviations from the template.
We must also change the way we model the continuum normalization (\S\ref{sec:norm}), given the high density of spectral lines and the consequent difficulty of finding a reliable continuum level to normalize against.
We therefore treat the unknown continuum normalization factor for each of the 14 epochs as a free parameter of the model, $f \sim \mathrm{Uniform}(0.3, 3.0)$; we also allow for a small, time-independent, additive offset $\Delta \sim \mathcal{N}(0, 0.1^2)$ for each of the four detectors.
We assume the same fiducial inclination $i = 70^\circ$, a rotational period $P = 4.9$ days, and allow the linear limb darkening coefficient and projected velocity to vary subject to $u_1 \sim \mathrm{Uniform}(0, 1)$ and $v \sin i \sim \mathcal{N}(26.1, 0.2^2)$ km/s, respectively, as in \citet{Crossfield2014}.
Finally, we place a uniform prior on the pixel intensities $\bvec{p} \sim \mathrm{Uniform}(0, 1)$ as in \S\ref{sec:twospec}, linearly transforming to an $l_\mathrm{max} = 10$ spherical harmonic representation of the map to compute the model at each step.
We must therefore solve for $2 \times (l_\mathrm{max} + 1)^2 = 242$ free parameters for the pixelized surface map, $3,748$ free parameters for the rest frame spectrum across each of the four detectors (equal to the size of our template spectrum after removing outliers), $18$ parameters describing the continuum normalization, and $2$ stellar parameters (the linear limb darkening coefficient and the projected velocity).
In total, our model has just over $4,000$ free parameters.

As in \S\ref{sec:twospec}, our choice of parametrization and priors precludes the use of the fast bilinear approach outlined in \S\ref{sec:inverse}.
We therefore follow the same approach as in \S\ref{sec:twospec}, optimizing our log-likelihood function using \texttt{Adam} with a learning rate of $10^{-3}$.
Since we have a very good guess for the template spectrum, this optimization is much faster than in \S\ref{sec:twospec}, converging within $1,000$ steps.
Because of our efficient gradient-based implementation (\S\ref{sec:starry}), the entire optimization takes only about 3 minutes of runtime on a 2018 Macbook laptop.

Our results are shown in Figures~\ref{fig:luhman16b:data_model}, \ref{fig:luhman16b:template_spectra}, and \ref{fig:luhman16b:maps}.
In Figure~\ref{fig:luhman16b:data_model}, our maximum \emph{a posteriori} (MAP) model for each of the 14 spectra is shown as the orange line; it agrees with the data to within the measurement uncertainties.
In Figure~\ref{fig:luhman16b:template_spectra} we show our MAP \emph{template} (rest frame) spectrum as the orange curve.
Since we used the \citet{Crossfield2014} template as a prior (blue curve), our solution is very similar to the input spectrum.
Nevertheless, we infer slightly different line depths for several of the lines at the level of a few percent; we also infer slightly different continuum levels in certain portions of the spectrum.
Note, importantly, since we only performed an optimization (rather than full posterior inference), we cannot quantify the actual significance of these differences.

The right panel of Figure~\ref{fig:luhman16b:maps} shows our principal result: the inferred solution for the surface brightness map of WISE 1049-5319B; compare to the right panel, showing the solution inferred by \citet{Crossfield2014}.
Broadly speaking, the two surface maps agree rather well: we infer the same circumpolar bright spot, of about the same size and at roughly the same phase, although the exact latitude is different.
We also infer a similar dark feature on the ``back'' side of the object (visible near the sub-observer point at $t = 2.4$ hours).
Bright equatorial regions also broadly match between both solutions.
There are some important differences, such as the behavior of the $t = 2.4$ hour dark region beyond $60^\circ$ of the visible pole.
In general, we find that different choices for our priors---in particular the regularization on the pixels and the template spectrum, the prior on the limb darkening coefficient, and the value of the inclination of the star---result in maps that differ somewhat.
We do, however, agree with \citet{Crossfield2014} in that the polar bright spot is the most robust of the inferred features, followed by the bottom portion of the dark spot seen at $t = 2.4$ hours.
Once again, since this is simply an \emph{optimization}, we are unable to place reliable uncertainties on our solutions.
These require a sampling scheme like MCMC (\S\ref{sec:discussion:priors}).

\section{Discussion}
\label{sec:discussion}

\subsection{WISE 1049-5319B}
\label{sec:discussion:luhman}

In \S\ref{sec:luhman16b} we validated our novel approach by reproducing the results of \citet{Crossfield2014}. 
Despite the fact that we obtained a qualitatively similar surface map for WISE 1049-5319B (Figure~\ref{fig:luhman16b:maps}), there are several aspects to our approach that could be improved in future work. 
First, we did not vary the inclination, but instead kept it fixed at the fiducial value $i = 70^\circ$ used in \citet{Crossfield2014}. 
The authors of that study argued that the SNR of the dataset was too low to constrain the inclination, and that their surface maps did not qualitatively change by varying the inclination between $60^\circ$ and $90^\circ$, the range consistent with the joint knowledge of the rotation period and $v\sin i$. 
We find a similar result, but we note that the likelihood of the solution improves marginally when we allow the inclination to vary.
A posterior inference scheme may thus be able to place stronger constraints on the value of $i$.
Second, neither our study nor that of \citet{Crossfield2014} accounted for the finite exposure time of the observations; given that each of the 14 spectra were collected over 20 minutes, the object rotates through $\frac{20\,\mathrm{min}}{4.9\,\mathrm{hr}}\times 360^\circ \approx 24^\circ$ during each exposure.
Nevertheless, this is comparable to the limiting resolution of our $l_\mathrm{max} = 10$ spherical harmonic expansion (equal to about $180^\circ / l_\mathrm{max} = 18^\circ$), so it is unlikely to substantially change our results.
Third, as in \citet{Crossfield2014}, we accounted for instrumental broadening in the template spectrum; in other words, the spectrum shown in Figure~\ref{fig:luhman16b:template_spectra}, the input into our Doppler imaging code, includes broadening due to the finite spectral resolution of CRIRES.
A better approach is to compute the Doppler-broadened spectrum based on the true rest frame spectrum and \emph{then} apply instrumental broadening (i.e., the order in which events take place in reality).
Finally, and most importantly, our results (like those of \citet{Crossfield2014}) are just a point estimate.
While the main features inferred in the map---the polar bright spot and the dark spot on the back side of the object---appear to be robust to changes in our priors, we are unable to quantify our confidence in the map solution.
Doing so with gradient-based posterior sampling methods is straightforward with the \starry implementation of our algorithm, but we defer it to future work.

\subsection{Regularization, priors, and posteriors}
\label{sec:discussion:priors}
%
%
Inferring surface maps from unresolved astronomical observations is a hard problem.
It is ``hard'' in the usual sense, requiring the development of complicated algorithms and the evaluation of computationally expensive models; but it is also ``hard'' in the sense that quite often the true answer is \emph{unknowable} because the data simply do not contain sufficient information about it.
This is especially the case in photometric studies, where one might seek to infer the distribution of spots on the surface of a star from its rotational light curve alone.
Recently, \citet{Luger2021a} showed that this problem is extremely ill-posed: only a very small fraction of the information about the two-dimensional intensity distribution on the stellar surface is encoded in the light curve \citep[see also][]{Basri2020}.
Even at infinite SNR, there exist an infinite number of solutions to the inverse problem, most of which look nothing alike.
In a photometric study, the \emph{only} way to obtain a unique solution is to supplement the dataset with external information, usually in the form of restrictive assumptions or explicit priors on the quantities of interest.
Most of the information in the solution to a light curve inversion problem comes from these prior assumptions, so it is extremely important to get them right.

In the case of Doppler imaging datasets, the problem is much more well-posed because of the availability of an entire new dimension of information (wavelength). 
In the limit of infinite SNR, spectral resolution, and temporal sampling, and if the rest frame spectrum is known exactly, in many cases there \emph{are} unique solutions to the visible portion of the stellar surface.%
\footnote{
    Recall that, except in the pathological case where $i = 90^\circ$, there is always a region on the ``back'' side of the star that is never in view at any phase. 
    No amount of statistical or algorithmic cleverness can ever tell us what that region looks like!%
}
In this limit, the data is sufficiently informative and changing one's prior does not substantially change the inferred surface map.
This is approximately the case in \S\ref{sec:spot_y1}, in which we inferred the true surface map given a very high SNR dataset and perfect knowledge of the rest frame spectrum (see Figure~\ref{fig:spot_infer_y}).
However, it is not the case in many of the other examples in that section, particularly when the rest frame spectrum is not known exactly and when the SNR of the dataset is finite.
In those cases, good solutions to the surface map (ones in which the \texttt{SPOT} feature is clearly legible in the inferred map) can be obtained only under certain choices of our priors (or the strength of the regularization term). 
In the sections below we discuss how to choose these priors in a principled way.

\subsubsection{Gaussian priors}
\label{sec:discussion:priors:gaussian}
In \S\ref{sec:spotstar} we took advantage of the bilinear nature of the problem to compute the uncertainty (\S\ref{sec:spot_y1}) and lower bounds on the uncertainty (\S\ref{sec:spot_y1b} and \S\ref{sec:spot_y1bs}) of the surface map solution. 
We found that the posterior uncertainty on the map ranges from about one percent at the north (visible) pole of the star to upwards of 10 percent at the south (hidden) pole (see Figures~\ref{fig:spot_infer_y}--\ref{fig:spot_infer_ybs}).
It is important to keep in mind that these values are entirely dependent on our choice of prior (namely, a zero-mean Gaussian prior on the spherical harmonic coefficients with constant variance equal to $10^{-4}$).
In particular, the value of the posterior uncertainty close to the south pole of the star---which is never in view, and whose true intensity cannot be constrained from the data---is determined exclusively from the prior variance we chose above (via Equation~\ref{eq:spotstar:uncert}).
Its value is therefore as arbitrary as our choice of prior.%
\footnote{While the exact \emph{value} of the posterior variance scales with the prior variance, the \emph{ratio} between the uncertainty in the north and the south does not. 
The gradient seen in the uncertainty maps of Figures~\ref{fig:spot_infer_y}--\ref{fig:spot_infer_ybs} \emph{is} therefore meaningful, as it directly encodes the spatial distribution of the information content of the data.}
In \S\ref{sec:spotstar} we chose a variance of $10^{-4}$ because we had access to the true maps and found, via trial and error, that this value yielded a good balance between underfitting and overfitting the synthetic data. 
In a realistic application, however, we could not do this, so we must naturally wonder: how should we go about choosing the prior variance in a principled way?

The Bayesian approach to this is to choose priors that truly encode our prior knowledge of the system. 
Absent prior observations of the system, these priors usually come from physics, such as our knowledge of the typical stellar photospheric temperatures or our understanding of the kinds of inhomogeneities we would expect on their surfaces (dark spots on stars, or patchy clouds on brown dwarfs). 
The most straightforward way to encode this information in our priors is to recognize that the prior variance on a spherical harmonic vector is closely related to a prior on its angular power spectrum \citep[see, e.g.,][]{Baldi2006,Wandelt2012}.
Our choice for prior covariance in \S\ref{sec:spotstar}, namely $\boldsymbol{\Lambda}_\bvec{y} = 10^{-4}\bvec{I}$, implies an isotropic, flat angular power spectrum prior on our surface maps with power equal to $10^{-4}$ in each spherical harmonic degree.
Increasing this value will in general lead to solutions with more structure on all scales; decreasing it will favor more homogeneous maps.
This may be an acceptable prior if truly nothing is known about the typical angular scale of features on the surface, but this is seldom the case.
Both observations of the Sun and magnetohydrodynamic simulations of stellar magnetic fields give us prior constraints on the typical scales of starspots \citep[e.g.,][]{Berdyugina2005,Solanki2006}, and climate models of brown dwarfs tell us about typical sizes of clouds and other atmospheric features \citep[e.g.,][]{Tan2019,Tan2020,Tan2021a,Tan2021b}.
These can be encoded into our Gaussian prior by choosing a prior variance that scales with spherical harmonic degree, as in Equation~(\ref{eq:isotropicprior}), where the variance at each degree is parametrized as, e.g.,
\begin{align}
    \label{eq:powerspec}
    \sigma_l^2(\alpha, \lambda, \tau) = \sigma_\lambda^2 \exp\left(-\frac{(l - \lambda)^2}{2\tau^2}\right)
    \quad,
\end{align}
where $\lambda$ is the typical spherical harmonic degree of features, $\sigma_\lambda^2$ is the prior variance at that degree, and $\tau$ controls the uncertainty in $\lambda$.
These hyperparameters can be either imposed or even \emph{learned} from the data; see \S\ref{sec:discussion:priors:mcmc}.

Sometimes, however, we may have even more information to encode in our priors, such as the typical \emph{location} of the features. 
This might be the case for solar-like or lower mass stars, which have specific active latitudes at which spots form. 
An isotropic prior of the form discussed above cannot encode this, but a dense covariance matrix (with structure in the off-diagonal elements) \emph{can}. 
\citet{Luger2021b} showed that it is possible to encode properties such as the distribution of spot latitudes, spot contrasts, and spot sizes into the prior covariance on the spherical harmonic coefficients; see \citet{Luger2021d} for an efficient implementation of this. 
Priors of this form can, once again, be imposed (if prior knowledge of the active latitudes or spot properties is available) or learned from the data by sampling over the corresponding hyperpriors. 
We discuss this in more detail in \S\ref{sec:discussion:priors:mcmc} below.

\subsubsection{Cross-validation}
\label{sec:discussion:priors:crossval}
The approach discussed above for choosing one's priors is distinctly Bayesian.
Alternatively, one may take a data-driven approach: to learn the ``optimal'' value of the priors from the data itself.
One common approach to this is \emph{cross-validation}, in which one's prior is optimized to yield the best predictive power for the data.
Typically, cross-validation consists of choosing a specific value for the prior, masking one (``leave-one-out-cross-validation'') or several (``leave-p-out-cross-validation'') data points, optimizing the parameters of the model given this smaller dataset, then using the optimized model to predict the values of the missing data \citep[e.g.,][]{Stone1974}.
The value of the prior is then chosen as the one with the highest predictive power, i.e., the smallest loss when fitting the missing data points \citep[for an example of this, see \S 3.7 in][]{Luger2018}.

In this context, the prior is usually referred to as a ``regularization'' term, since this procedure is decidedly not Bayesian.
Specifically, it regularizes the model to ensure a balance between underfitting (the model doesn't fit the data well) and overfitting (the model fits the data so well that it's also fitting the noise, leading to spurious results).
In addition to helping one choose the value for (say) the prior covariance on the surface map, cross-validation can be used to optimize most of the numerical choices one makes when fitting a dataset, such as the degree $l_\mathrm{max}$ of the spherical harmonic expansion (or, in a discrete basis, the number of pixels in the surface grid) and the number of map components (\S\ref{sec:twospec}).

\subsubsection{Other priors}
\label{sec:discussion:priors:nongaussian}

While we used Gaussian priors in most of our examples (\S\ref{sec:spotstar}), other kinds of priors might make more sense in different contexts. 
In particular, in \S\ref{sec:luhman16b} we enforced a uniform prior in the range $[0, 1]$ on the pixel intensities. 
This encodes our seemingly trivial but surprisingly restrictive \citep[see][]{Fienup1982} belief that stellar intensities must be non-negative, which a Gaussian prior cannot do.
Priors like this one break the analyticity of the MAP solution (\S\ref{sec:inverse}), but they can be especially important in the limit that the data is not sufficiently constraining.
Traditionally, many Doppler imaging studies impose an additional \emph{maximum entropy} constraint \citep{Vogt1987}, whose objective is to ensure a similar balance between underfitting and overfitting as discussed above.
These can also easily be incorporated into our formalism, although we recommend using a method such as cross-validation to choose the regularization strength in this context.

\subsubsection{Posterior sampling}
\label{sec:discussion:priors:mcmc}
In the limit of finite SNR, and when things like the true stellar spectrum, the stellar inclination, and the limb darkening profile are not well known (which is usually the case in Doppler imaging studies!), it is important to recognize the limitation of point estimates like those obtained in this work.
The maps of WISE 1049-5319B presented in Figure~\ref{fig:luhman16b:maps}, for instance, are just that: point estimates that carry no sense of the uncertainty associated with any of those features.
Given the low SNR of the dataset (Figure~\ref{fig:luhman16b:data_model}), the unknown inclination of the star, and the uncertainty in parameters like the rotation period, $v\sin i$, and the limb darkening profile, the maps shown in Figure~\ref{fig:luhman16b:maps} are almost certainly \emph{not} what WISE 1049-5319B actually looks like in reality.
The fact that we obtained a solution that is qualitatively similar to that of \citet{Crossfield2014} is encouraging, but it is not sufficient.
Doppler imaging studies are notorious for producing features that are artificially elongated in latitude, just like the dark spot seen in the figure \citep[see, e.g.,][]{Unruh1995}.
The Doppler imaging method is also less sensitive to azimuthally symmetric features, like the zonal bands that may be present on brown dwarfs as on Jupiter.
Effects like these can bias the point estimates obtained in Doppler imaging studies, but they \emph{can} be captured if one fully propagates uncertainties to the final result.
This requires computing the posterior distribution of the model parameters (including the parametrization of the surface map), which is absolutely critical when one wishes to \emph{interpret} a Doppler map.

In all but the simplest cases, maximum likelihood estimators are typically biased estimators of the true parameters \citep[e.g.,][]{Lehman1998}, meaning point estimates like those shown in Figure~\ref{fig:luhman16b:maps}, as well as in many other Doppler imaging studies, are also biased.
If the goal of a study is \emph{only} to produce a map of the stellar surface, this may not be too much of an issue, especially because it is generally understood that Doppler maps are subject to artifacts like latitudinal elongation and suppression of azimuthally-symmetric features like bands, as mentioned above.
Some of the features in Figure~\ref{fig:luhman16b:maps}, for instance, are likely to be real, such as the bright polar spot and the dark spot on the back side of the object; but their exact properties are uncertain, and many of the other features in the maps may be spurious.
However, generating a surface map is often \emph{not} the ultimate end goal of Doppler imaging studies.
Rather, we are usually interested in a more fundamental question: what does our inferred map tell us about the star's properties, or even about stellar physics in general?
And therein lies the problem: as soon as we attempt to interpret the maps we've generated in terms of metadata such as the typical contrast of the star spots, their typical latitudes, or  the rate of differential rotation (based on comparing Doppler imaging maps taken at different epochs), we are subject to substantial bias due to the point estimate problem \citep[for an example of this same problem in the context of inferring the exoplanet eccentricity distribution, see][]{Hogg2010}.

For this reason, it is crucial to treat Doppler imaging as a posterior inference problem, when at all possible; see \citet{AsensioRamos2021} for recent work on a Doppler imaging model that harnesses approximate Bayesian techniques to perform posterior inference. 
In some restricted cases, we showed that is possible to obtain the full surface map posterior analytically with our method (\S\ref{sec:inverse}), but only when all priors are Gaussian and the inclination, limb darkening profile, and template spectrum are known exactly. 
In all other cases, one must compute an approximation to the posterior distribution over all these quantities numerically, such as with MCMC.
This is typically a computationally expensive problem, given the intricate degeneracies at play (which makes posterior sampling difficult), but also the sheer number of parameters, as we must sample over all the parameters representing our surface map (typically on the order of several hundred). 
When the input spectrum is unknown, the dimension of our problem can easily exceed a few thousand (\S\ref{sec:luhman16b}).

Fortunately, our model is computationally efficient to evaluate and, given its analyticity, trivial to differentiate using an autodifferentiation scheme (see \S\ref{sec:starry}).
It can therefore be used out-of-the-box within gradient-based sampling schemes such as \texttt{pymc3} \citep{Salvatier2016}, in particular the No-U-Turn Sampler (NUTS), a variant of Hamiltonian Monte Carlo that harnesses gradient information to greatly speed up the convergence of the MCMC chains \citep[for a recent application of HMC in the context of spectroscopy, see][]{Kawahara2021}.
Our model may also be useful in the context of variational inference (VI), and in particular automatic differentiation variational inference \citep[ADVI;][]{Kucukelbir2016}, which casts posterior inference as an optimization problem; both VI and ADVI are also implemented in \texttt{pymc3}.
The script used to infer the MAP solution for the map of WISE 1049-5319B (see Figure~\ref{fig:luhman16b:maps}) uses \texttt{pymc3} to perform the optimization; extending it to perform posterior inference is straightforward.
However, doing so is beyond the scope of the present paper, given that this task is still likely to be challenging given the very large number of parameters (${\sim}4{,}000$). 
We will therefore revisit this system in the context of posterior estimation in a future paper.

One final word regarding posterior inference: given that our end goal is often to infer something about the physics of stellar atmospheres rather than exactly what they look like, posterior sampling can be employed to explicitly marginalize over all possible surface maps to directly constrain our posterior beliefs about the physics.
An example of this is the model suggested in \S\ref{sec:discussion:priors:gaussian} above, where we express our prior covariance in terms of hyperparameters controlling the amplitude of the surface features and their typical lengthscale (Equation~\ref{eq:powerspec}) or in terms of the typical size, contrast, and latitude of spots \citep{Luger2021b,Luger2021d}.
When the value of those hyperparameters is not known \emph{a priori}, we can include them as free variables in our model and perform MCMC to compute their marginal distribution (i.e., marginalized over everything else, including all possible surface maps that are consistent with the data).
We can even imagine doing this in an ensemble setting under a fully hierarchical model, in which we attempt to constrain these properties from the analysis of many Doppler imaging datasets of different stars simultaneously.
This procedure is likely computationally prohibitive, but the linearity of our model (\S\ref{sec:linear}) allows us to extend the Gaussian process formalism of \citet{Luger2021b} to Doppler imaging datasets, allowing us to marginalize over the surface maps consistent with the data \emph{analytically}; see \S\ref{sec:gp} below.

\subsection{Assumptions}
\label{sec:discussion:assumptions}

Both the derivation and application of our model make several assumptions, many of which we have already discussed. Nevertheless, we present a detailed list of these assumptions here for clarity:

\begin{itemize}
    \item We assume there is a well-defined stellar ``surface'', an effective layer from which the star emits its light. 
    This is certainly an oversimplification, as stars are fundamentally three-dimensional objects with complex vertical temperature gradients.
    \item We assume the emission is isotropic. 
    We relax this assumption by allowing for limb darkening (\S\ref{sec:ld}), but this is treated as a suppression of flux toward the limb rather than by self-consistently solving the equations of radiative transfer.
    \item We assume the radial velocity field of the stellar surface is a simple dipole, controlled only by the rigid body rotation of the star. This ignores effects like convective blueshift and differential rotation. We discuss incorporating these into our work in \S\ref{sec:discussion:caveats} below.
    \item We assume the surface map is constant over time, and that all time variability comes from stellar rotation. 
    We discuss modeling temporal surface variability in the context of a stochastic process in \S\ref{sec:gp} below.
    \item We assume that the surface map may be well modeled by a spherical harmonic expansion truncated at some degree $l_\mathrm{max}$. 
    The spherical harmonic basis is complete, meaning this assumption is rigorously correct in the limit  $l_\mathrm{max} \rightarrow \infty$.
    In our experiments, we chose $l_\mathrm{max} = 10-20$, which allows us to model surface features on scales larger than about $18^\circ$ and $9^\circ$, respectively.
    \item We assume the rest frame spectrum is defined on a grid with sufficiently high resolution that the convolution with the Doppler kernel may be approximated as a discrete linear convolution (Equation~\ref{eq:linear:convolution_def}).
    \item In our base case (\S\ref{sec:the_solution}), we assume the star has a single rest frame spectrum, and that spatial changes are strictly changes in overall intensity. 
    We relax this assumption in \S\ref{sec:eigen}, where we assume that variations in the spectrum with location can be expressed as the outer product of a set of $J$ spectral principal components $\boldsymbol{\mathbb{S}}$ and a set of $J$ spatial principal components $\boldsymbol{\mathbb{Y}}$.
    If we choose $J = (l_\mathrm{max} + 1)^2$ (the number of spherical harmonic coefficients in the expansion of each component), this allows us to model \emph{any} spectral-spatial configuration on the surface of the star, band-limited to degree $l_\mathrm{max}$.
    Truncation of this basis at smaller values of $J$ limits our ability to model arbitrary spatial variability in the spectrum.
\end{itemize}

All of the assumptions listed above relate to our \emph{forward} model (\S\ref{sec:linear}).
Throughout the paper, we make several additional assumptions when solving the \emph{inverse} problem, that is, solving for the map and/or spectrum given a dataset.
None of these assumptions are necessary for the application of our model to other datasets, but we include them here for completeness:

\begin{itemize}
    \item We assume the data are correctly calibrated and free of significant outliers, and that the noise model (e.g., Gaussian errors with known standard deviation) is known.
    \item We assume there are either no telluric lines, or that these have been removed in a pre-processing step.
    These may also be modeled simultaneously, but that is beyond the scope of the present work.
    \item We assume the SNR of the data is high enough that the maximum \emph{a posteriori} solution to the surface map is meaningful.
    In general, however, we stress the importance of performing posterior inference to constrain the uncertainties and one's degree of confidence in the solution.
    \item In all of our examples, we assume the stellar rotation period and inclination are known exactly.
    In practice, this is rarely ever the case, and one should optimize for (or, ideall, sample over) these variables.
\end{itemize}

\subsection{Limitations and future work}
\label{sec:discussion:caveats}

There are a few important limitations of the algorithm presented here and its implementation in \starry that we will address in future papers in this series.
First, we do not currently address the effects of differential rotation and convective blueshift.
Both are straightforward to account for by modifying the path $\mathcal{C}(\lnlam,\bvec{x})$ of the line integral in Equation~(\ref{eq:deconv:kT}) to correspond to the isovelocity lines in the velocity field accounting for these two effects.
Under the common assumption of linear differential rotation, in which the shear is proportional to the square of the sine of the latitude, the velocity field may be written exactly as a cubic polynomial in the Cartesian coordinates of the stellar surface \citep{Short2018}, therefore admitting an exact representation in terms of an $l_\mathrm{max} = 3$ spherical harmonic expansion.\footnote{\url{https://starry.readthedocs.io/en/latest/notebooks/RVDerivation}}
Similarly, the effect of convective blueshift on the velocity field can be modeled with a low degree spherical harmonic expansion consisting only of radial ($m = 0$) modes.
In the Appendix, we showed that in the absence of these effects, the curves $\mathcal{C}(\lnlam,\bvec{x})$ are straight lines, so the line integral in Equation~(\ref{eq:deconv:kT}) is trivial to solve.
In the presence of differential rotation and convective blueshift, these curves become polynomials, making the integration trickier.
While we were unable to find simple closed-form expressions for the resulting line integrals, they may still exist; we will investigate this in future work.
Even if a closed-form solution is unavailable, it is likely still possible to efficiently evaluate the integrals numerically.

Another effect we do not presently model is stellar oblateness.
Many of the Doppler imaging targets are massive stars, which tend to be fast-rotating and therefore oblate.
We recently expanded the \starry framework to model stellar oblateness \citet{Dholakia2021}, and will similarly adapt our Doppler imaging framework in an upcoming paper. 
In the absence of differential rotation and convective blueshift, oblateness can be accounted for by simply weighting the integration limits in Equation~(\ref{eq:appendix:kT}) by the semi-minor axis of the elliptical boundary of the star in projection. 
Accounting for those effects may require numerical integration, as discussed above.

In a similar vein, our present model involves an integration over the entire stellar disk. 
This is the desired approach in Doppler imaging studies, but does not apply in the context of Doppler \emph{tomography} \citep[e.g.,][]{Johnson2017}, in which the star is observed during a transit of an orbiting companion to infer, e.g., the degree of misalignment between the companion's orbit and the stellar spin axis.
Our formalism can be adapted to Doppler tomography by changing the limits of integration in Equation~(\ref{eq:appendix:kT}) to exclude the region occulted by the companion; we will also investigate this in upcoming work.
We will also explore an extension of the method to the case of Zeeman-Doppler imaging \citep{Marsh1988}, which makes use of spectropolarimetric observations to map the stellar magnetic field topology.

Regarding systematic effects, we do not explicitly account for instrumental broadening. 
The observed spectrum is a convolution of the Doppler-shifted, disk-integrated stellar spectrum (whose model our algorithm calculates) with the point spread function (PSF) of the spectograph.
This convolution can be modeled as a dot product of a Toeplitz matrix with the Doppler design matrix $\bvec{D}$, much in the same way as Equation~(\ref{eq:linear:convolution_def}).
In a similar vein, our formalism assumes the data has been corrected for telluric contamination. 
This is typically done by removing a telluric model (based on a template spectrum) from the data in a pre-processing step \citep[as in][]{Crossfield2014}, but in the data-driven approach of this paper, they could also be learned from the data, similar to the approach in \citet{Bedell2019}.
We will investigate this in future work.

\section{A Spectral-Temporal Gaussian Process}
\label{sec:gp}

\begin{figure}[t!]
    \begin{centering}
        \includegraphics[width=\linewidth]{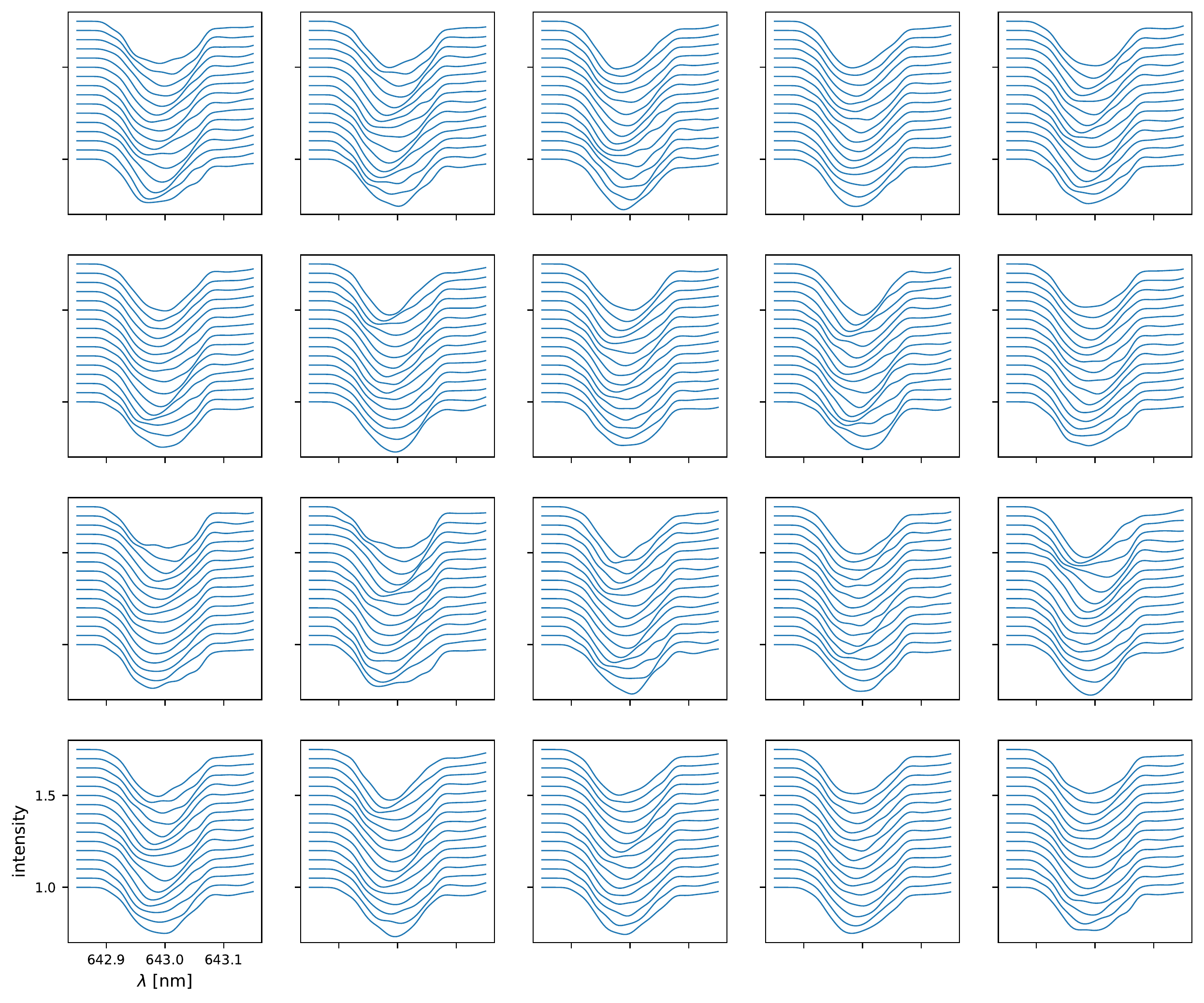}
        \caption{%
            20 samples from a spectral-temporal Gaussian process constructed from the algorithm presented here.
            Each panel shows a different sample, each of which consists of 16 epochs, offset vertically for visibility.
            The rest frame spectrum is the same as in Figure~\ref{fig:spot_setup} and the star is inclined at $i = 40^\circ$ with equatorial velocity $v = 50$ km/s.
            The star is assumed to have spots with radius $r = 15^\circ$, 25 percent contrast, and active latitude $\phi = 30^\circ \pm 5^\circ$.
        }
        \label{fig:gp}
    \end{centering}
\end{figure}

There is one immediate extension of the formalism developed in this work that merits a more in-depth discussion: the adaption of the Doppler imaging method into a Gaussian process (GP) for stellar spectroscopy.
While the goal of Doppler imaging is often to infer the actual surface map of the star, sometimes the goal may be to infer higher-level statistical information, such as the active latitude of spots or their typical sizes.
And sometimes, the goal of a spectroscopic analysis may not be to infer something about the star at all, such as in the case of radial velocity surveys of exoplanets.
In both of these cases, one does not really care about what the \emph{actual} map of the star is irrelevant; rather, the goal is to \emph{marginalize} over all the possible stellar surfaces consistent with the data to obtain the marginal posterior of the quantities of interest, such as the active latitude or the properties of an orbiting exoplanet.
In \S\ref{sec:discussion:priors:mcmc}, we discussed how one might perform this marginalization via MCMC sampling.
However, thanks to the linearity of the model developed here, we can use the formalism developed in \citet{Luger2021b} to do this \emph{analytically}.

\citet{Luger2021b} presented an interpretable GP for stellar photometric variability, based on the fact that it is possible to analytically compute the covariance matrix $\boldsymbol{\Lambda}_\bvec{y}$ describing the distribution of spherical harmonic coefficients characterizing a given stellar surface conditioned on certain bulk spot properties $\boldsymbol{\theta}$, such as their radius, latitude, and contrast distributions.%
\footnote{In \citet{Luger2021b}, this covariance matrix is denoted $\boldsymbol{\Sigma}_\bvec{y}$.}
Since the model for the (photometric) flux is linear in the spherical harmonic coefficients \citep{Luger2019}, the covariance of the GP for stellar light curves is then simply
\begin{align}
    \boldsymbol{\Sigma}_\mathrm{photo} = \mathbf{A} \, \boldsymbol{\Lambda}_\bvec{y} (\boldsymbol{\theta}) \, \mathbf{A}^\top
    \quad,
\end{align}
where $\mathbf{A}$ is the design matrix that transforms from spherical harmonic coefficients to flux, which implicitly depends on the stellar inclination, rotation period, etc.

Conditioned on a fixed spectrum, the Doppler imaging problem is also linear in $\bvec{y}$ (Equation~\ref{eq:inverse:fy}), so we can write the covariance of the GP for a spectroscopic timeseries dataset as
\begin{align}
    \label{eq:discussion:gpcov}
    \boldsymbol{\Sigma}_\mathrm{Doppler} = \mathbf{D} \, \mathbf{S} \, \boldsymbol{\Lambda}_\bvec{y} (\boldsymbol{\theta}) \, \mathbf{S}^\top \, \mathbf{D}^\top
    \quad,
\end{align}
where $\bvec{S}$ is given by Equation~(\ref{eq:inverse:S}) and $\bvec{D}$ is the Doppler design matrix (Equation~\ref{eq:linear:D}).
The likelihood of the Gaussian with covariance given by Equation~(\ref{eq:discussion:gpcov}) is the probability of the observed spectra conditioned on spot hyperparameters $\boldsymbol{\theta}$, \emph{marginalized} over all possible stellar surface maps.
Since this marginalization is done analytically, one need only sample over the parameters of interest (the components of $\boldsymbol{\theta}$, as well as the properties of the orbiting planet, in the case of a radial velocity study) in an inference problem.
Thus, this GP formalism can reduce the dimensionality of one's model from several hundred (the number of parameters used to model the surface map) to a handful (the hyperparameters of the GP, plus the parameters of the planet model, if applicable).

As an example, Figure~\ref{fig:gp} shows twenty samples from the GP described above. 
We condition the GP on the same spectrum used in \S\ref{sec:spotstar} and on a stellar inclination $i = 40^\circ$ and equatorial velocity $v = 50{,}000$ km/s, and choose hyperparameters $\boldsymbol{\theta}$ corresponding to spots with mean radius $r = 15^\circ$, mean fractional contrast $c = 0.25$, and active latitude $\phi = 30^\circ \pm 5^\circ$.
Each panel shows a full spectral-temporal dataset sampled from this GP, where individual spectra are vertically offset from each other for visibility.

The main application of our GP, of course, will not be for drawing samples, but for doing inference. 
In particular, we expect this GP to be useful for extreme precision radial velocity (EPRV) planet searches, which must contend with stellar variability signals that are often orders of magnitude stronger than the planetary signals of interest.
Our GP can be used as a principled, physically-motivated way to marginalize over the stellar signal to robustly detect the Keplerian signal of an orbiting planet.
It is even straightforward to account for time variability in the stellar surface with this formalism. \citet{Luger2021b} discussed how to incorporate a stochastic time-variable component into the photometric variability GP, which can be directly incorporated into Equation~(\ref{eq:discussion:gpcov}) via the inclusion of a decoherence timescale $\tau$ in the set of GP hyperparameters $\boldsymbol{\theta}$.

The next paper in the \emph{Mapping Stellar Surfaces} series will focus on further development of this GP, including accounting for uncertainty in the spectral template, accounting for the nonlinearity introduced by the continuum normalization, marginalizing over unknowns such as the inclination and limb darkening parameters, and rigorously testing the formalism on both synthetic and real datasets.

\section{Implementation in \starry}
\label{sec:starry}

We implement the Doppler imaging algorithm presented here in the \starry package \citep{Luger2019,Luger2021c}, an open source \texttt{Python} library for stellar and planetary light curve analysis. 
All of the figures in this paper were generated via the Doppler imaging API in \starry. 
The scripts used to generate them are available on GitHub\footnote{\url{https://github.com/rodluger/paparazzi}} and can be directly accessed via the GitHub icons next to each figure caption.
These are well-commented and can be used as a starting point for other Doppler imaging projects.

The algorithm is implemented using \texttt{theano}/\texttt{aesara} \citep{Bergstra2010,Willard2021}, a just-in-time compiling framework for \texttt{Python} with powerful autodifferentiation and optimization capabilities.
As such, the algorithm is fully differentiable and thus easy to use within gradient-based optimization and/or inference schemes such as Hamiltonian Monte Carlo.
The analysis presented in Figures~\ref{fig:spot_infer_ybs} and \ref{fig:luhman16b:data_model}---\ref{fig:luhman16b:maps} harnesses this capability within the \texttt{pymc3} modeling framework; refer to the corresponding scripts for details.

\begin{figure}[t!]
    \begin{centering}
        \includegraphics[width=\linewidth]{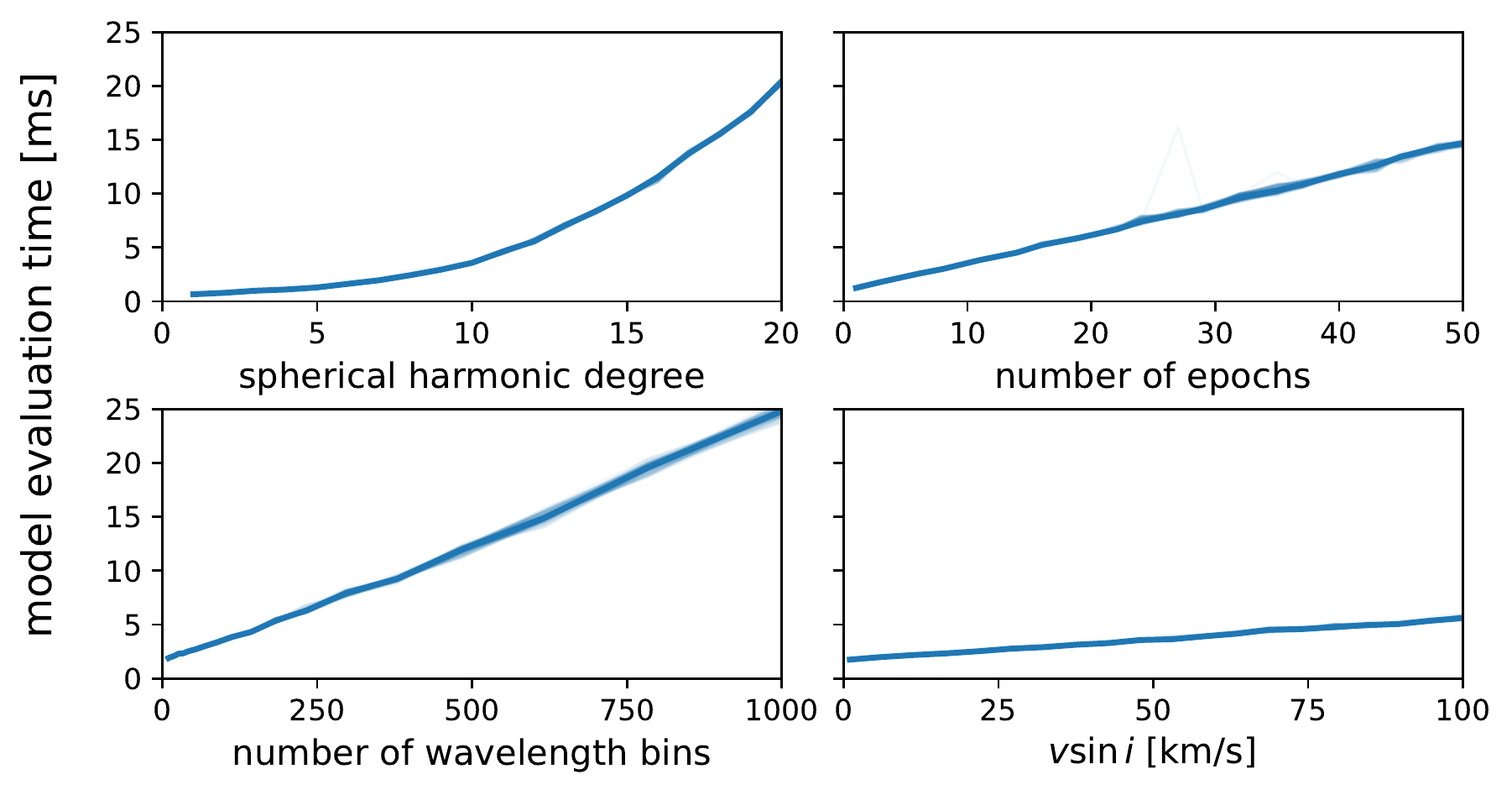}
        \caption{%
            Model evaluation time in milliseconds as a function of various parameters, for spherical harmonic degree $l_\mathrm{max} = 10$, $M = 10$ spectral epochs, $K = 200$ wavelength bins, and $v\sin i = 50$ km/s.
            Thin lines correspond to individual trials; the thick lines show the median evaluation time.
            Timing tests were performed on a 2-core Linux virtual GitHub Actions machine.
        }
        \label{fig:runtime}
    \end{centering}
\end{figure}

The implementation of the algorithm in \starry is highly optimized.
Even though we derived the model for the flux as a purely linear operation on the spectral/spatial expansion of the stellar surface (Equation~\ref{eq:deconv:F}), it is more  efficient in practice to compute it as a convolution using the highly optimized \texttt{conv2d} operations in \texttt{theano}/\texttt{aesara}.
Typical runtimes and scalings for the \starry algorithm are shown in Figure~\ref{fig:runtime}, where we plot the evaluation of the full forward model for all spectra in a dataset as a function of several parameters.
Equation~(\ref{eq:deconv:F}) requires first the rotation of the $(\lmax + 1)^2$ $\kT$ kernels (each of length $W$) to the correct stellar rotational phase, an operation that scales approximately as $W \lmax^3$, followed by $(\lmax + 1)^2$ linear convolutions of the rotated $\kT$ kernels with the spectral components (each of length $K$), which scales approximately as $W K \lmax^2$. 
Since typically $K \gg \lmax$, the latter computation dominates and the entire evaluation of the model for $M$ spectral epochs takes $\mathcal{O}(M W K \lmax^2)$ operations. 
This is consistent with the relationships shown in Figure~\ref{fig:runtime}: the runtime is linearly proportional to the number of spectral epochs $M$, the number of wavelength bins $K$, and the width of the convolution kernel $W \propto v\sin i$; and it scales quadratically with the degree of the spherical harmonic expansion $l_\mathrm{max}$.
In general, evalution of the model for a typical dataset takes on the order of 10 or a few tens of milliseconds.

\section{Conclusions}
\label{sec:conclusions}

We have derived a novel method to infer the surface brightness maps of stars from high resolution spectroscopic timeseries data.
Our method is based on that of \citet{Vogt1987}, but it explicitly exposes the linear dependence of the observed spectrum on both the representation of the surface map (expressed as a vector of spherical harmonic coefficients) and the template (rest frame) spectrum, making it fast, differentiable, and, in certain cases, trivially invertible.
It also naturally extends to stellar surfaces with multiple spectral components, such as a star whose spectrum differs between spot and photosphere.
Ours is fundamentally a data-driven approach to the problem of Doppler imaging, where one can not only impose knowledge of the template spectrum but also \emph{learn} it from the data.
This is particularly useful in the case where reliable spectral templates are not available, such as in certain wavelength regimes or for certain objects, like the brown dwarf considered in \S\ref{sec:luhman16b}.
Importantly, we derive a model for the \emph{entire observed spectrum}.
Most Doppler imaging studies rely on techniques like least squares deconvolution \citep[LSD;][]{Donati1997} to infer a mean spectral line shape at each epoch, which is then treated as the input to the Doppler imaging step.
These studies usually focus on a few distinct, unblended, and well-understood spectral lines, which are transformed into a single mean profile evaluated over a relatively small number of wavelength bins.
Our approach, however, is to model the entire observed spectrum at full resolution, including line blends and wavelength regimes where the template spectrum may be unreliable.
This is made computationally possible given our highly optimized \texttt{Python} implementation within the \starry stellar modeling framework \citep{Luger2019}, which is openly available on GitHub with ample documentation.
By using the entire dataset and learning the template spectrum in regions where it is unavailable, we are able to maximize the information content of the inputs.
Our approach therefore has potential applications to slow rotators, low signal-to-noise spectra, and spectra of objects, such as brown dwarfs and directly imaged exoplanets, whose templates are not well known \emph{a priori}.

As a proof-of-concept, we qualitatively reproduced the Doppler map of the brown dwarf WISE 1049-5319B (Luhman 16B) inferred in \citet{Crossfield2014} from VLT/CRIRES observations; see Figure~\ref{fig:luhman16b:maps}.
We also tested our model's ability to accurately infer the surface maps and rest frame spectra from synthetic high SNR spectral timeseries observations, even in the limit of minimally informative priors on both quantities. 
Finally, we showed that we can infer the spectrum within spots when it is significantly different from that of the surrounding photosphere, again in the limit of high SNR and uninformative priors.

Given the linearity of our model, under certain choices of prior one can obtain the posterior distribution of surface maps analytically and in under a second for most datasets. 
However, given the usual uncertainty on stellar properties such as the inclination angle, the equatorial velocity, and the limb darkening profile, as well as the common need for restrictive priors on the surface map, most realistic applications of Doppler imaging require the use of computationally expensive numerical methods to infer the posterior distribution of the parameters.
For this reason, Doppler imaging is traditionally performed in an optimization setting: most studies thus far yield only point estimates of the inferred surface maps, with no formal uncertainties.
As we argued in \S\ref{sec:discussion:priors:mcmc}, this can lead to significant biases when these maps are interpreted to infer higher-level properties of stars, such as the typical sizes and active latitudes of star spots, or the average spot coverage fraction.
These biases are due to both the use of shrinkage priors (such as the commonly used maximum entropy constraint) and the fact that point estimates usually include spurious artifacts (like latititudinally elongated features) while suppressing certain surface modes (like azimuthally-symmetric features).
To accurately infer spot properties from Doppler maps, one must treat the problem probabilistically, by explicitly marginalizing over unknown quantities to obtain a posterior \emph{distribution} over the quantities of interest.
The model presented here, as well as its implementation in \starry, is easy to use in a posterior inference setting.
In particular, our implementation harnesses the autodifferentation capabilities of the \texttt{theano}/\texttt{aesara} \citep{Bergstra2010,Willard2021} package within the \texttt{pymc3} \citep{Salvatier2016} probabilistic modeling framework, allowing for efficient posterior inference using Hamiltonian Monte Carlo (HMC) techniques.
We also present a sketch of how to incorporate our model into an interpretable, physically-motivated Gaussian process (GP) formalism for stellar spectroscopic timeseries datasets, a method that is highly relevant to extreme precision radial velocity (EPRV) searches for exoplanets. We will explore this application in more detail in the next paper in this series.

While Doppler imaging has classically been used to infer the surfaces of giant or luminous pre-main sequence stars, the study of \citet{Crossfield2014} used the method to infer the first Doppler map of a brown dwarf, which we reproduced in this work.
Future studies could help us understand the intricate atmospheric dynamics of brown dwarfs \citep{Tan2021a,Tan2021b}, as well as the global properties of exoplanets, which may be amenable to Doppler imaging given the advent of a new generation of extremely large ground-based telescopes \citep{Crossfield2014b}.
A likely first target will be the directly-imaged giant planet $\beta$ Pictoris b, which is only an order of magnitude dimmer than WISE 1049-5319B; \citet{Snellen2014} showed that future observations with the planned Extremely Large Telecope (ELT) should be able to produce a global map of its surface.
Given the expected low SNR of such observations and the considerable uncertainty in the rest-frame spectra of directly imaged exoplanets, an efficient, probabilistic Doppler imaging approach like the one developed in this work will be imperative to maximizing the scientific output of those studies.

\vspace{1em}

This paper was prepared using the \showyourwork open source scientific article workflow\footnote{\url{https://github.com/rodluger/showyourwork}}.
The PDF was generated automatically from a GitHub repository\footnote{\url{https://github.com/rodluger/paparazzi}} using a frozen \texttt{conda} environment on GitHub Actions CI\footnote{\url{https://github.com/features/actions}}, ensuring that all the results are exactly reproducible and easily extensible with minimal effort from the user.
The algorithm derived in this paper is fully implemented in the \starry \texttt{Python} package, which is installable via \texttt{pip} or directly from GitHub\footnote{\url{https://github.com/rodluger/starry}}, with ample documentation, examples, and tutorials.\footnote{\url{https://starry.readthedocs.io}}

\vspace{1em}

We would like to thank Eric Agol, Christina Hedges, Rachael Roettenbacher, and the Astronomical Data Group at the Center for Computational Astrophysics (CCA) for useful comments, suggestions, and insights.
RL and LLZ are supported by Flatiron Fellowships at the CCA, a division of the Simons Foundation.

\clearpage
\appendix

\section{Solving the $\kT$ integral}
\label{app:kT}
In this section we will show how to compute the integral in Equation~(\ref{eq:kT:kT}), yielding the terms of the convolution kernel $\kT$. 
It is convenient to first change basis from spherical harmonics to polynomials on the sphere:
\begin{align}
    \label{eq:appendix:kT}
    \kT(\lnlam)
     & =
    \int\limits_{-\sqrt{1 - x^2}}^
    {\sqrt{1 - x^2}}
    \pbasis
    \Big(x, y\Big)
    \mathrm{d}y
    \,
    \AOne
    \nonumber \\[0.5em]
     & \equiv
    \rhoT(\lnlam)
    \,
    \AOne
    \quad,
\end{align}
where $\AOne$ is the change of basis matrix from spherical harmonics to polynomials \citep[Equation B11 in][]{Luger2019} and
\begin{align}
    \pbasis(\x) \equiv
    \Big(
    1 \quad\quad\quad\quad\quad\quad x \quad\quad\quad\quad\quad\quad z \quad\quad\quad\quad\quad\quad y \quad\quad\quad\quad\quad\quad x^2 \quad\quad\quad\quad\quad\quad xz \quad\quad\quad\quad\quad\quad xy \quad\quad\quad\quad\quad\quad yz \quad\quad\quad\quad\quad\quad y^2 \quad\quad\quad\quad\quad\quad
    ...
    \Big)^\top
    \quad,
\end{align}
is the polynomial basis \citep[Equation 7][]{Luger2019}.
The $n^\mathrm{th}$ term of the polynomial basis is a polynomial in $x$, $y$, and $z = \sqrt{1 - x^2 - y^2}$:
\begin{align}
    \pbasisn
    &=
    x^i y^j z^k
\end{align}
where $i, j, k$ are integers given by
\begin{align}
    \label{eq:kT:lm}
    i &= \floor*{\Lambda - \Delta}
    \nonumber \\[0.5em]
    j &= \floor*{\Delta}
    \nonumber \\[0.5em]
    k &= \ceil*{\Delta} - \floor*{\Delta}
\end{align}
with
\begin{align}
    \Lambda &= \floor*{\sqrt{n}}
    \nonumber \\[0.5em]
    \Delta &= \frac{n - \Lambda^2}{2}
    \quad .
\end{align}
We may now explicitly write down the $n^\mathrm{th}$ term of the vector $\rhoT(\lnlam)$ in Equation~(\ref{eq:appendix:kT}),
\begin{align}
    \label{eq:kT:sTexpression}
    \rho_n(\lnlam)
     & =
    \rho_{i,\,j,\,k}(\lnlam)
    =
    x^i
    \int\limits_{-\sqrt{1 - x^2}}^
    {\sqrt{1 - x^2}}
    y^j
    \left(1 - x^2 - y^2\right)^{\frac{k}{2}} \,
    \mathrm{d}y
    \quad .
\end{align}
When $i = k = 0$, the integral is easy to evaluate:
\begin{align}
    \rho_{i,\,j,\,0}(\lnlam_0)
     & =
    \begin{cases}
        2 \left( 1 - x^2 \right)^\frac{j + 1}{2}
        \quad\quad\quad\quad\quad\quad\quad\quad\quad\quad\quad\quad
          & j \, \mathrm{even}        \\
        0 & j \, \mathrm{odd} \quad .
    \end{cases}
\end{align}
The general term may be computed from the recurrence relation
\begin{align}
    \rho_{0,\,j,\,0}(\lnlam_0) = \frac{j - 1}{j + 1} \big(1 - x^2\big) \rho_{0,\,j-2,\,0}
\end{align}
provided
\begin{align}
    \rho_{0,\,0,\,0} & = 2 \sqrt{1-x^2} \nonumber \\
    \rho_{0,\,1,\,0} & = 0 \quad.
\end{align}
When $i = 0$ and $k = 1$, we may substitute $y = \sin\psi\sqrt{1 - x^2}$ in the integrand to obtain
\begin{align}
    \rho_{0,\,j,\,1}(\lnlam_0)
     & =
    (1 - x^2)^{\frac{j + 2}{2}}
    \int\limits_{-\frac{\pi}{2}}^{\frac{\pi}{2}}
    \sin^j\psi
    \cos^2\psi \,
    \mathrm{d}\psi
    \quad.
\end{align}
We may solve this integral using the recurrence relation
\begin{align}
    \int
    \sin^j\psi
    \cos^m\psi \,
    \mathrm{d}\psi
     & =
    -\frac{\sin^{j-1}\psi \cos^{m+1}\psi}{j + m}
    +
    \frac{j - 1}{j + m}\int\sin^{j-2}\psi \cos^m\psi \mathrm{d}\psi
    \quad ,
\end{align}
which, in our case, simplifies to
\begin{align}
    \int\limits_{-\frac{\pi}{2}}^{\frac{\pi}{2}}
    \sin^j\psi
    \cos^2\psi \,
    \mathrm{d}\psi
     & =
    \frac{j - 1}{j + 2}\int\limits_{-\frac{\pi}{2}}^
    {\frac{\pi}{2}}\sin^{j-2}\psi \cos^2\psi \mathrm{d}\psi
    \quad.
\end{align}
We thus arrive at the recurrence relation
\begin{align}
    \rho_{0,\,j,\,1} & = \frac{j - 1}{j + 2} \big(1 - x^2\big) \rho_{0,\,j-2,\,1}
\end{align}
with initial conditions
\begin{align}
    \rho_{0,\,0,\,1} & = \frac{\pi}{2} \big(1-x^2\big) \nonumber \\
    \rho_{0,\,1,\,1} & = 0 \quad.
\end{align}
Given these expressions, recursing upward in $i$ is easy:
\begin{align}
    \rho_{i,\,j,\,k} & = x \, \rho_{i-1,\,j,\,k} \quad.
\end{align}
This completes our derivation. 
To summarize, given the boundary conditions
\begin{align}
    \rho_{0,\,0,\,0} & = 2 \sqrt{1-x^2} \nonumber                \\
    \rho_{0,\,0,\,1} & = \frac{\pi}{2} \big(1-x^2\big) \nonumber \\
    \rho_{0,\,1,\,k} & = 0 \quad ,
\end{align}
we may compute any term in $\rhoT$ via the expressions
\begin{align}
    \label{eq:kT:sTrecurrence}
    \rho_{0,\,j,\,k} & = \frac{j - 1}{j + 1 + k} \big(1 - x^2\big) \rho_{0,\,j-2,\,k} \nonumber \\
    \rho_{i,\,j,\,k} & = x \, \rho_{i-1,\,j,\,k} \quad.
\end{align}
The terms of the convolution kernel $\kT$ are then obtained by dotting this vector into the change of basis matrix $\AOne$, as discussed above.

\clearpage
\begin{center}
    \begin{longtable}{cll}
        \caption{Common notation used in this paper}
        \label{tab:notation}                                                                                                                                            \\
        \toprule
        \multicolumn{1}{c}{\textbf{Symbol}}                 &
        \multicolumn{1}{c}{\textbf{Description}}            &
        \multicolumn{1}{c}{\textbf{Reference}}                                                                                                                          \\
        \midrule
        \endfirsthead
        \multicolumn{3}{c}%
        {{\bfseries \tablename\ \thetable{}}. (continued from previous page)}                                                                                           \\[0.5em]
        \toprule
        \multicolumn{1}{c}{\textbf{Symbol}}                 &
        \multicolumn{1}{c}{\textbf{Definition}}             &
        \multicolumn{1}{c}{\textbf{Reference}}                                                                                                                          \\
        \midrule
        \endhead
        \bottomrule
        \endfoot
        \endlastfoot
        \midrule
        \multicolumn{3}{c}{\emph{Variables}}                                                                                                                            \\
        \midrule
        $a_{lm}$                                            & spectral spherical harmonic coefficient                      & Equation~(\ref{eq:deconv:Ixi0})            \\
        $\almt$                                              & vector of spectral spherical harmonic coefficients          & Equation~(\ref{eq:deconv:almt})             \\
        $\bvec{A}$                                          & matrix form of $\bvec{a}$                                    & Equation~(\ref{eq:inverse:alm})          \\
        $\bvec{B}$                                          & baseline design matrix                                       & \S\ref{sec:norm}                 \\
        $c$                                                 & speed of light                                               & Equation~(\ref{eq:kT:beta})                \\
        $\Curve$                                            & curve of constant radial velocity                            & Equation~(\ref{eq:deconv:kT})              \\
        $\D$                                                & logarithmic Doppler shift                                    & Equation~(\ref{eq:the_problem:D})          \\
        $\Doppler$                                          & Doppler design matrix                                        & Equation~(\ref{eq:linear:f})               \\
        $\bvec{f}$                                          & vector of observed flux values                               & Equation~(\ref{eq:linear:f})               \\
        $F$                                                 & flux                                                         & Equation~(\ref{eq:the_problem:F})          \\
        $i$                                                 & inclination                                                  & Equation~(\ref{eq:kT:beta})                \\
        $I$                                                 & specific intensity                                           & Equation~(\ref{eq:the_problem:Ixi})        \\
        $\bvec{I}$                                          & identity matrix                                              &                                            \\
        $l$                                                 & spherical harmonic degree                                    &                                            \\
        $\bvec{L}$                                          & limb darkening operator                                      & Equation~(\ref{eq:ld:L})                   \\
        $m$                                                 & spherical harmonic order                                     &                                            \\
        $\bvec{P}$                                          & Change of basis matrix from $Y_{lm}$ to pixels               & Equation~(\ref{eq:spotstar:uncert})        \\
        $\bvec{R}$                                          & Wigner rotation matrix                                       & Equation~(\ref{eq:deconv:R})               \\
        $\bvec{s}$                                          & vector representation of the spectrum                        & Equation~(\ref{eq:inverse:alm})          \\
        $\bvec{S}$                                          & block matrix representation of $\bvec{s}$                    & Equation~(\ref{eq:inverse:S})              \\
        $\boldsymbol{\mathbb{S}}$                           & matrix of $J$ spectral components                            & Equation~(\ref{eq:eigen:alm})            \\
        $\Surf$                                             & projected visible area of the star                           & Equation~(\ref{eq:the_problem:F})          \\
        $t$                                                 & time                                                         & \S\ref{sec:the_problem}                    \\
        $T$                                                 & artificial temperature in bilinear solver                    & \S\ref{sec:spotstar}                       \\
        $T_0$                                               & initial artificial temperature setting                       & \S\ref{sec:spotstar}                       \\
        $\Delta\log T$                                      & artificial temperature step size                             & \S\ref{sec:spotstar}                       \\
        $u$                                                 & limb darkening coefficient                                   & \S\ref{sec:ld}                             \\
        $\x$                                                & Cartesian position on the sky                                & \S\ref{sec:the_problem}                    \\
        $y_{lm}$                                            & spherical harmonic coefficient                               & \S\ref{sec:inverse}                        \\
        $\bvec{y}$                                          & vector of spherical harmonic coefficients                    & Equation~(\ref{eq:inverse:alm})          \\
        $Y_{lm}$                                            & spherical harmonic                                           & Equation~(\ref{eq:deconv:Ixi0})            \\
        $\bvec{Y}$                                          & block matrix representation of $\bvec{y}$                    & Equation~(\ref{eq:inverse:Y})              \\
        $\boldsymbol{\mathbb{Y}}$                           & matrix of $J$ spherical harmonic components                  & Equation~(\ref{eq:eigen:alm})            \\
        \pagebreak
        \midrule
        \multicolumn{3}{c}{\emph{Variables (Greek)}}                                                                                                                    \\
        \midrule
        $\beta$                                             & relativistic parameter                                       & Equation~(\ref{eq:the_problem:D})          \\
        $\delta$                                            & Dirac delta function                                         & Equation~(\ref{eq:deconv:convolution})     \\
        $\kT$                                               & convolution kernel basis vector                              & Equation~(\ref{eq:deconv:kT})              \\
        $\lambda$                                           & wavelength                                                   & \S\ref{sec:the_problem}                    \\
        $\lambda_\mathrm{r}$                                & reference wavelength                                         & \S\ref{sec:the_problem}                    \\
        $\lnlam$                                            & $\ln\frac{\lambda}{\lambda_\mathrm{r}}$ (observer frame)     & \S\ref{sec:the_problem}                    \\
        $\lnlam_0$                                          & $\ln\frac{\lambda}{\lambda_\mathrm{r}}$ (rest frame)         & \S\ref{sec:the_problem}                    \\
        $\boldsymbol{\Lambda}$                              & prior covariance matrix                                      &                                            \\
        $\boldsymbol{\mu}$                                  & prior mean vector                                            &                                            \\
        $\ylmbasis$                                         & spherical harmonic basis vector                              & Equation~(\ref{eq:deconv:ylmbasis})        \\
        $\boldsymbol{\Sigma}$                               & covariance matrix                                            &                                            \\
        $\omega$                                            & angular velocity                                             & Equation~(\ref{eq:kT:beta})                \\
        \midrule
        \multicolumn{3}{c}{\emph{Dimensions}}                                                                                                                           \\
        \midrule
        $J$                                                 & \# of spectral components                                    & \S\ref{sec:eigen}                          \\
        $K$                                                 & \# of wavelength bins in the broadened spectrum              & \S\ref{sec:linear}                         \\
        $K'$                                                & \# of wavelength bins in the latent spectrum                 & \S\ref{sec:linear}                         \\
        $K_\mathrm{obs}$                                    & \# of wavelength bins in the observed spectrum               & \S\ref{sec:spotstar}                       \\
        $M$                                                 & \# of epochs observed                                        & \S\ref{sec:linear}                         \\
        $N$                                                 & \# of spherical harmonic coefficients                        & \S\ref{sec:linear}                         \\
        $N'$                                                & \# of spherical harmonic coefficients after limb darkening   & \S\ref{sec:ld}                             \\
        $W$                                                 & \# of wavelength bins in the convolution kernel              & \S\ref{sec:linear}                         \\
        \midrule
        \multicolumn{3}{c}{\emph{Operators \& Other symbols}}                                                                                                           \\
        \midrule
        $\mathrm{diag}$                                     & diagonal matricization                                       & Equation~(\ref{eq:ld:Dmwav})               \\
        $\mathrm{vec}$                                      & vectorization                                                & Equation~(\ref{eq:inverse:alm})          \\
        $*$                                                 & convolution                                                  & Equation~(\ref{eq:deconv:convolution_def}) \\
        $\circ$                                             & element-wise product                                         & \S\ref{sec:norm}           \\
        $\otimes$                                           & Kronecker product                                            & Equation~(\ref{eq:linear:Dm})              \\
        $\ceil*{\,}$                                        & ceiling                                                      & Equation~(\ref{eq:kT:lm})                  \\
        $\floor*{\,}$                                       & floor                                                        & Equation~(\ref{eq:kT:lm})                  \\
        $\bvec{0}$                                          & the zero vector, $(0 \quadquad 0 \quadquad \cdots \quadquad 0)^\top$ &                                                                                                           \\
        $\bvec{1}$                                          & the one vector, $(1 \quadquad 1 \quadquad \cdots \quadquad 1)^\top$ &                                                                                                           \\
    \end{longtable}
\end{center}

\clearpage
\bibliography{bib}

\end{document}

%% file: style.tex
\usepackage{xifthen}
\usepackage{array}
\usepackage{upgreek}
\usepackage[bbgreekl]{mathbbol}
\usepackage{afterpage}
\usepackage[bb=boondox]{mathalpha}
\usepackage{tipa}
\usepackage{booktabs}

\newcommand{\noop}[1]{}

\newcommand{\starry}{\textsf{starry}\xspace}

\newcommand{\bvec}[1]{{\ensuremath{\mathbf{#1}}}}

\DeclarePairedDelimiter\floor{\lfloor}{\rfloor}
\DeclarePairedDelimiter\ceil{\lceil}{\rceil}

\renewcommand{\quad}{\hskip 0.33em}
\newcommand{\quadquad}{\quad\quad\quad\quad}
\newcommand{\R}{\bvec{R}}
\newcommand{\AOne}{\bvec{A_1}}
\newcommand{\alm}{\bvec{a}}
\newcommand{\x}{\bvec{x}}
\newcommand{\D}{D}
\newcommand{\Doppler}{\bvec{D}}
\newcommand{\Surf}{\mathcal{S}}
\newcommand{\Curve}{\mathcal{C}}

\newcommand{\lmax}{\ensuremath{l_\mathrm{max}}}
\newcommand{\spot}{\texttt{SPOT}\xspace}

\newcommand{\kT}{\boldsymbol{\kappa}^\top}
\newcommand{\rhoT}{\boldsymbol{\rho}^\top}
\newcommand{\ylmbasis}{\boldsymbol{\psi}^\top}
\newcommand{\pbasis}{\boldsymbol{\phi}^\top}
\newcommand{\pbasisn}{\ensuremath{\phi_n}}
\newcommand{\almt}{\ensuremath{\bvec{a}}}
\newcommand{\lnlam}{\mbox{\textipa{\textcrlambda}}}
\newcommand{\fnorm}{\tilde{\bvec{f}\hspace{0.2em}}}